\documentclass[11pt]{article}

\usepackage[T1]{fontenc}
\usepackage[utf8]{inputenc}
\usepackage{arxiv}
\usepackage{newtxtext,newtxmath}
\usepackage{graphicx}
\usepackage{amsmath,bm}
\usepackage{booktabs,tabularx}
\usepackage{placeins}
\usepackage{float}
\usepackage{xcolor}
\usepackage{microtype}
\usepackage{authblk}
\usepackage{orcidlink}
\usepackage[authoryear,round]{natbib}
\usepackage{url}
\usepackage{hyperref}

\makeatletter
\renewcommand{\fnum@figure}{\textbf{Figure \thefigure}}
\renewcommand{\fnum@table}{\textbf{Table \thetable}}
\makeatother

\hypersetup{
  colorlinks=true,
  linkcolor={red!50!black},
  citecolor={blue!50!black},
  urlcolor={blue!80!black},
  pdftitle={Coupled multiscale paleoclimate reconstruction with four-dimensional variational data assimilation},
  pdfauthor={Zilu Meng, Gregory J. Hakim, Julien Emile-Geay, Tanaya Gondhalekar, and Eric J. Steig}
}

\let\cite\citep
\graphicspath{{figures/}}

\newcolumntype{Y}{>{\raggedright\arraybackslash}X}

\newcommand{\state}{\mathbf{x}}
\newcommand{\obs}{\mathbf{y}}

\newcommand{\Qk}{\mathbf{Q}}
\newcommand{\Rk}{\mathbf{R}}
\newcommand{\sisection}[2]{\section{#1}\label{#2}}

\renewcommand{\headeright}{Meng et al. 2026}
\renewcommand{\undertitle}{}
\renewcommand{\shorttitle}{LMR4D-Var}

\title{Coupled multiscale paleoclimate reconstruction with four-dimensional variational data assimilation}
\author[1,*]{Zilu~Meng\,\orcidlink{0000-0001-5706-592X}}
\author[1]{Gregory~J.~Hakim\,\orcidlink{0000-0001-8486-9739}}
\author[2]{Julien~Emile-Geay\,\orcidlink{0000-0001-5920-4751}}
\author[2]{Tanaya~Gondhalekar}
\author[3,1]{Eric~J.~Steig\,\orcidlink{0000-0002-8191-5549}}
\affil[1]{Department of Atmospheric and Climate Science, University of Washington, Seattle, WA 98195, USA}
\affil[2]{Department of Earth Sciences, University of Southern California, Los Angeles, CA 90089, USA}
\affil[3]{Department of Earth and Space Sciences, University of Washington, Seattle, WA 98195, USA}
\affil[*]{Corresponding author: zilumeng@uw.edu}
\date{November 12, 2026}

\usepackage{ragged2e}
\usepackage[export]{adjustbox}
\let\citeA\citet
\renewcommand{\theHfigure}{main.\arabic{figure}}
\renewcommand{\theHtable}{main.\arabic{table}}

\begin{document}
\maketitle
\begin{abstract}
Paleoclimate archives extend climate knowledge beyond the instrumental era, registering different seasons, variables, time averages, and memory lengths. A longstanding problem is to integrate these heterogeneous sources of information within a unified methodology. Here we present a new data-assimilation framework, Last Millennium Reanalysis 4D-Var (LMR4D-Var), which reconstructs climate trajectories from these heterogeneous datasets while balancing errors in the model, observations, and initial conditions. We compare results using LMR4D-Var to assimilate proxies from PAGES2k, Temp12k, and borehole temperature profiles without treating them as instantaneous equivalents. Instrumental validation demonstrates that LMR4D-Var achieves the highest skill in reconstructing global mean temperature among existing reconstructions. Borehole assimilation preserves skill against withheld annually resolved records, increases agreement between reconstructed 300--2000-m ocean heat content (OHC) and independent estimates, and yields a cooler reconstructed Little Ice Age ocean. Modern 130-year OHC trends in the 0--300-m and 300--2000-m layers exceed their respective reconstructed pre-1870 Common Era trend distributions.
Results for Temp12k demonstrate assimilation of decadal-to-millennial records and the potential for Holocene and deeper-time applications with suitable emulators.

\end{abstract}

\section*{Plain Language Summary}
Evidence of past climate comes from natural archives that record different aspects of climate over very different periods. Tree rings record individual seasons or years, while sediments can record averages over decades to centuries. Temperatures measured in boreholes retain a memory of past surface temperatures. Methods that reconstruct climate by moving forward through time without revisiting earlier estimates cannot use later observations to revise those estimates. This limits their ability to use data, such as borehole temperature profiles, that simultaneously contain information about present conditions and preceding decades or centuries. We developed LMR4D-Var to reconstruct an entire climate history jointly, combining observations with a simplified climate model while accounting for uncertainty in both. The method combines tree-ring, sediment, and borehole records to reconstruct seasonal changes in the atmosphere, ocean, and sea ice. Adding boreholes improves estimates of slowly changing deeper-ocean heat storage without reducing agreement with records that resolve individual years. In our reconstruction, ocean warming from 1870 to 2000 was faster than during any equally long interval between 500 BCE and 1869 CE. Experiments with older records suggest that this approach could reconstruct climate over the past 12,000 years with models suited to those conditions.

\section{Introduction}
\label{sec:introduction}

Instrumental observations of recent climate change in the atmosphere and the ocean are too short to define the full range of natural variability and distinguish it from forced climate variability \cite{IPCC_AR6_SYR_SPM_2023}. Paleoclimate archives provide indirect climate information, but different types of archives record different aspects of the climate system, and are relevant to different timescales. For example, tree rings and corals \cite{PAGES2k2017global} resolve annual or seasonal temperature and precipitation, many sediment and pollen records \cite{kaufman2020global} resolve temperature changes averaged over decades or longer, and borehole temperature profiles contain a diffused memory of past surface temperature changes over a wide range of time scales as a function of depth \cite{bradley1999paleoclimatology,PollackHuang:2000,national2007surface,Jones_Holocene09,christiansen2017challenges,IPCC_AR6_SYR_SPM_2023}. A central challenge in climate reconstruction is to combine this information that is heterogeneous in space, time, and climate variable, while accounting for errors in all sources \cite{tierney2025advances,zhang2025paleoclimate}. An additional complication is that many archives integrate climate information over time, so an observation may constrain an extended history rather than a climate state at a single point in time. If the temporal averaging or memory represented by these records is ignored, the reconstruction may misrepresent climate variability across different timescales.

Recently, paleoclimate data assimilation has emerged as a promising framework for climate reconstruction \cite{hakim2016last,tierney2025advances,ning2026progress}. Most paleoclimate data-assimilation reconstructions use time-local or sequential filtering updates \cite{hakim2016last,tardif2019last,franke2017monthly,Tierney:Nature2020,osman2021globally,vallerUpdatedGlobalAtmospheric2022,erb2022reconstructing,valler2024mode}. ``Offline'' approaches estimate each analysis time without dynamically propagating the preceding analyzed state \cite{okazaki2021revisiting,Emile-Geay:2025,sun2022analog,sun2024hybrid}, while ``online'' approaches instead propagate the analyzed state forward with a climate model or emulator, carrying dynamical memory between successive analyses \cite{perkins2017reconstructing,okazaki2021revisiting,meng2024reconstructing,sun2025online,meng2025coupled}. In sequential filtering, observations are assimilated in chronological order, and the state at each analysis time is estimated using observations up to that time. Later observations do not revise earlier analyzed states \cite{sun2026ensemble}. Smoothing instead uses observations from the full assimilation window to revise past states, making it well suited to retrospective paleoclimate reconstruction. Four-dimensional variational data assimilation (4D-Var) provides a variational formulation of smoothing. This distinction is particularly important for temporally nonlocal archives: for example, borehole profiles \cite{huang2000temperature} encode a diffusive memory of centuries of surface-temperature history, while sediment and pollen records commonly constrain multidecadal to centennial averages \cite{kaufman2020global}. Treating such records as updates at a single analysis time discards part of their temporal information and makes it difficult to combine their low-frequency constraints with the short-timescale information carried by annually resolved records.

This limitation is especially consequential for slowly responding components of the climate system. Global mean temperature (GMT) is already difficult to reconstruct prior to the instrumental era because paleoclimate observations (``proxies'') are unevenly distributed in space and time, physically selective, and are sparse, especially in many oceanic and high-latitude regions \cite{Jones_Holocene09,PAGES2k2017global}. Ocean heat content (OHC) provides an even greater challenge because it integrates energy uptake over decades to centuries while ocean circulation redistributes heat. Modern ocean reconstructions and energy-budget estimates constrain recent changes \cite{zannaGlobalReconstructionHistorical2019a,ishii2017accuracy,wu2025energybudget}, but the instrumental interval is too short to establish the full Common Era distribution of ocean heat storage, and multicentennial estimates remain sensitive to assumptions about boundary forcing and ocean circulation \cite{gebbie2019little}. A central unresolved question is therefore whether the modern rate of ocean heat accumulation is distinguishable from OHC changes over comparable intervals before the instrumental era. Addressing this question requires a layer-resolved reconstruction long enough to define historical trend distributions. Such comparisons must also recognize the strong temporal persistence of OHC and changes in observational constraints. Moreover, most paleoclimate proxies respond primarily to surface conditions rather than to subsurface ocean temperature. Reconstructing OHC separately for the 0--300-m and 300--2000-m layers therefore provides a stringent test of whether surface-sensitive archives can constrain slow ocean heat storage through an evolving coupled climate state. Such an OHC reconstruction may be viewed as a dynamically constrained estimate rather than as directly constrained by observations.

\citeA{steiger2016multitimescale} demonstrated that assimilating proxy records at multiple timescales can improve the reconstruction of low-frequency climate variability. However, extending ensemble-based assimilation to full climate trajectories requires estimating cross-time covariances, which can be difficult with finite ensembles \cite{anderson2012localization}. Therefore, we formulate paleoclimate reconstruction as a trajectory-smoothing problem using 4D-Var, in which the full climate trajectory is estimated conditional on all observations within the assimilation window \cite{LeDimetTalagrand1986,TalagrandCourtier1987,fisherKalmanSmoothing2005,evensen2022data}. Each observation operator acts on the portion of the trajectory represented by an archive's seasonality, averaging interval, or physical memory, and observations later in calendar time can revise earlier states through the modeled temporal relationships. This formulation allows annually resolved records to constrain short-timescale variability while low-resolution and long-memory archives constrain slower changes within the same evolving climate trajectory, without forcing these distinct observations to behave as instantaneous equivalents.
Rather than analyzing each observation time-independently, 4D-Var determines the model trajectory that best fits observations distributed across the entire assimilation window \cite{LeDimetTalagrand1986,TalagrandCourtier1987,Courtier1994,evensen2022data}. ``Strong-constraint'' 4D-Var assumes that the model is perfect, so that the best-fit trajectory is defined by the initial condition \cite{Lorenc2003}. ``Weak-constraint'' 4D-Var, which we use here, instead estimates model-error increments together with the initial state, allowing observations to correct an imperfectly modeled trajectory \cite{tremoletAccountingImperfectModel2006,evensen2022data}. The variational objective explicitly balances uncertainty in the initial condition, departures from the model trajectory, and mismatch to the observations. The associated error covariances determine the relative influence of these three terms, allowing proxy records with different error characteristics to be weighted within a common optimization. This distinction is important for paleoclimate applications because simplified climate models and long climate-model integrations can exhibit drift, structural bias, and model-dependent low-frequency variability, particularly in the ocean \cite{o2016scenario}.

We develop Last Millennium Reanalysis 4D-Var (\texttt{LMR4D-Var}), a coupled weak-constraint framework designed to combine archives with different temporal resolutions and memory lengths. The present implementation uses a seasonal linear inverse model (LIM) to represent the coupled evolution of the atmosphere, ocean, and sea ice. The initial state and a sequence of model-error increments are optimized over the full assimilation window. Each proxy archive is connected to the candidate climate trajectory through an observation operator that represents what the archive records and over what interval. The framework combines PAGES2k records (version 2.0.0, see \cite{PAGES2k2017global}), many of which are seasonally or annually resolved, with Temp12k records \cite{kaufman2020global} spanning decadal to millennial resolution. Our framework also includes terrestrial borehole temperature profiles that integrate surface-temperature history through thermal diffusion \cite{cuestaValero2021longterm}.
In this way, each record constrains the climate trajectory according to its seasonal sensitivity, temporal averaging, or memory of past climate, rather than being treated as an instantaneous observation.
The differentiable implementation provides a practical route for adding nonlinear, time-averaged, or history-dependent proxy system models \cite{Evans_QSR13,dolman2018sedproxy}.

We evaluate LMR4D-Var using three complementary strategies: a controlled pseudo-proxy experiment, verification against withheld observations, and comparisons with external estimates of GMT and layer-resolved OHC. The pseudo-proxy experiment tests whether sparse, surface-sensitive observations can recover a known but sparsely sampled coupled climate trajectory, including OHC, that is not directly assimilated. Experiments with real proxies over the instrumental era are used to evaluate whether the reconstructed GMT and layer-resolved OHC are consistent with independent observational and reconstruction products. The withholding experiments test whether low-frequency archives add information without compromising skill against annually resolved records. Finally, Temp12k-only sensitivity experiments examine how the balance between observational constraints and an imperfect reduced-order prior changes over Holocene-length windows when the climate emulator is known to have missing physics. Together, this evaluation strategy distinguishes validation of the assimilation framework from evaluation of the resulting climate reconstruction and identifies where indirect OHC constraints and long-window applications require particular caution.

\section{Materials and Methods}
\label{sec:methods}
\subsection{Linear Inverse Model}
\label{sec:lim}

To make optimization over long assimilation windows computationally tractable, we use the LIM as a reduced-order climate emulator (Fig.~\ref{fig:framework}a). It retains the coupled covariance structure linking the atmosphere, upper ocean, and sea ice. LIMs are estimated directly from lagged covariance statistics of a training ensemble and provide a compact stochastic approximation to coupled climate dynamics \cite{penland1993prediction,penland1994balance,perkinsLinearInverseModeling2020}. In the present application, the LIM is not intended to replace a full climate model; rather, it provides the dynamically consistent backbone on which the weak-constraint reconstruction is built.

The reconstructed climate trajectory is represented in a reduced coupled state space derived from long climate-model integrations. Large-scale surface, sea-ice, and upper-ocean variables are projected onto a truncated empirical orthogonal function (EOF) basis using the ``Statistic Analysis for Climate data in Python" (SACPY) package \cite{meng2023sacpy}, and the time-evolving state is represented by the corresponding principal-component (PC) amplitudes. The reduced state includes PCs for 2-meter air temperature, sea-surface temperature, ocean heat content in the upper 300 m (OHC(0--300 m)), ocean heat content over 300--2000 m (OHC(300--2000 m)), and Northern Hemisphere sea-ice concentration and thickness. This reduced-order representation is designed to retain the dominant covariance structure of the coupled system while keeping the control dimension tractable over long assimilation windows. Separating OHC(0--300 m) and OHC(300--2000 m) allows the LIM to capture the different adjustment timescales of the upper and deeper ocean, which is important for reconstructing the slow evolution of ocean heat storage \cite{perkinsLinearInverseModeling2020}.

Previous work has shown the importance of leveraging several model priors \cite{parsons2021multi,stiller2026top}, so we train a separate LIM for each parent climate-model product. We use five parent products: the CESM Last Millennium Ensemble (LME) \cite{OttoBliesner_etal2016}, together with PMIP simulations from MPI-ESM-P (PMIP3) \cite{giorgetta2013mpi}, MPI-ESM1.2-LR (PMIP4) \cite{mauritsen2019mpi}, MRI-ESM2-0 (PMIP4) \cite{yukimoto2019mri}, and CCSM4 (PMIP3) \cite{landrum2013last}. Each of these products yields its own LIM, for a total of five LIM priors. This design broadens the range of coupled dynamics available to the reconstruction and reduces reliance on any single model's low-frequency behavior. Each LIM is trained on seasonal-mean anomalies relative to the corresponding seasonal climatology, with linear trends removed over the training interval. Long-term OHC drifts differ substantially among the parent simulations, whereas their detrended variability is more comparable (Fig.~\ref{si:fig:si_model_ohc_changes}). Linear detrending prevents unrealistically persistent drift or slow adjustment from initial-condition imbalance from being absorbed into the fitted linear dynamics. The LIM formulation, state-vector construction, detrending procedure, and training-data summary are given in SI Section~\ref{si:sec:si_lim_derivation} and Table~\ref{si:tab:si_models}.

The LIM is discretized on the same seasonal calendar used throughout the reconstruction, with four analysis steps per year corresponding to March--April--May (MAM), June--July--August (JJA), September--October--November (SON), and December--January--February (DJF). In discrete form,
\begin{equation}
\state_{k+1} = \mathbf{G}\state_k + \mathbf{n}_k,
\label{eq:lim_dynamics}
\end{equation}
where $\state_k$ is the reduced climate state at analysis step $k$, $\mathbf{G}$ is the LIM propagator, and $\mathbf{n}_k$ is an additive model-error increment. For each LIM, $\Qk$ is estimated from the stationary covariance balance \cite{penland1994balance} and defines the Gaussian prior $\mathbf{n}_k\sim\mathcal{N}(0,\Qk)$ (SI Section~\ref{si:sec:si_lim_derivation}). For each LIM, $\mathbf{G}$ and $\Qk$ are time independent and are used at every seasonal analysis step.

\subsection{Weak-constraint 4D-Var implementation}
\label{sec:wc4dvar}

\begin{figure}[!tbp]
\centering
\includegraphics[width=\textwidth,max height=0.65\textheight,keepaspectratio]{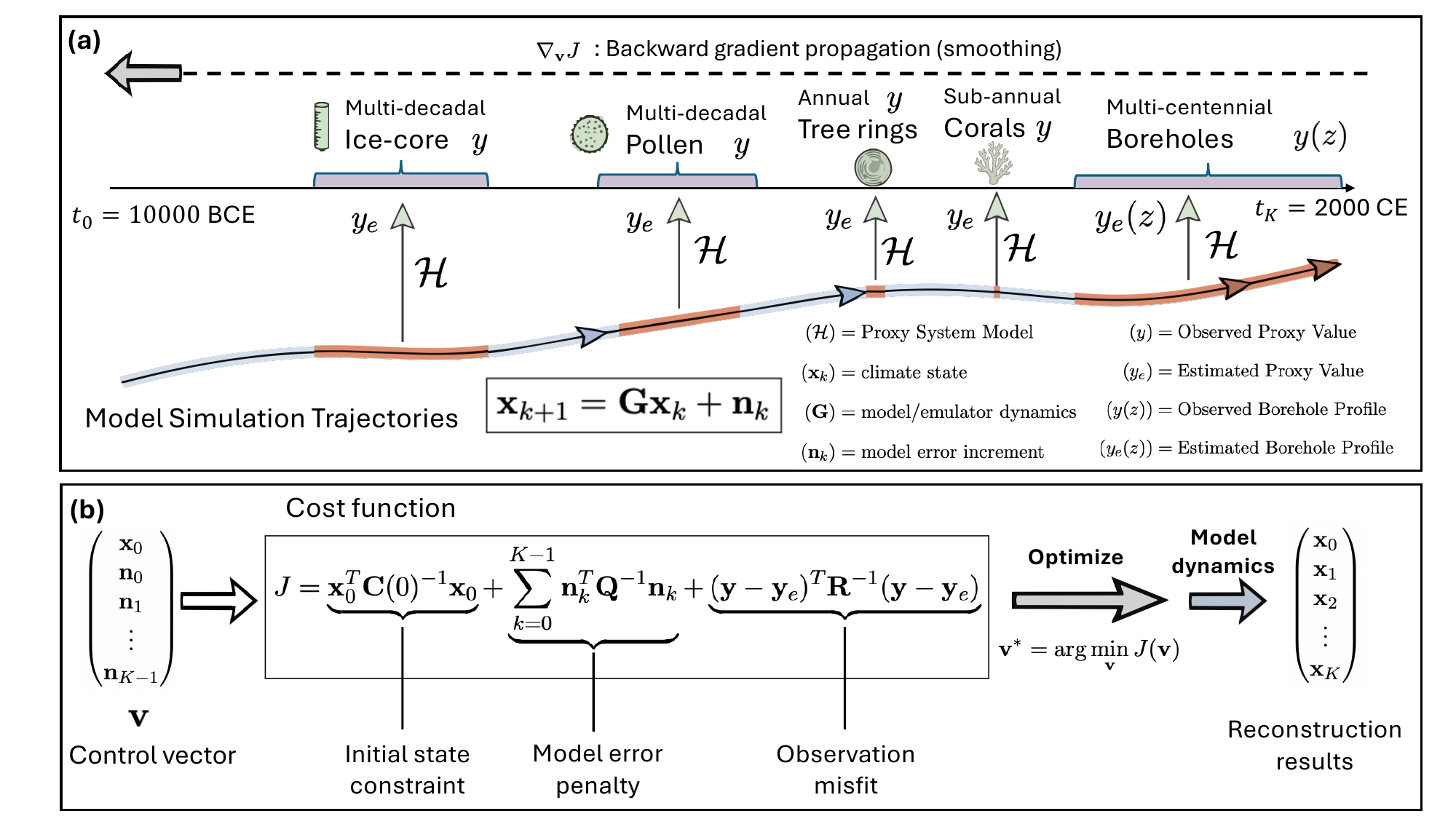}
\caption{\textbf{Multiscale weak-constraint 4D-Var framework for reconstructing past climate trajectories.} (a) A schematic assimilation window extends from 10{,}000 BCE to 2000 CE. The reduced-order climate model emulator propagates the trajectory according to $\state_{k+1}=\mathbf{G}\state_k+\mathbf{n}_k$, and archive-specific observation operators $\mathcal{H}$ map the evolving climate state to proxy-equivalent values. Selected proxy examples illustrate different temporal sensitivities. Horizontal bars indicate the periods of climate history represented by each proxy and are not drawn to scale. Right-pointing arrows denote forward model propagation, and the dashed left-pointing arrow denotes backward gradient propagation during optimization. Joint optimization allows later observations to revise earlier states and model-error increments. (b) The control vector $\mathbf{v}$ comprises the initial state $\state_0$ and model-error increments $\mathbf{n}_k$ for $k=0,\ldots,K-1$. The cost function balances the initial-state penalty, model-error penalty, and observation misfit. Here, $\mathbf{y}_e=\mathcal{H}(\mathbf{U}\mathbf{v})$ denotes the proxy-equivalent values calculated from the trajectory. The common factor of $1/2$ is omitted in the schematic without changing the minimizing solution. Propagating the optimized control vector through the emulator yields the reconstructed trajectory.}
\label{fig:framework}
\end{figure}

The assimilation problem is formulated in weak-constraint four-dimensional variational form, which extends trajectory-level variational assimilation by including model-error increments in the control vector \cite{Sasaki1970,LeDimetTalagrand1986,TalagrandCourtier1987,fisherKalmanSmoothing2005,tremoletAccountingImperfectModel2006,fisherWeakConstraintLongWindow2011,evensen2022data}. Unlike a stochastic LIM forecast, which samples model-error increments, LMR4D-Var treats both these increments and the initial state as unknowns to be optimized jointly. Using the seasonal LIM grid defined above, we reconstruct a sequence of seasonal-mean anomaly states, with four states per year relative to the corresponding seasonal climatologies. The mean annual cycle is not explicitly simulated. As illustrated in Fig.~\ref{fig:framework}b, the control vector contains the initial reduced state together with the sequence of model-error increments. Following the forcing or control-variable formulation of weak-constraint 4D-Var \cite{tremoletAccountingImperfectModel2006,fisherWeakConstraintLongWindow2011}, and the notation used in SI Section~\ref{si:sec:si_wc4dvar}, the control vector is $\mathbf{v}=(\state_0^{T},\mathbf{n}_0^{T},\mathbf{n}_1^{T},\ldots,\mathbf{n}_{K-1}^{T})^{T}$, where $\state_0$ is the initial reduced state and $\mathbf{n}_k$ are the additive model-error increments in Eq.~\ref{eq:lim_dynamics}. Each index $k$ denotes one seasonal analysis step. $\Qk$ specifies the variances and covariances among state components within each seasonal model-error increment. Increments at different seasonal steps are assumed uncorrelated a priori. The full trajectory follows from the recursive LIM dynamics, or equivalently from the linear transformation $\state=\mathbf{U}\mathbf{v}$ with
\begin{equation}
\mathbf{U}=
\begin{pmatrix}
\mathbf{I} & \mathbf{0} & \mathbf{0} & \cdots & \mathbf{0} \\
\mathbf{G} & \mathbf{I} & \mathbf{0} & \cdots & \mathbf{0} \\
\mathbf{G}^{2} & \mathbf{G} & \mathbf{I} & \cdots & \mathbf{0} \\
\vdots & \vdots & \vdots & \ddots & \vdots \\
\mathbf{G}^{K} & \mathbf{G}^{K-1} & \mathbf{G}^{K-2} & \cdots & \mathbf{I}
\end{pmatrix},
\end{equation}
so that the initial state and all subsequent model-error increments determine a single dynamically consistent trajectory. The temporal-independence assumption for model-error increments, together with the within-step covariance represented by $\Qk$, follows standard weak-constraint formulations \cite{tremoletAccountingImperfectModel2006,fisherWeakConstraintLongWindow2011,evensen2022data}. The prior covariance of the initial state and model-error increments is then compactly written as
\begin{equation}
\mathbf{B}_{v}=\operatorname{diag}\!\left(\mathbf{C}(0),\Qk,\Qk,\ldots,\Qk\right),
\end{equation}
with the stationary covariance $\mathbf{C}(0)$ in the first block and the discrete model-error covariance $\Qk$ in the remaining blocks, so explicit construction of the full prior covariance matrix of climate states across all times is avoided.

Assuming Gaussian initial-state, model, and observation errors, the cost function takes the standard quadratic variational form \cite{LeDimetTalagrand1986,fisherKalmanSmoothing2005,tremoletAccountingImperfectModel2006,evensen2022data}. It can be written compactly in control space and expanded directly into the initial-state, model-error, and observation terms, corresponding to the three-way balance sketched in Fig.~\ref{fig:framework}b:
\begin{equation}
\begin{aligned}
J(\mathbf{v})
&=
\underbrace{\frac{1}{2}\mathbf{v}^{T}\mathbf{B}_{v}^{-1}\mathbf{v}}_{\text{prior penalty}}
+ \underbrace{\frac{1}{2}\left[\obs-\mathcal{H}(\mathbf{U}\mathbf{v})\right]^{T}
\mathbf{R}^{-1}
\left[\obs-\mathcal{H}(\mathbf{U}\mathbf{v})\right]}_{\text{proxy misfit}} \\
&=
\underbrace{\frac{1}{2}\state_0^{T}\mathbf{C}(0)^{-1}\state_0}_{\text{initial-state penalty}}
+ \underbrace{\frac{1}{2}\sum_{k=0}^{K-1}\mathbf{n}_k^{T}\Qk^{-1}\mathbf{n}_k}_{\text{model-error penalty}}
+ \underbrace{\frac{1}{2}\left[\obs-\mathcal{H}(\mathbf{U}\mathbf{v})\right]^{T}
\mathbf{R}^{-1}
\left[\obs-\mathcal{H}(\mathbf{U}\mathbf{v})\right]}_{\text{proxy misfit}}.
\end{aligned}
\end{equation}
Here $\mathbf{B}_{v}$ is the block-diagonal prior covariance in control space, $\mathbf{C}(0)$ is the stationary covariance of the reduced LIM state, $\Qk$ is the discrete model-error covariance, $\mathcal{H}$ denotes the collection of proxy observation operators, and $\Rk$ is the observation-error covariance. In this form the three-way balance among the initial-state constraint, model-error penalty, and proxy misfit is explicit, while the optimization remains computationally tractable over long assimilation windows. The inverse error covariances weight deviations according to their prescribed uncertainties, with larger variances allowing greater departures from the initial-state prior, model dynamics, or observations. Under the assumed Gaussian prior and error models, minimizing $J$ yields a maximum a posteriori (MAP) estimate of the control vector \cite{evensen2022data}. For windows much longer than the LIM's memory, the initial-state penalty mainly influences the beginning of the reconstruction.

For sensitivity experiments, and in particular for the Temp12k-only experiments, we introduce a scalar model-error weight \(J_m\) that multiplies the model-error penalty (\(J_m \cdot \frac{1}{2}\sum_{k=0}^{K-1}\mathbf{n}_k^{T}\Qk^{-1}\mathbf{n}_k \)). Changing \(J_m\) controls the relative weight assigned to the reduced-order model prior during optimization. Larger \(J_m\) penalizes model-error increments more strongly and keeps the trajectory closer to the LIM-implied evolution, whereas smaller \(J_m\) lowers the effective model weight and allows the trajectory to depart more freely from the prior. We do not interpret \(J_m\) as a directly observable model-error amplitude; rather, it is a sensitivity parameter that controls how strongly the optimized trajectory is penalized for departing from the reduced-order model.

Optimization is performed using the limited-memory Broyden–Fletcher–Goldfarb–Shanno (L-BFGS) algorithm \cite{liu1989limited}, a gradient-based quasi-Newton method, as the default solver. In classical 4D-Var, an adjoint computes gradients by propagating cost-function sensitivities backward through the linearized model and observation operators \cite{TalagrandCourtier1987,widmann2010using}. Gradients are obtained here through reverse-mode automatic differentiation of the reduced-order trajectory model and differentiable proxy operators using PyTorch \cite{paszke2019pytorch}. This applies the chain rule through the differentiable forward code. The gradient-based implementation requires differentiable forward calculations for the model dynamics and proxy observation operators, but does not require either to be invertible. This gradient-based implementation makes LMR4D-Var modular with respect to the observation operators: conventional short-memory proxy system models, history-dependent operators such as boreholes, and future nonlinear proxy models can be introduced through differentiable forward operators without separately deriving and maintaining hand-written adjoint code for each archive type. Practical diagnostics include cost-function reduction, gradient norms, solver convergence, and the time dependence of inferred model-error penalties. Computational costs are summarized in Table~\ref{si:tab:si_computational_costs}.

\subsection{Proxy observations and observation operators}
\label{sec:proxy_observations}

\begin{figure}[!tbp]
\centering
\includegraphics[width=\textwidth,max height=0.65\textheight,keepaspectratio]{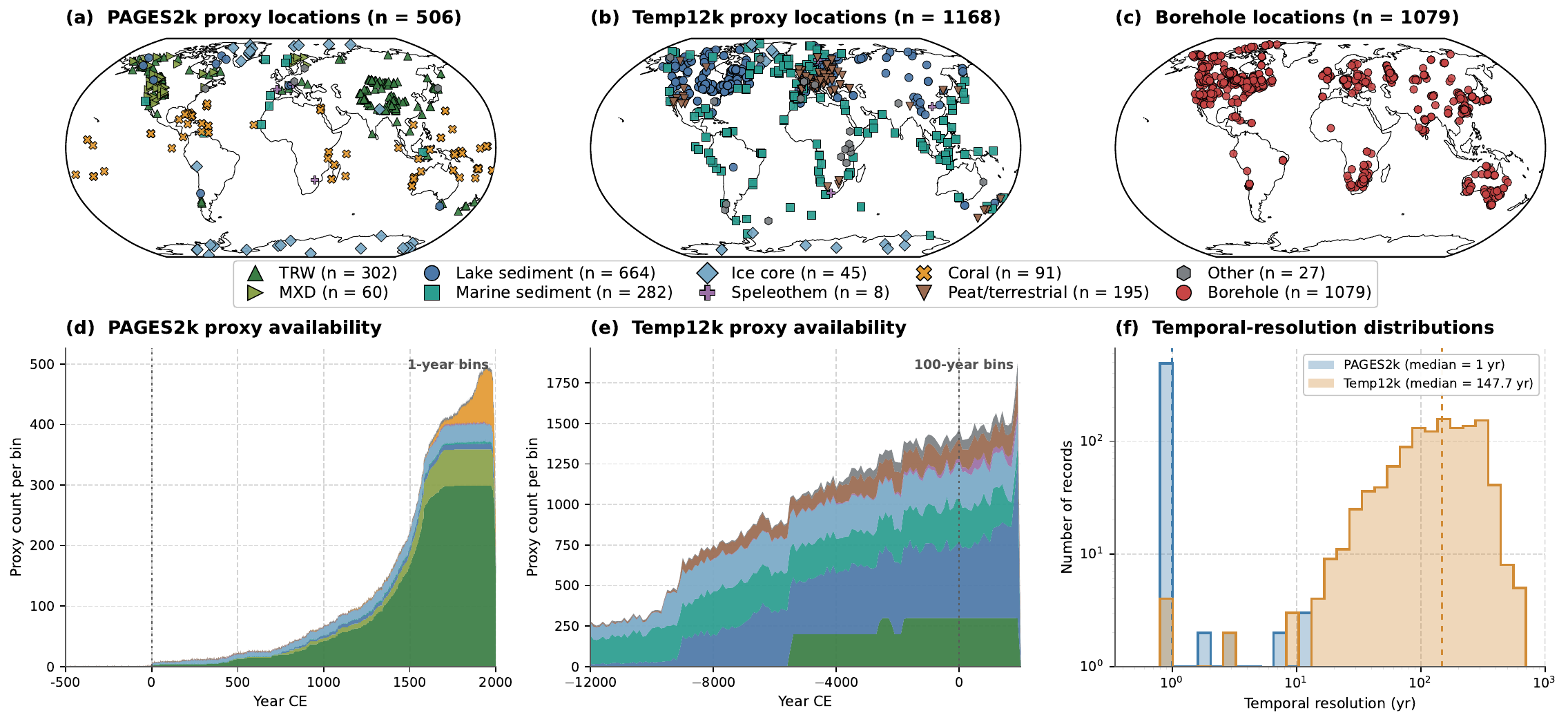}
\caption{\textbf{Spatial distribution, temporal availability, and resolution of the assimilated paleoclimate records.} (a to c) Locations of the PAGES2k records \cite{PAGES2k2017global} (\(n=506\)), Temp12k records \cite{kaufman2020global} (\(n=1168\)), and terrestrial boreholes \cite{cuestaValero2021longterm} (\(n=1079\)), respectively. Record types in the PAGES2k and Temp12k networks include tree-ring width (TRW; \(n=302\)), maximum latewood density (MXD; \(n=60\)), lake sediment (\(n=664\)), marine sediment (\(n=282\)), ice core (\(n=45\)), speleothem (\(n=8\)), coral (\(n=91\)), peat and other terrestrial archives (\(n=195\)), and other archives (\(n=27\)); borehole locations are counted separately. (d and e) Temporal availability of PAGES2k and Temp12k records, shown in 1-year and 100-year bins, respectively, with colors indicating archive type. (f) Distributions of nominal temporal resolution for PAGES2k and Temp12k records. Dashed vertical lines indicate the median resolutions of 1 year for PAGES2k and 147.7 years for Temp12k. Together, the networks provide annually resolved, temporally averaged, and long-memory observations that constrain complementary components of climate variability across seasonal to millennial timescales.}
\label{fig:proxy_network}
\end{figure}

The baseline observation network consists of temperature-sensitive PAGES2k records \cite{PAGES2k2017global}, which provide the densest annually resolved information over the Common Era. Two additional proxy classes are used to introduce lower-frequency constraints: the Temp12k temperature compilation \cite{kaufman2020global} and a recent compilation of terrestrial borehole temperature profiles \cite{cuestaValero2021longterm}. As summarized schematically in Fig.~\ref{fig:framework}a and documented empirically in Fig.~\ref{fig:proxy_network}, these archives differ substantially in temporal footprint, temporal coverage, and the physical meaning of the observation, and they therefore require different observation operators within the same variational objective. For PAGES2k, Temp12k, and boreholes, we assume a diagonal $\Rk$, neglecting observation-error correlations between records and among observations within each record \cite{hakim2016last,meng2025coupled}. Although such correlations may arise from shared environmental influences or nearby records, their covariances are difficult to estimate reliably from the available proxy data.

For annually resolved PAGES2k records, we use univariate linear proxy system models (PSMs; cf  \citeA{Evans_QSR13}) calibrated in the reduced EOF-truncated climate space. Calibration targets are local GISTEMP v4 land temperature \cite{lenssen2019improvements} or ERSSTv5 sea-surface temperature \cite{huang2017extended}. Seasonality is selected by the best calibration fit. As in LMR Seasonal \cite{meng2025coupled}, we exclude records with calibration correlation below 0.05 or one-year lag residual autocorrelation above 0.90. The calibration interval, candidate averaging windows, and screening limitations are detailed in SI Section~\ref{si:sec:si_pages2k_psm}.

Temp12k is treated differently: because the compilation already provides temperature anomalies and associated uncertainty estimates \cite{kaufman2020global}, it is assimilated directly as a low-frequency temperature target rather than through a separately calibrated PSM. Temp12k anomalies are referenced to the 0--2000 CE climatology so that they are consistent with the anomaly convention used for the Common Era reconstruction.

For terrestrial boreholes, the observation operator maps a reconstructed surface-temperature history directly to a present-day subsurface-temperature-anomaly profile through one-dimensional conductive diffusion in a semi-infinite medium \cite{huang2000temperature,cuestaValero2021longterm}. In this setting, the transient temperature field satisfies
\begin{equation}
\frac{\partial T}{\partial t}=\kappa\frac{\partial^2 T}{\partial z^2},
\end{equation}
where $\kappa$ is thermal diffusivity and $z$ is depth. If $T_s(t)$ denotes the reconstructed surface-temperature history, the predicted borehole anomaly profile is written schematically as $\mathbf{y}_{\mathrm{bh}}=\mathcal{H}_{\mathrm{bh}}[T_s(t)]$, or in discrete form as $\mathbf{y}_{\mathrm{bh}}=\mathbf{M}\mathbf{T}_s$, where $\mathbf{T}_s$ is the discretized surface-temperature history and $\mathbf{M}$ contains the conductive kernels determined by depth, thermal diffusivity, and the temporal discretization of the surface history. Because the borehole response depends on the integrated thermal history rather than a single analysis time, this proxy class is particularly well suited to a variational treatment with explicit time dependence. The conductive timescale increases approximately as $t \sim Z^2/(4\kappa)$, so temperatures deeper than 100 m preferentially record multidecadal to centennial variability; accordingly, the borehole comparison excludes the shallower part of each profile and retains only depths between 100 and 300 m (see SI Section~\ref{si:sec:si_borehole_operator}), which limits their influence to approximately 1300--2000 CE. The 300 m cutoff follows the standardized depth range used in the source compilation \cite{cuestaValero2021longterm}. Excluding the upper 100 m emphasizes longer-timescale information and reduces the contribution of more recent, higher-frequency surface variability. Thermal diffusion reduces temporal resolution further back in time, while finite borehole depth limits sensitivity to older temperature changes. Borehole observation uncertainty is represented with an effective depth-dependent error model derived from an inverse-plus-forward ensemble consistency test. The resulting residual variance absorbs unresolved variance associated with both site-specific environmental disturbance and the simplified conductive forward model, and defines the borehole error variances.

\subsection{Experiments and validation strategy}
\label{sec:experiments}

We compare five reconstruction experiments that differ in the proxy classes admitted to the variational objective: P2k (PAGES2k only), P2k\_BH (PAGES2k plus boreholes), P2k\_T12k (PAGES2k plus Temp12k), P2k\_BH\_T12k (PAGES2k plus boreholes and Temp12k), and T12k (Temp12k only). This design is intended to isolate the roles of the two added low-frequency proxy classes. In particular, P2k\_BH versus P2k isolates the influence of long-memory borehole constraints on Common Era temperature and ocean heat content, whereas P2k\_T12k versus P2k isolates the effect of assimilating Temp12k as an additional low-frequency temperature target. P2k\_BH\_T12k tests whether all three proxy classes can be assimilated simultaneously, and T12k tests Holocene-scale Temp12k assimilation and sensitivity to the model-error weight. The proxy composition of each experiment is summarized in Table~\ref{si:tab:si_experiments}. For each proxy configuration, we optimize a seasonal trajectory separately with each LIM using its own dynamics and covariances. The reconstructions are combined only after optimization. The five real-proxy experiments use the same five model-specific LIM priors, the same seasonal assimilation grid, and the same optimization strategy. Reconstructions are therefore obtained first as seasonal analyses with four time steps per year, and all annual or lower-frequency diagnostics shown in the main text are derived from these seasonal solutions after optimization. Differences among the five experiments can thus be attributed primarily to the observational constraints rather than to changes in model structure or solver configuration. The pseudo-proxy experiment described below is the exception: it uses four priors because the entire CESM-LME model family is excluded.

\textbf{Independent validation.} For each Common Era proxy configuration, we perform 20 proxy-withholding repetitions with each of the five LIMs. In each repetition, 20\% of available records from each proxy family (PAGES2k, Temp12k, and boreholes) are withheld for validation, following \citeA{hakim2016last}. Withheld PAGES2k and Temp12k records are evaluated using correlation over the relevant Common Era intervals, and borehole profiles using root mean square error. Because withheld boreholes may share regional climate signals with nearby assimilated sites, random withholding does not establish predictive skill in spatially separated regions.

\textbf{Uncertainty quantification.} The same withholding experiments yield 100 reconstruction trajectories per configuration. Their pointwise 5th--95th percentiles define empirical posterior uncertainty ranges following \citeA{tardif2019last} and \citeA{perkins2021coupled}.

\textbf{Pseudo-proxy experiment.} The pseudo-proxy experiment provides a controlled test in which the true climate trajectory is known. We use CESM-LME \cite{OttoBliesner_etal2016} member 001 as the target trajectory and generate synthetic proxy observations at the same locations, seasons, and temporal resolutions as the real proxy network. For each proxy record, the corresponding observation operator is applied to the target simulation to produce a noiseless pseudo-proxy value: annually or seasonally resolved proxy system models sample the target near-surface temperature field at the calibrated season and location, Temp12k records use the same temporal averaging implied by their archive resolution, and borehole pseudo-profiles are generated by passing the target surface-temperature history through the conductive borehole forward model. Observation error is then added in observation space by drawing independent Gaussian perturbations with variance given by the corresponding diagonal element of the observation-error covariance \(\mathbf{R}\); for boreholes, the perturbations use the depth-dependent borehole error variances described above. We adopt a leave-one-model-family-out design: all CESM-LME simulations are excluded from every stage of prior construction, including estimation of the EOF basis, lagged covariances, LIM propagator \(\mathbf{G}\), and model-error covariance \(\mathbf{Q}\). The target trajectory is reconstructed using the four remaining model-specific LIM priors, so neither member 001 nor another member of the same model family supplies dynamical information to the reconstruction. This choice emulates the structural uncertainty inherent in using a coupled general circulation model for assimilating real-world observations \cite{amrhein2020quantifying}. Skill is evaluated by comparing the reconstructed trajectory with the known target fields for GMT, OHC(0--300 m), and OHC(300--2000 m), none of which is directly prescribed as an assimilated OHC observation. Additional details and diagnostics are given in SI Section~\ref{si:sec:si_ppe}.

\textbf{External GMT and OHC validation.} The external OHC products differ in their surface-temperature histories and ocean-transport constraints. \citeA{gebbie2019little} combine a tracer-constrained ocean circulation model with Common Era surface-temperature histories and Challenger--WOCE hydrographic constraints to reconstruct long-term ocean adjustment. \citeA{zannaGlobalReconstructionHistorical2019a} propagate observed instrumental SST anomalies into the ocean interior using Green's functions derived from the ECCO-GODAE circulation estimate. \citeA{wu2025energybudget} use a related Green's-function approach with additional tracer constraints on the transport kernels and revised surface excess-temperature boundary conditions. Comparisons of reconstructed GMT and OHC with instrumental and external estimates are evaluated using correlation and coefficient of efficiency (CoE) over their common intervals. Following \citeA{nash1970river}, CoE is defined as
\begin{equation}
\mathrm{CoE} = 1-\frac{\sum_t\left(\hat{y}_t-y_t\right)^2}{\sum_t\left(y_t-\bar{y}\right)^2},
\label{eq:coe}
\end{equation}
where \(y_t\) is the validation target, \(\hat{y}_t\) is the corresponding reconstruction estimate, and \(\bar{y}\) is the temporal mean of the validation target over the evaluation interval. A value of \(\mathrm{CoE}=1\) indicates perfect agreement, whereas \(\mathrm{CoE}<0\) indicates less skill than using \(\bar{y}\) as the prediction.

\textbf{OHC trend comparison.} We compare reconstructed 130-year historical OHC trends with the modern 1870--2000 CE trends using intervals of equal duration. Each interval spans 130 years and contains 131 annual values, including both endpoints. Linear trends are calculated separately from the 100 unfiltered annual P2k\_BH\_T12k reconstruction trajectories. For each trajectory and OHC layer, we calculate trends over every complete historical window within 500 BCE--1869 CE. These trajectory-specific trends are pooled to form the historical distribution. Modern uncertainty intervals are defined by the 5th--95th percentiles of the trajectory-specific modern trends. Sensitivity diagnostics use non-overlapping historical windows and highlight four pre-1870 intervals beginning after 1300 CE (Fig.~\ref{si:fig:si_ohc_trend_nonoverlap}). Because windows and reconstruction trajectories are not independent replicates, these distributions provide a descriptive comparison with reconstructed history rather than an independent-sample significance test. The window construction and additional pseudo-proxy trend diagnostics are detailed in SI Section~\ref{si:sec:si_ohc_trend_diagnostics}.

\section{Results and Discussion}
\label{sec:results}

In the leave-one-model-family-out pseudo-proxy experiment, the pointwise median across the four emulator-specific P2k\_BH\_T12k reconstructions captures GMT variability ($r=0.71$, $\mathrm{CoE}=0.51$) and detrended OHC variability in the 0--300-m ($r=0.69$, $\mathrm{CoE}=0.45$) and 300--2000-m ($r=0.71$, $\mathrm{CoE}=0.49$) layers over 1000--2000 CE (Fig.~\ref{si:fig:si_pseudoproxy_indices}). The reconstruction recovers several major low-frequency departures in the target simulation, including responses associated with volcanic forcing. A comparison of P2k and P2k\_BH further tests the contribution of boreholes to deep-ocean recovery against the known target (Fig.~\ref{si:fig:si_pseudoproxy_boreholes}). For annual detrended OHC(300--2000 m), adding boreholes increases $r$ from 0.67 to 0.71 and CoE from 0.46 to 0.49. For the 31-year running medians, $r$ increases from 0.68 to 0.71 and CoE from 0.46 to 0.48. These results show a modest improvement in recovering deep-ocean variability in this controlled experiment, although some multicentennial excursions remain underestimated. We therefore use OHC as an informative but demanding coupled diagnostic, and interpret it below with external comparisons and the limitations of indirect constraints.

\begin{figure}[p]
\centering
\includegraphics[width=\textwidth,max height=0.65\textheight,keepaspectratio]{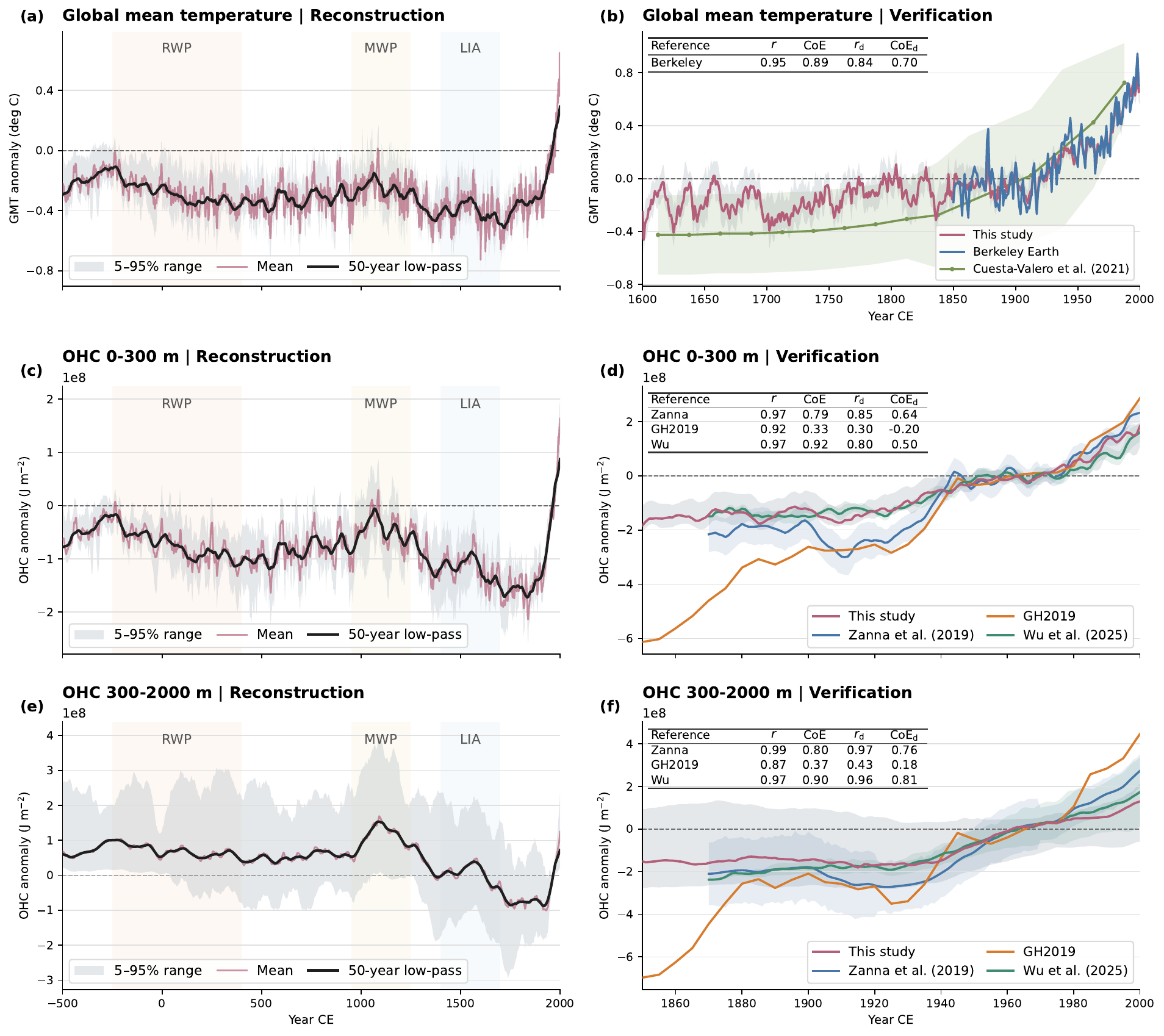}
\caption{\textbf{Reconstructions and external verification for global mean temperature and layer-resolved upper-ocean heat content.}
Results are from the P2k\_BH\_T12k experiment. (a) Reconstructed global mean temperature (GMT) anomaly from 500 BCE to 2000 CE. (b) Comparison with Berkeley Earth \cite{rohde2020berkeley} over the instrumental era and the geothermal reconstruction of \citeA{cuestaValero2021longterm} since 1600 CE. (c and d) Reconstruction and external verification, respectively, of global ocean heat content (OHC) anomalies in the 0--300-m layer. (e and f) Corresponding reconstruction and external verification for the 300--2000-m layer. OHC estimates are compared with the external reconstructions of \citeA{zannaGlobalReconstructionHistorical2019a}, \citeA{gebbie2019little} (GH2019), and \citeA{wu2025energybudget}. Muted rose curves show the mean across the five model-emulator-specific reconstructions. Gray shading denotes the pointwise 5th--95th percentile range across the 100 reconstruction trajectories. Black curves in the reconstruction panels show the 50-year low-pass-filtered mean. In the verification panels, Berkeley Earth and \citeA{zannaGlobalReconstructionHistorical2019a} are shown in blue, \citeA{cuestaValero2021longterm} in olive green, \citeA{gebbie2019little} in orange, and \citeA{wu2025energybudget} in green. Shading around the external estimates indicates reported uncertainties where available. Inset tables in (b), (d), and (f) report Pearson correlation coefficients (\(r\)), coefficients of efficiency (CoE), and their detrended counterparts (\(r_{\mathrm{d}}\) and \(\mathrm{CoE}_{\mathrm{d}}\)). Detrended metrics are calculated after linearly detrending each reconstruction-reference pair over its common interval. Roman Warm Period (RWP; ca.~250 BCE--400 CE), Medieval Warm Period (MWP; ca.~950--1250 CE), and Little Ice Age (LIA; ca.~1400--1700 CE) intervals are marked in (a), (c), and (e).}
\label{fig:figuref1a}
\end{figure}

With this controlled test as context, the P2k\_BH\_T12k reconstruction provides the main Common Era application of the framework. From 500 BCE to 2000 CE, reconstructed GMT and upper-2000-m OHC both show coherent low-frequency variability, including warm conditions during the Roman Warm Period (RWP, ca. 250 BCE--400 CE) and Medieval Warm Period (MWP, ca. 950--1250 CE) and colder conditions during the Little Ice Age (LIA, ca. 1400--1700 CE) (Fig.~\ref{fig:figuref1a}). Over the instrumental era, reconstructed GMT agrees closely with Berkeley Earth \cite{rohde2020berkeley} ($r=0.95$, $\mathrm{CoE}=0.89$), and over the last four centuries it is also broadly consistent with the geothermal reconstruction of \citeA{cuestaValero2021longterm}, despite the different treatment of borehole information in the two approaches. After linear detrending, GMT remains consistent with Berkeley Earth ($r=0.84$, $\mathrm{CoE}=0.70$), showing that the agreement is not solely attributable to the common long-term warming trend.

Comparison with earlier paleoclimate data-assimilation products further contextualizes the instrumental-era GMT skill (Table~\ref{tab:gmt_berkeley_skill}). To reduce differences arising from the choice of climate model, rather than proxy information, we emphasize the CCSM4 LMR4D-Var experiments because LMR v2.1, LMR Seasonal, and LMR Online also use CCSM4-based prior information \cite{tardif2019last,meng2025coupled,perkins2021coupled}. Holding the parent model family in common makes these comparisons more directly interpretable. Measuring performance in terms of GMT within this CCSM4-based comparison, P2k gives $r=0.94$ and $\mathrm{CoE}=0.87$, while adding boreholes in P2k\_BH leaves the correlation unchanged and increases CoE only slightly to 0.88. P2k\_BH\_T12k with the CCSM4 emulator gives the same $r=0.94$ and $\mathrm{CoE}=0.88$. The five-model P2k\_BH\_T12k mean, shown separately to evaluate the effect of aggregating model-specific reconstructions, gives the highest central-estimate scores ($r=0.95$, $\mathrm{CoE}=0.89$). The earlier reconstructions have correlations of 0.90--0.93 and CoE values of 0.77--0.80. Relative to previous reconstructions, LMR4D-Var increases CoE from 0.80 to 0.89, corresponding to an approximately 45\% reduction in MSE. After linear detrending, the CCSM4-based LMR4D-Var experiments retain substantial agreement with Berkeley Earth ($r_{\mathrm{d}}=0.82$--$0.83$, $\mathrm{CoE}_{\mathrm{d}}=0.67$--$0.68$), comparable to LMR Seasonal ($r_{\mathrm{d}}=0.82$, $\mathrm{CoE}_{\mathrm{d}}=0.67$). The five-model P2k\_BH\_T12k mean achieves the highest detrended scores among the products compared here ($r_{\mathrm{d}}=0.84$, $\mathrm{CoE}_{\mathrm{d}}=0.70$; Table~\ref{tab:gmt_berkeley_skill}). These results show that reconstruction skill extends beyond the shared linear warming trend. These central-estimate comparisons indicate that LMR4D-Var preserves, and in this evaluation modestly improves, instrumental-era GMT skill while adding long-memory observations and coupled ocean variables. The shared CCSM4 basis reduces one source of potential difference and provides a more controlled shared-prior comparison, although the products still differ in their observation networks, calibration choices, and assimilation details.

\begin{table}[!tbp]
\centering
\caption{\textbf{Instrumental-era global mean temperature skill relative to Berkeley Earth \cite{rohde2020berkeley}.} Pearson correlation coefficients ($r$) and coefficients of efficiency (CoE) compare annual GMT anomalies from each reconstruction with Berkeley Earth over 1850--2000 CE after removing the 1950--1980 climatology. Metrics are reported for the original series ($r$, CoE) and after separately removing a linear trend from each reconstruction and Berkeley Earth over the comparison interval ($r_{\mathrm{d}}$, $\mathrm{CoE}_{\mathrm{d}}$). Previously published products are listed in chronological order: PHYDA \cite{steiger2018reconstruction}, LMR v2.1 \cite{tardif2019last}, LMR Online \cite{perkins2021coupled}, and LMR Seasonal \cite{meng2025coupled, mengNoEvidenceStatistically2026}, followed by the LMR4D-Var results from this study. Assimilation methods include offline and online ensemble Kalman filtering (EnKF) and weak-constraint 4D-Var. The update interval is given for the online methods, and the model/emulator column identifies the climate-model information used by each reconstruction. Comparisons among the CCSM4-based rows provide the closest assessment of differences among assimilation frameworks, although the observation networks and implementation details are not identical. Within LMR4D-Var, P2k, P2k\_BH, P2k\_T12k, and P2k\_BH\_T12k use the CCSM4-trained emulator and differ in the observations assimilated; P2k\_BH\_T12k mean averages the five model-emulator-specific reconstructions. PHYDA uses CESM-based model information.}
\label{tab:gmt_berkeley_skill}
\begingroup
\footnotesize
\newcommand{\skillrow}[7]{\noindent\parbox[t]{0.23\linewidth}{\raggedright #1}\parbox[t]{0.25\linewidth}{\raggedright #2}\parbox[t]{0.18\linewidth}{\raggedright #3}\parbox[t]{0.08\linewidth}{\centering #4}\parbox[t]{0.08\linewidth}{\centering #5}\parbox[t]{0.08\linewidth}{\centering #6}\parbox[t]{0.08\linewidth}{\centering #7}\par}
\hrule\smallskip
\skillrow{\textbf{Series}}{\textbf{Assimilation method}}{\textbf{Model/emulator}}{\textbf{$r$}}{CoE}{$r_{\mathrm{d}}$}{$\mathrm{CoE}_{\mathrm{d}}$}
\smallskip\hrule\smallskip
\skillrow{PHYDA}{Offline EnKF}{CESM}{0.90}{0.78}{0.73}{0.43}
\skillrow{LMR v2.1}{Offline EnKF}{CCSM4}{0.93}{0.80}{0.80}{0.63}
\skillrow{LMR Online}{Online EnKF (annual)}{CCSM4}{0.90}{0.80}{0.70}{0.45}
\skillrow{LMR Seasonal}{Online EnKF (seasonal)}{CCSM4}{0.93}{0.77}{0.82}{0.67}
\skillrow{P2k}{Weak-constraint 4D-Var}{CCSM4}{0.94}{0.87}{0.82}{0.67}
\skillrow{P2k\_BH}{Weak-constraint 4D-Var}{CCSM4}{0.94}{0.88}{0.82}{0.67}
\skillrow{P2k\_T12k}{Weak-constraint 4D-Var}{CCSM4}{0.94}{0.87}{0.82}{0.68}
\skillrow{P2k\_BH\_T12k}{Weak-constraint 4D-Var}{CCSM4}{0.94}{0.88}{0.83}{0.68}
\skillrow{P2k\_BH\_T12k mean}{Weak-constraint 4D-Var}{Five models}{\textbf{0.95}}{\textbf{0.89}}{\textbf{0.84}}{\textbf{0.70}}
\smallskip\hrule
\endgroup
\end{table}

\begin{figure}[!tbp]
\centering
\includegraphics[width=\linewidth,max height=0.65\textheight,keepaspectratio]{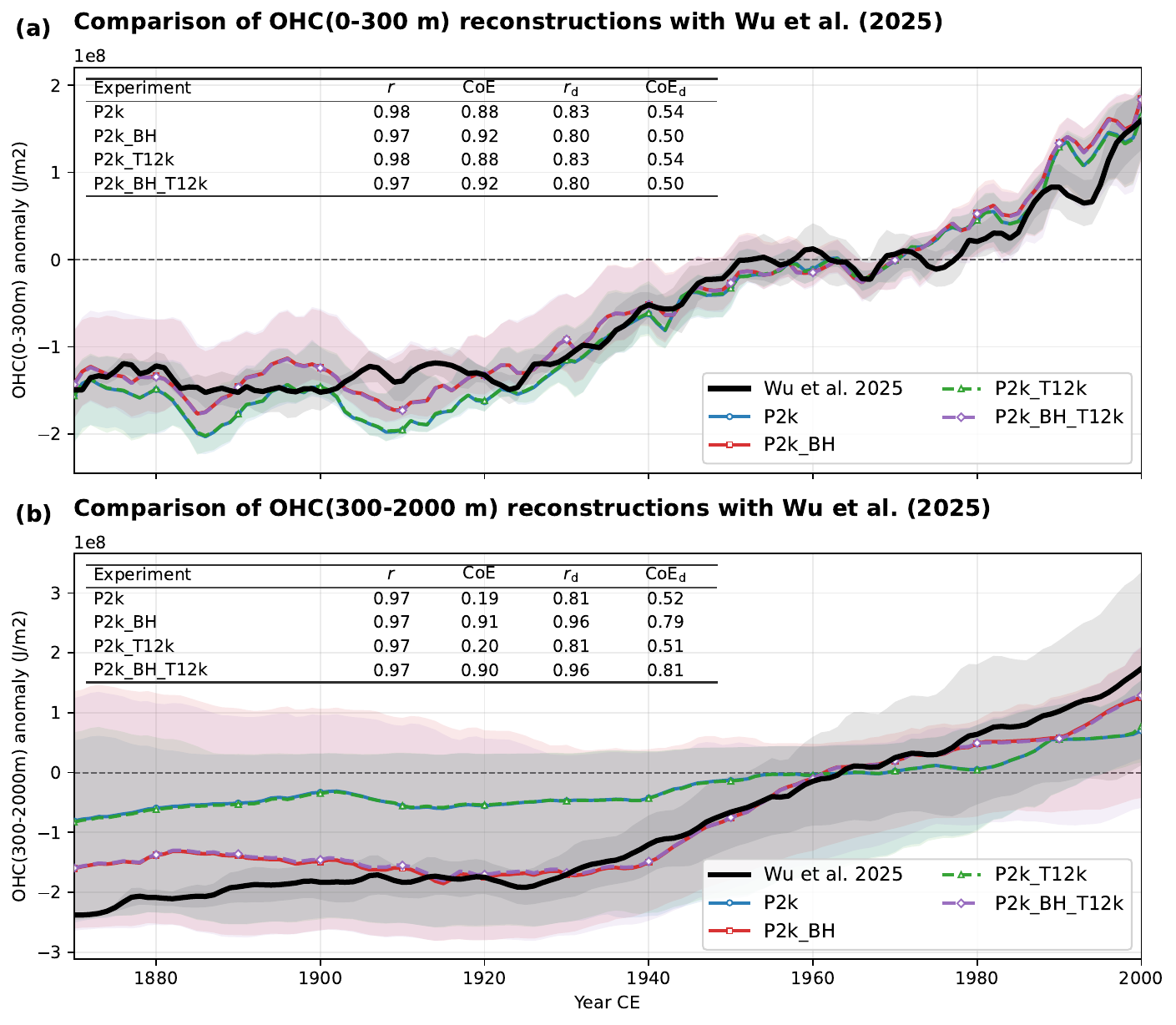}
\caption{\textbf{Borehole assimilation improves agreement with an external reconstruction of deeper-ocean heat content.}
Global-mean ocean heat content (OHC) anomalies from four LMR4D-Var experiments compared with the estimate of \citeA{wu2025energybudget} over 1870--2000 CE. (a) OHC anomalies for the 0--300-m layer. (b) OHC anomalies for the 300--2000-m layer. Colored lines show the mean across the five model-emulator-specific reconstructions for the P2k, P2k\_BH, P2k\_T12k, and P2k\_BH\_T12k experiments. Colored shading denotes the pointwise 5th--95th percentile range across the 100 reconstruction trajectories for each experiment. In panel (a) and (b), P2k closely overlaps P2k\_T12k, whereas P2k\_BH closely overlaps P2k\_BH\_T12k, producing two visually distinct pairs. Black lines show the external estimate, with gray shading indicating its reported uncertainty. All anomalies are referenced to the 1950--1980 CE climatology. Pearson correlation coefficients (\(r\)) and coefficients of efficiency (CoE), and their detrended counterparts (\(r_{\mathrm{d}}\) and \(\mathrm{CoE}_{\mathrm{d}}\)), reported in the tables, are calculated against the same external estimate over 1870--2000 CE. Detrended metrics are calculated after linearly detrending each reconstruction-reference pair over its common interval.}
\label{fig:ohc_wu_experiments}
\end{figure}

Validating OHC tests the algorithm's ability to reconstruct an indirect field, because the assimilated proxies primarily constrain surface climate. Results show that reconstructed OHC late historical evolution is consistent in magnitude and trend with independent estimates based on historical ocean reconstruction and energy-budget methods \cite{zannaGlobalReconstructionHistorical2019a,wu2025energybudget}. For OHC(0--300 m), the comparison with \citeA{wu2025energybudget} yields $r=0.97$ and $\mathrm{CoE}=0.92$, and for OHC(300--2000 m) it yields $r=0.97$ and $\mathrm{CoE}=0.90$. To distinguish agreement in the historical trend from agreement in residual variability, we also compute $r$ and CoE after linear detrending. The detrended comparison with \citeA{wu2025energybudget} remains strong in the 300--2000-m layer ($r=0.96$, $\mathrm{CoE}=0.81$), but is weaker in the 0--300-m layer ($r=0.80$, $\mathrm{CoE}=0.50$), indicating that the common warming trend contributes more to the upper-layer scores. Detrended agreement with \citeA{zannaGlobalReconstructionHistorical2019a} also remains high after detrending for both layers ($r=0.85$, $\mathrm{CoE}=0.64$ for 0--300 m; $r=0.97$, $\mathrm{CoE}=0.76$ for 300--2000 m), whereas the corresponding comparison with \citeA{gebbie2019little} is substantially weaker ($r=0.30$, $\mathrm{CoE}=-0.20$; $r=0.43$, $\mathrm{CoE}=0.18$). Layer-resolved spatial comparisons with \citeA{wu2025energybudget} show stronger positive correlations over the Southern Ocean and the South Pacific, Indian, and South Atlantic sectors, but weaker correlations over the North Atlantic and Pacific (Fig.~\ref{si:fig:si_ohc_wu_corr}). These spatial correlations reflect consistent trend patterns: the 0--300-m trends agree most strongly over the Southern Ocean, Indian Ocean, and Atlantic sectors, whereas the 300--2000-m trends are most consistent over the Southern Hemisphere, especially the extratropical Southern Ocean (Fig.~\ref{si:fig:si_ohc_wu_trend_compare}). Comparisons across the four proxy configurations show how each archive class affects agreement with \citeA{wu2025energybudget} (Fig.~\ref{fig:ohc_wu_experiments}). In the 0--300-m layer, the reconstructed trajectories are nearly identical, with CoE ranging from 0.88 to 0.92 and detrended CoE from 0.50 to 0.54. The borehole effect is much larger in the 300--2000-m layer: CoE increases from 0.19--0.20 without boreholes to 0.90--0.91 with boreholes, while detrended CoE increases from 0.51--0.52 to 0.79--0.81. Adding Temp12k produces little additional change over the instrumental interval, as P2k overlaps P2k\_T12k and P2k\_BH overlaps P2k\_BH\_T12k. The borehole-related improvement is also evident against \citeA{gebbie2019little}, with CoE increasing from 0.18 to 0.60--0.61 and detrended CoE from 0.25 to 0.47. Against \citeA{zannaGlobalReconstructionHistorical2019a}, CoE increases from 0.12--0.13 to 0.80--0.81 and detrended CoE from 0.27--0.28 to 0.76--0.77 (Fig.~\ref{si:fig:si_ohc_external_experiments}). These comparisons associate the improved deep-ocean agreement primarily with long-memory borehole information transmitted through the coupled emulator covariance. Support is therefore strongest for large-scale OHC evolution, while regional and depth-dependent patterns remain less certain.

\begin{figure}[!tbp]
\centering
\includegraphics[width=0.86\textwidth,max height=0.65\textheight,keepaspectratio]{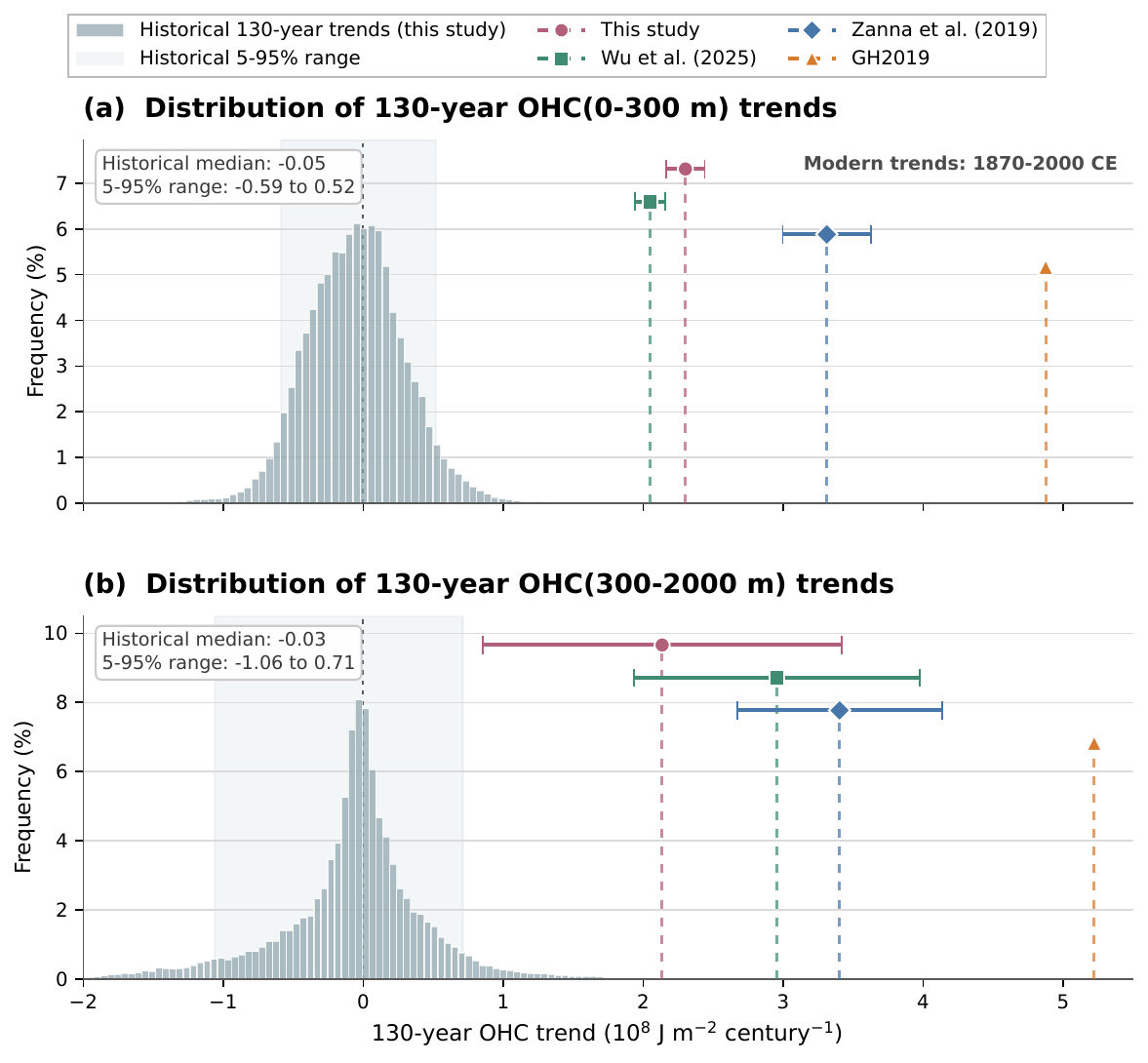}
\caption{\textbf{Modern layer-resolved ocean heat-content trends exceed reconstructed pre-1870 Common Era trends.} Histograms pool 130-year linear trends calculated separately from the 100 unfiltered annual P2k\_BH\_T12k reconstruction trajectories for (a) 0--300 m and (b) 300--2000 m. Each trajectory contributes all complete 130-year windows within 500 BCE--1869 CE. Gray shading indicates the 5th--95th percentile ranges of the pooled historical trends, and gray dotted lines indicate their medians. Colored markers indicate the 1870--2000 CE trends from this study (magenta circles), \citeA{wu2025energybudget} (green squares), \citeA{zannaGlobalReconstructionHistorical2019a} (blue diamonds), and \citeA{gebbie2019little} (orange triangles). Dashed vertical lines connect the estimates to the horizontal axis; marker heights are used only for visual separation. For this study, horizontal error bars show the 5th--95th percentile range across the 100 trajectory-specific modern trends; external error bars show reported 90\% uncertainty intervals where available. The historical ranges describe the pooled trend distribution, rather than confidence intervals for its mean. Trends are expressed in units of \(10^{8}\) J m\(^{-2}\) century\(^{-1}\).}
\label{fig:figuref2}
\end{figure}

The reconstructed 1870--2000 CE OHC trend estimates exceed the pre-1870 historical trend distributions in both ocean layers (Fig.~\ref{fig:figuref2}). The modern trends are \(2.302\times10^{8}\) J m\(^{-2}\) century\(^{-1}\) for OHC(0--300 m) and \(2.136\times10^{8}\) J m\(^{-2}\) century\(^{-1}\) for OHC(300--2000 m). The modern--historical separation persists when non-overlapping windows are used, including the four highlighted historical intervals beginning after 1300 CE (Fig.~\ref{si:fig:si_ohc_trend_nonoverlap}). Nevertheless, it remains conditional on the reconstructed histories and does not eliminate uncertainty associated with changing proxy coverage or attenuated low-frequency variability.

The modern trends are also consistent with the estimates of \citeA{zannaGlobalReconstructionHistorical2019a} and \citeA{wu2025energybudget}. By contrast, the trends implied by \citeA{gebbie2019little} are much larger in both layers. The layer-resolved comparison with \citeA{gebbie2019little} (Fig.~\ref{si:fig:si_ohc_external}) highlights the remaining uncertainty in multicentennial OHC amplitude: that reconstruction implies variability that is approximately an order of magnitude larger than our estimate in both layers, likely reflecting differences in boundary forcing assumptions, circulation constraints, and the propagation of surface anomalies into subsurface heat storage. One likely contributor to the smaller amplitude in our reconstruction is the prior itself. The last-millennium model simulations used to train the climate emulators contain relatively small global-mean OHC variability, on the order of \(10^8\) J m\(^{-2}\), and also contain strong long-term drifts, especially in OHC(300--2000 m) (Fig.~\ref{si:fig:si_model_ohc_changes}). Such drifts may arise from incomplete equilibration of the initial deep-ocean state and the long adjustment timescale of the deeper ocean, consistent with the importance of deep-ocean initial conditions for low-frequency climate variability proposed by ~\citeA{zhu2019climate}. Because the emulators are trained after linear detrending, the procedure removes unrealistic drift but may also remove multicentennial variability. In addition, the emulators represent coupled climate evolution using a time-invariant linear propagator estimated in a truncated EOF space. The linear dynamics may not fully capture state-dependent changes in ocean circulation, mixing, and heat uptake when strong external forcing drives the climate outside the range of variability represented in the last-millennium training simulations. Moreover, although EOF truncation retains the dominant modes of variability, it may omit lower-variance spatial or vertical structures that contribute to the forced OHC response, particularly in the deeper ocean. These approximations may attenuate the amplitude or alter the spatial and vertical distribution of reconstructed OHC, although their relative contributions have not been quantified here. Even with these limitations, the historical-period OHC trends remain close to the estimates of \citeA{wu2025energybudget}, indicating that the framework retains skill for recent forced changes while leaving substantial uncertainty in multicentennial OHC variability.

Within LMR4D-Var, the MWP-to-LIA OHC(300--2000 m) trend shows pronounced North Pacific cooling (Fig.~\ref{si:fig:si_ohc_mwp_lia_trend}b). By contrast, the reconstructed 1880--2000 CE trend shows warming in parts of the Southern Ocean, without the same prominent North Pacific structure (Fig.~\ref{si:fig:si_ohc_wu_trend_compare}c). One possible explanation is that the two intervals reflect different forcing regimes. The MWP-to-LIA transition includes strong volcanic forcing and associated dynamical responses \cite{mann2009global,toohey2017volcanic}, whereas the modern increase is dominated by anthropogenic greenhouse-gas forcing \cite{gleckler2012human,IPCC2021WGI}. Different forced-response patterns can project onto different modes of ocean circulation and heat redistribution, which may in turn produce distinct OHC patterns. We therefore treat this contrast as a hypothesis-generating result rather than formal attribution, because targeted forced-simulation experiments are needed to test whether the inferred pattern difference reflects volcanic forcing, greenhouse-gas forcing, internal variability, or model-dependent covariance structure.

Additionally, pairwise comparisons among the reconstruction experiments show that the different proxy classes affect the trajectory in ways that are consistent with their temporal information content (Figs.~\ref{si:fig:figure14a} and \ref{si:fig:figure14b}). Reconstructions that include PAGES2k share very similar GMT evolution, with pairwise correlations generally exceeding 0.95. Adding boreholes or Temp12k therefore does not substantially change the main Common Era surface-temperature variability. OHC comparisons are also highly correlated across P2k, P2k\_BH, P2k\_T12k, and P2k\_BH\_T12k, but show larger amplitude differences, especially when boreholes are included, consistent with OHC being more sensitive to low-frequency constraints. In contrast, the T12k reconstruction has substantially lower correlations with the PAGES2k-constrained experiments ($r$ is around $0.5$), indicating that Temp12k captures broad low-frequency structure but does not replace the higher-resolution Common Era constraints provided by PAGES2k.

\begin{figure}[!tbp]
\centering
\includegraphics[width=\textwidth,max height=0.65\textheight,keepaspectratio]{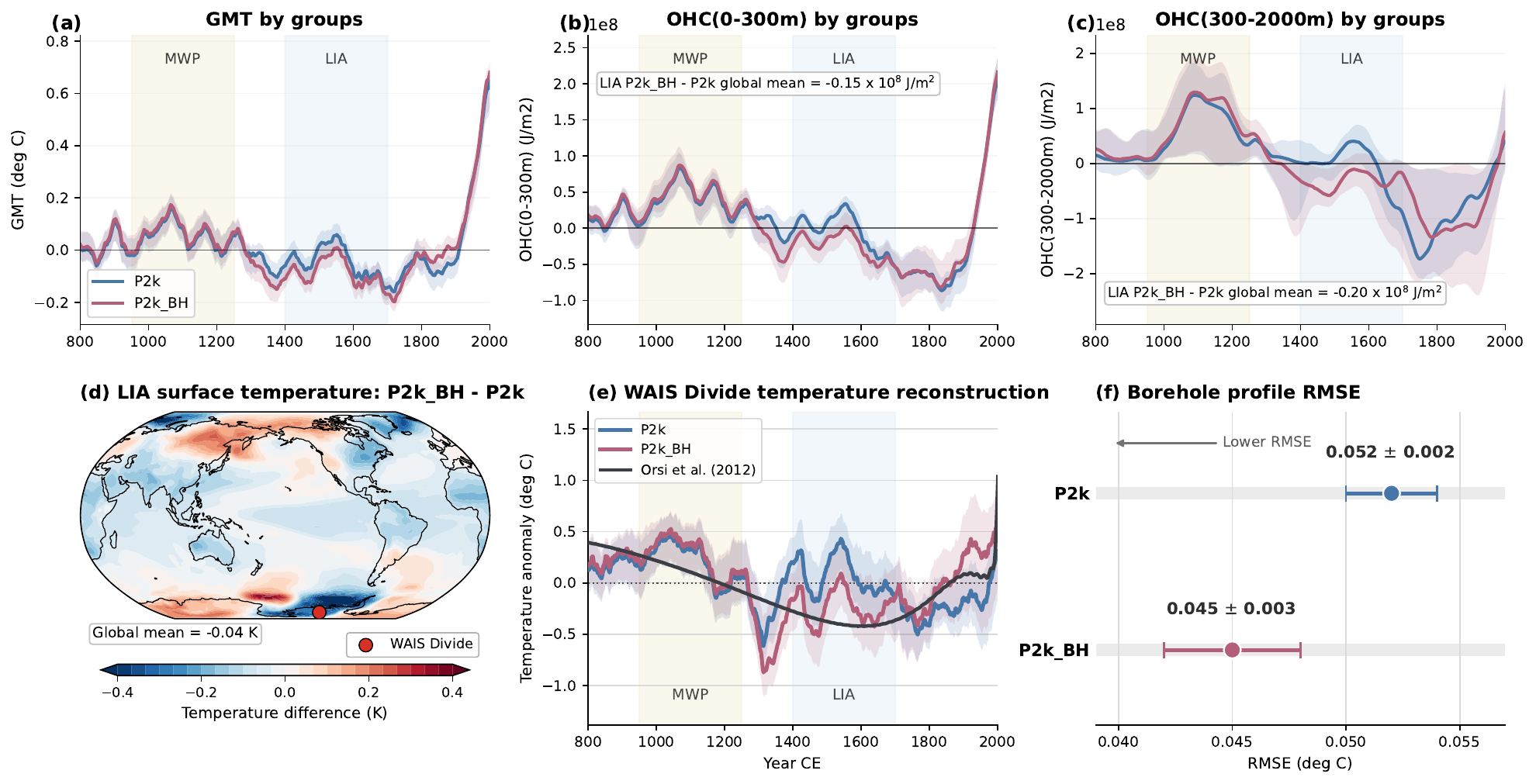}
\caption{\textbf{Borehole assimilation modifies the long-term evolution of temperature and ocean heat content and improves agreement with an independent West Antarctic borehole constraint.}
(a to c) Reconstructed global mean temperature (GMT), ocean heat content (OHC) in the 0--300-m layer, and OHC in the 300--2000-m layer for P2k (PAGES2k only) and P2k\_BH (PAGES2k plus terrestrial boreholes), respectively. Lines show the mean across five model-emulator-specific reconstructions. Colored shading denotes the pointwise 5th--95th percentile range across the 100 reconstruction trajectories for each experiment. Anomalies are referenced to the 0--2000 CE climatology. Values reported in (b) and (c) give the global-mean P2k\_BH minus P2k differences during the Little Ice Age (LIA). (d) Spatial distribution of the LIA surface-temperature difference between P2k\_BH and P2k, calculated as the 1400--1700 CE mean. The global-mean difference is reported in the panel, and the WAIS Divide site is marked by a red dot. (e) Reconstructed temperature anomalies at WAIS Divide for P2k and P2k\_BH compared with the independent borehole reconstruction of \citeA{orsi2012little}. All experimental time series in (a) to (c) and (e) are smoothed with a 50-year low-pass filter. Colored shading in (e) likewise denotes the pointwise 5th--95th percentile range across the 100 reconstruction trajectories for each experiment. (f) Root-mean-square error (RMSE) of the reconstructed borehole temperature profiles at WAIS Divide. Markers show the mean RMSE, and horizontal error bars denote the 5th--95th percentile range across the 100 reconstruction trajectories for each experiment. Lower RMSE indicates closer agreement with the observed borehole profile. Shaded intervals mark the MWP (950--1250 CE) and LIA (1400--1700 CE).}
\label{fig:figure5b}
\end{figure}

The borehole experiments provide the clearest example of how long-memory archives alter the reconstruction (Fig.~\ref{fig:figure5b}). Relative to the P2k baseline, P2k\_BH produces an additional global-mean LIA cooling of approximately 0.04 K. Borehole assimilation produces a more pronounced response in ocean heat storage: global-mean LIA OHC decreases by approximately \(0.15\times10^{8}\) J m\(^{-2}\) in the 0--300-m layer and \(0.20\times10^{8}\) J m\(^{-2}\) in the 300--2000-m layer. The pronounced OHC response is consistent with the role of boreholes as long-memory constraints: because borehole profiles integrate surface-temperature history over decades to centuries, they are especially informative about slowly varying quantities such as OHC. We note that this response in the reconstruction arises naturally, without any explicit linkage between borehole proxies and OHC.

The borehole influence is not purely local. The P2k\_BH minus P2k LIA surface-temperature pattern has a La Ni\~na-like structure and reaches its largest amplitude over West Antarctica. We do not assimilate borehole temperature profiles measured in ice, because doing so would require accounting for ice advection. An independent ice-core borehole record from West Antarctica \cite{orsi2012little,steigorsi2013} therefore provides an out-of-sample check. When the reconstructed temperature histories are passed through the borehole forward model \cite{orsi2012little} at that site, the mean RMSE decreases from 0.052 K in P2k to 0.045 K in P2k\_BH (Fig.~\ref{fig:figure5b}). Although this single-site comparison does not provide a formal significance test, the reduction in RMSE provides an independent consistency check: assimilating boreholes makes the LIA temperature reconstruction more consistent with an independent long-memory constraint, supporting the interpretation that the large-scale cooling is not simply an overfit to local borehole information.

The corresponding LIA OHC differences show that the long-memory surface constraint also projects onto OHC. Borehole assimilation produces a cooler global-mean LIA ocean state in both OHC layers (Fig.~\ref{fig:figure5b}).
Because the detailed regional OHC anomalies are inferred indirectly through emulator-derived cross-variable covariance and lack independent validation, we focus on the sign and magnitude of the global-mean response rather than interpreting the spatial pattern as a robust dynamical fingerprint. Together, these results help reconcile earlier differences between borehole-based reconstructions, which tend to imply greater LIA cooling than reconstructions based primarily on annually resolved natural proxies.

\begin{figure}[!tbp]
\centering
\includegraphics[width=\textwidth,max height=0.65\textheight,keepaspectratio]{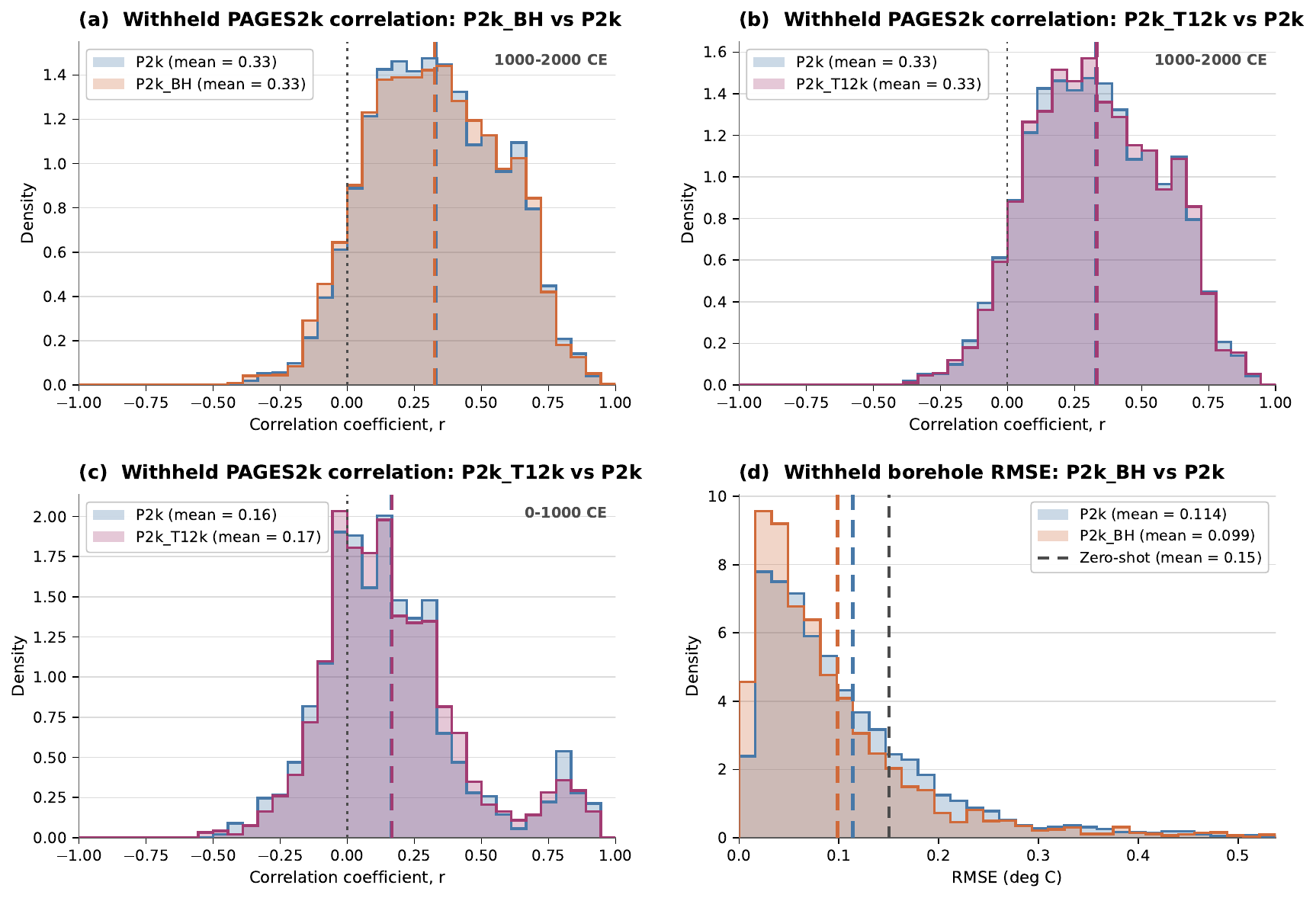}
\caption{\textbf{Verification against proxy records withheld from assimilation.} Probability-density histograms summarize proxy-level skill across 20 proxy-withholding repetitions with each of the five LIMs, withholding 20\% of records from each proxy family. (a) Correlations between observed values of the withheld PAGES2k records and values predicted from the P2k and P2k\_BH reconstructions over 1000--2000 CE. (b and c) Corresponding correlations for the P2k and P2k\_T12k reconstructions over 1000--2000 CE and 0--1000 CE, respectively. (d) Root-mean-square errors (RMSEs) between withheld borehole temperature profiles and profiles predicted from the P2k and P2k\_BH reconstructions; lower RMSE indicates better agreement. The zero-proxy benchmark in (d), labeled ``Zero-shot,'' uses a free LIM simulation without assimilating any proxy observations. Colored dashed lines indicate the mean of each reconstruction distribution, the gray dashed line in (d) indicates the mean zero-proxy RMSE, and the dotted vertical lines in (a) to (c) mark zero correlation.}
\label{fig:figure6a}
\end{figure}

Independent verification tests whether the changes produced by the added proxy classes generalize to records excluded from assimilation, rather than reflecting only closer agreement with the records used in the optimization. We perform 20 proxy-withholding repetitions with each of the five LIMs, withholding 20\% of records from each proxy family. Adding boreholes does not degrade correlation skill against withheld PAGES2k records over 1000--2000 CE, and adding Temp12k does not degrade PAGES2k skill over either 1000--2000 CE or 0--1000 CE (Fig.~\ref{fig:figure6a}). Thus, multiscale assimilation does not sacrifice skill for annually resolved proxies. The strongest improvement occurs in the borehole validation: assimilating PAGES2k already reduces held-out borehole RMSE relative to the zero-proxy case, and adding boreholes further reduces the error.

\begin{figure}[!tbp]
\centering
\includegraphics[width=\textwidth,max height=0.65\textheight,keepaspectratio]{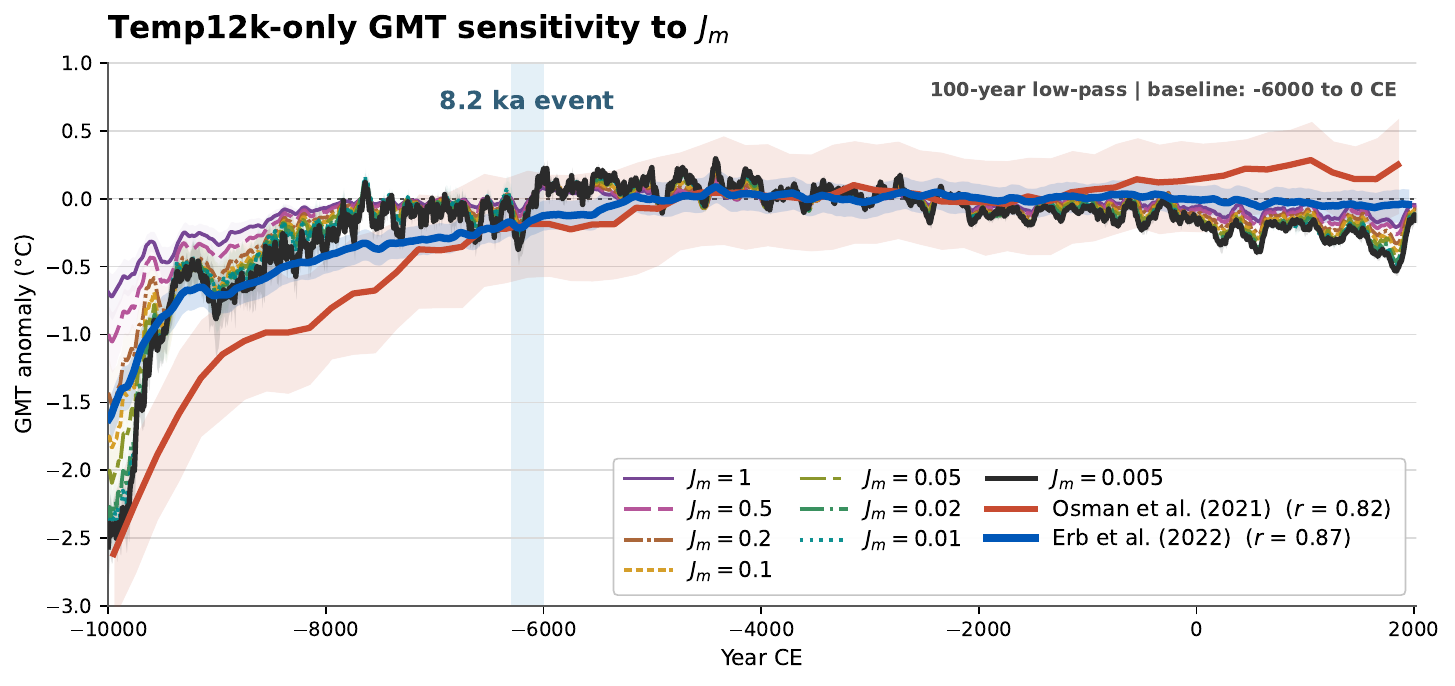}
\caption{\textbf{Sensitivity of Temp12k-only global mean temperature reconstructions to the model-error weight \(J_m\).} Curves show reconstructed global mean temperature (GMT) anomalies for Temp12k-only experiments with different values of \(J_m\). For each \(J_m\), the plotted curve is the average across the available model-emulator-specific reconstructions, and shading indicates \(\pm 1\) standard deviation across model-emulator-specific reconstructions. All series are referenced to the \(-6000\) to \(0\) CE climatology and smoothed with a 100-year low-pass filter. Temperature reconstructions from \citeA{osman2021globally} and \citeA{erb2022reconstructing} are shown for comparison, with shading indicating their reported uncertainty ranges. The 8.2 ka event is highlighted by a light blue band. Correlations are computed separately between the median Temp12k-only reconstruction with \(J_m=0.005\) and the reconstructions of \citeA{osman2021globally} and \citeA{erb2022reconstructing} over their respective overlapping intervals.}
\label{fig:figures12}
\end{figure}

Finally, the T12k sensitivity experiments test whether the weak-constraint 4D-Var framework can assimilate proxy information across a much broader range of timescales. Temp12k records span decadal to millennial timescales, making them a useful test case for multiscale assimilation beyond the Common Era setting, which here uses proxies at seasonal to annual resolution. In these Temp12k-only reconstructions, the model-error weight parameter \(J_m\) controls the balance between the model emulator and the low-frequency proxy constraints (Fig.~\ref{fig:figures12}). Because the LIM prior is trained mainly on last-millennium simulations, it does not fully represent Holocene-scale dynamics, particularly the deglaciation during the early Holocene. Holocene changes in the seasonal distribution of insolation may also produce seasonal temperature responses that are not represented by the last-millennium-trained LIM.  Reducing \(J_m\) therefore lowers the effective weight on the model emulator, allowing the optimized trajectory to respond more strongly to the Temp12k observations. As \(J_m\) decreases, the reconstructed GMT variability increases, particularly in the early Holocene, while remaining broadly consistent with existing Holocene reconstructions, including the 8.2 ka event and the overall amplitude of Holocene temperature variability. The low-\(J_m\) reconstruction also shows a mid-Holocene thermal maximum near 6 ka and a late-Holocene cooling trend, consistent with the Temp12k-based reconstruction of \citeA{kaufman2020holocene}. This behavior differs from the reconstructions of \citeA{osman2021globally} and \citeA{erb2022reconstructing}, which do not show the same late-Holocene cooling in the global mean. We therefore interpret the \(J_m\) experiments primarily as a sensitivity analysis of the model--data balance, rather than a quantitative Holocene reconstruction. These results show that LMR4D-Var framework can assimilate proxy records with strongly heterogeneous temporal resolution to yield a 12,000-year-long reconstruction at seasonal resolution. Realistic Holocene-scale applications will however require careful consideration of the model used to perform the optimization.

\FloatBarrier

\section{Conclusions and Outlook}
\label{sec:discussion}

The LMR4D-Var framework enables trajectory-level paleoclimate data assimilation across archives that operate on different temporal clocks. This framework combines weak-constraint 4D-Var with a coupled seasonal reduced-order emulator, allowing seasonal, annual, multidecadal, and multicentennial observations to constrain a single evolving climate history. Here, coupling means more than reconstructing several variables simultaneously. Surface air temperature, sea-surface temperature, sea ice, and OHC in the 0--300-m and 300--2000-m layers are represented within one multivariate state and advanced by the same climate-model emulator.  Contemporaneous and lagged cross-variable covariances provide the pathway by which observations that directly constrain surface climate can also inform ocean heat storage and sea ice, including regions without assimilated observations. Optimizing the full trajectory propagates these adjustments across variables and through time, while model-error increments allow flexibility in the  dynamics that fit the observations. This coupled pathway underlies the recovery of layer-resolved OHC in the pseudo-proxy experiment and the spatially distributed temperature and OHC responses to borehole assimilation.

The OHC reconstruction illustrates how surface-sensitive records can inform slow oceanic variables through the coupled state. The pseudo-proxy experiment supports this pathway under controlled model-world conditions, while historical OHC trends are consistent with external estimates, especially \citeA{wu2025energybudget}. The modern rate of heat accumulation in both layers lies beyond the reconstructed pre-1870 Common Era trend distribution. All four experiments track the modern OHC increase, but borehole assimilation improves agreement with \citeA{wu2025energybudget} in the 300--2000-m layer, including after detrending (Fig.~\ref{fig:ohc_wu_experiments}).
Including terrestrial boreholes also slightly cools LIA global mean temperature but produces a larger coherent OHC response in both layers. This behavior is consistent with the expectation that long-memory archives should be especially informative for slow climate variables. The influence of borehole assimilation is felt globally, and modestly improves validation during the LIA in an independent West Antarctic ice-core borehole record \cite{orsi2012little}. In this sense, LMR4D-Var helps reconcile the cooler low-frequency temperatures often inferred from boreholes with reconstructions based mainly on annually resolved natural proxies \cite{huang2000temperature,national2007surface,christiansen2017challenges}: the archives are treated as complementary constraints on one coupled trajectory rather than as competing estimates of the same instantaneous quantity.

The Temp12k experiments extend the same logic to longer timescales, but they should only be interpreted as a framework-extensibility proof of concept. Temp12k records span decadal to millennial resolutions, so they test whether the framework can assimilate low-frequency temperature information beyond the Common Era. The sensitivity to model error, \(J_m\), shows the importance of allowing for model error in online reconstructions. Lowering the effective weight assigned to the last-millennium-trained emulator allows a larger reconstructed temperature range while retaining consistency with existing Holocene reconstructions. Future work could allow model-error covariances to vary with time or climate state.

There are three areas that could further improve upon the results reported here.  First, the model emulator is trained on detrended last-millennium simulations in a small state space. This makes coupled seasonal optimization over multicentennial to millennial windows tractable, but comes at the cost of potentially damped multicentennial OHC variability and has no representation of nonlinear ocean adjustment under strong forcing. Structural model errors are not represented separately, and $\Qk$ is not independently calibrated to characterize them. Model-error increments may compensate for some deficiencies but cannot restore variability absent from the training simulations. Full Holocene reconstructions will require models that incorporate forcing and a wider range of variability.  Second, the observation network primarily constrains surface climate, samples a limited set of archive classes, assumes a diagonal observation-error covariance, and simplifies chronological and temporal-resolution uncertainty. Temporally correlated proxy errors may make low-frequency reconstruction skill lower than in our white-noise pseudo-proxy tests. The empirical posterior uncertainty ranges from the 100 reconstruction trajectories do not fully characterize the posterior distribution. Sparse early proxy coverage can increase reliance on the emulator, affecting comparisons of reconstructed variance, trends, and spatial patterns across periods. Third, the current state vector excludes precipitation and other hydroclimate variables, limiting the assimilation and reconstruction of Hydro2k and related records. Extending the state vector and differentiable observation operators to precipitation, soil moisture, and other moisture-balance variables would broaden the framework, but would require careful treatment of intermittent and non-Gaussian variability, proxy-system-model uncertainty, and model-dependent hydroclimate covariance.

These caveats notwithstanding, LMR4D-Var provides a flexible platform for reconstructing physically consistent climate trajectories from heterogeneous paleoclimate evidence. The key practical advantage is that any proxy that can be forward-modeled (climate to observation) with an error estimate can be included. This makes the framework well suited for future assimilation of complex proxies whose sensitivities depend on seasonality, archive physics, time averaging, age uncertainty, or the full climate history, including sedimentary proxies, isotope-enabled records, accumulation- or diffusion-affected ice-core signals, and other nonlinear proxy system models \cite{dolman2018sedproxy}. The broader contribution is therefore not a single Common Era OHC estimate alone, but a general framework for bringing diverse paleoclimate archives into one reconstruction problem.
Beyond reconstruction, this framework may also support long transient simulations. With an appropriate climate model as the dynamical prior, proxy-constrained trajectory optimization could provide better-balanced initial conditions and reduce drift associated with incompletely equilibrated ocean states.

\section*{Acknowledgments}
The authors thank Francisco Jose Cuesta-Valero from St. Francis Xavier University for his conversations related to the borehole forward modeling. We also thank Zhiqiang Lyu at the University at Albany, State University of New York (SUNY Albany), for providing the ice-borehole forward-modeling code used in this study. We thank Rapha\"el H\'ebert and two anonymous reviewers for their constructive comments and suggestions.
This work was supported in part by NSF awards 2402475, 2202526, and 2105805, and Heising-Simons Foundation award 2023-4715.

\section*{Author Contributions}
Z.M. conceived the study, developed the methodology, performed the analyses, and drafted the manuscript. G.J.H. and E.J.S. supervised the research. G.J.H., E.J.S., and J.E.-G. acquired funding. J.E.-G., E.J.S., and Z.M. contributed to the borehole modeling. Z.M., G.J.H., J.E.-G., E.J.S., and T.G. interpreted the results and revised the manuscript.

\section*{Conflict of Interest}
The authors declare no conflicts of interest relevant to this study.

\section*{Open Research Statement}
The plotting and EOF package SACPY \cite{meng2023sacpy,meng2024pacific} is available at \url{https://github.com/ZiluM/sacpy}. The LMR4D-Var source code, configuration templates, and workflow documentation are publicly available at \url{https://github.com/ZiluM/LMR-4DVar}. Proxy data products are derived from publicly available source compilations, including PAGES~2k v2.0.0, Temp12k, and published borehole-temperature data sets cited in the main text. The repository documentation describes the input-data requirements, preprocessing procedures, assimilation settings, and reconstruction and verification workflows.

\clearpage
\setcounter{section}{0}
\setcounter{subsection}{0}
\setcounter{figure}{0}
\setcounter{table}{0}
\setcounter{equation}{0}
\renewcommand{\theequation}{S\arabic{equation}}
\renewcommand{\theHequation}{supp.\arabic{equation}}
\renewcommand{\theHsection}{supp.\arabic{section}}
\renewcommand{\theHfigure}{supp.\arabic{figure}}
\renewcommand{\theHtable}{supp.\arabic{table}}
\section*{Supporting Information}
\noindent Additional methods, validation results, 16 supporting figures, and 6 supporting tables.
\renewcommand{\thefigure}{S\arabic{figure}}
\renewcommand{\thetable}{S\arabic{table}}
\renewcommand{\thesection}{\arabic{section}}
\setcounter{secnumdepth}{1}
\setcounter{tocdepth}{1}

\begin{center}
{\large\bfseries Contents}
\end{center}
\noindent\hyperref[si:sec:si_abbreviations]{\ref*{si:sec:si_abbreviations}. Abbreviations used throughout the manuscript}\dotfill\pageref{si:sec:si_abbreviations}\\
\noindent\hyperref[si:sec:si_lim_derivation]{\ref*{si:sec:si_lim_derivation}. Construction of the reduced-order LIM prior}\dotfill\pageref{si:sec:si_lim_derivation}\\
\noindent\hyperref[si:sec:si_wc4dvar]{\ref*{si:sec:si_wc4dvar}. Weak-constraint 4D-Var in control-variable form}\dotfill\pageref{si:sec:si_wc4dvar}\\
\noindent\hyperref[si:sec:si_pages2k_psm]{\ref*{si:sec:si_pages2k_psm}. PAGES2k proxy system models in the reduced LIM space}\dotfill\pageref{si:sec:si_pages2k_psm}\\
\noindent\hyperref[si:sec:si_borehole_operator]{\ref*{si:sec:si_borehole_operator}. Borehole forward operator}\dotfill\pageref{si:sec:si_borehole_operator}\\
\noindent\hyperref[si:sec:si_borehole_error_cov]{\ref*{si:sec:si_borehole_error_cov}. Borehole observation-error covariance}\dotfill\pageref{si:sec:si_borehole_error_cov}\\
\noindent\hyperref[si:sec:si_trend_sig]{\ref*{si:sec:si_trend_sig}. Significance testing of spatial OHC trends}\dotfill\pageref{si:sec:si_trend_sig}\\
\noindent\hyperref[si:sec:si_ppe]{\ref*{si:sec:si_ppe}. Pseudo-proxy experiment}\dotfill\pageref{si:sec:si_ppe}\\
\noindent\hyperref[si:sec:si_ohc_diagnostics]{\ref*{si:sec:si_ohc_diagnostics}. Additional Common Era reconstruction and validation diagnostics}\dotfill\pageref{si:sec:si_ohc_diagnostics}\\
\noindent\hyperref[si:sec:si_ohc_trend_diagnostics]{\ref*{si:sec:si_ohc_trend_diagnostics}. Sensitivity diagnostics for 130-year OHC trends}\dotfill\pageref{si:sec:si_ohc_trend_diagnostics}\\
\clearpage

\begin{center}
{\large\bfseries List of Tables}
\end{center}
\noindent\hyperref[si:tab:si_abbreviations]{Table~\ref*{si:tab:si_abbreviations}. Common abbreviations used in the manuscript}\dotfill\pageref{si:tab:si_abbreviations}\\
\noindent\hyperref[si:tab:si_models]{Table~\ref*{si:tab:si_models}. Model simulations used to train the LIM priors}\dotfill\pageref{si:tab:si_models}\\
\noindent\hyperref[si:tab:si_eof_variance]{Table~\ref*{si:tab:si_eof_variance}. Retained EOF modes and cumulative explained variance}\dotfill\pageref{si:tab:si_eof_variance}\\
\noindent\hyperref[si:tab:si_experiments]{Table~\ref*{si:tab:si_experiments}. Reconstruction experiments analyzed in the main text}\dotfill\pageref{si:tab:si_experiments}\\
\noindent\hyperref[si:tab:si_computational_costs]{Table~\ref*{si:tab:si_computational_costs}. Computational resources and runtime}\dotfill\pageref{si:tab:si_computational_costs}\\
\noindent\hyperref[si:tab:si_pages2k_selection]{Table~\ref*{si:tab:si_pages2k_selection}. Archive composition of the full and retained PAGES2k networks}\dotfill\pageref{si:tab:si_pages2k_selection}\\

\begin{figure}[p]
\centering
\includegraphics[width=\textwidth,max height=0.65\textheight,keepaspectratio]{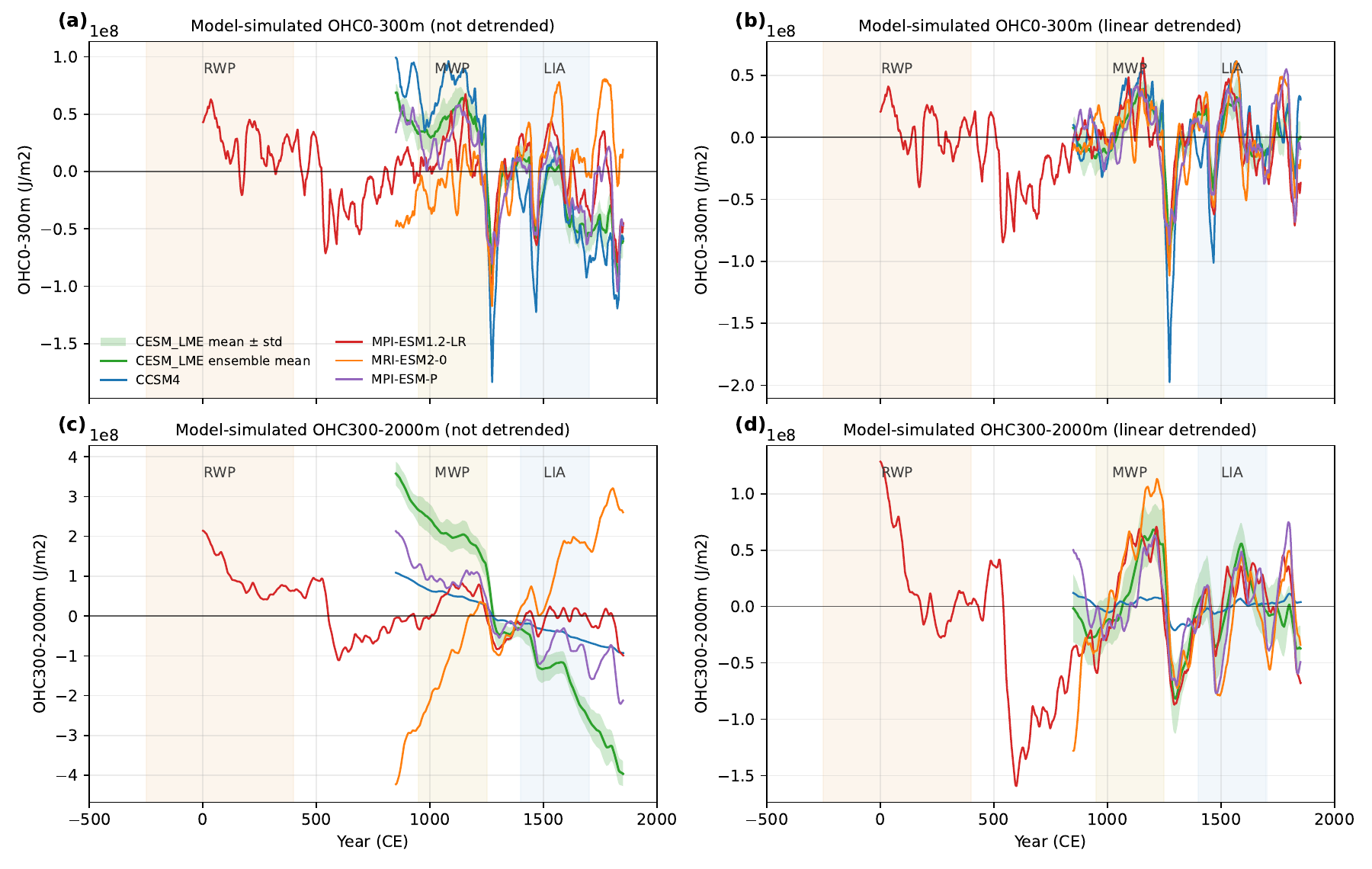}
\caption[Figure S1]{\textbf{Layer-resolved model-simulated ocean heat-content changes.} Upper panels show OHC(0--300 m), and lower panels show OHC(300--2000 m). Left panels show the original model-simulated global-mean OHC series, and right panels show the same series after linear detrending. All series are anomalies relative to the 850--1850 CE mean and are smoothed with a 30-year centered running mean. CESM-LME is shown as the ensemble mean, with shading indicating \(\pm 1\) standard deviation; other model simulations are shown as individual lines. Shaded vertical bands mark the Roman Warm Period (RWP), Medieval Warm Period (MWP), and Little Ice Age (LIA).}
\label{si:fig:si_model_ohc_changes}
\end{figure}

\begin{figure}[!p]
\centering
\includegraphics[width=\textwidth,max height=0.65\textheight,keepaspectratio]{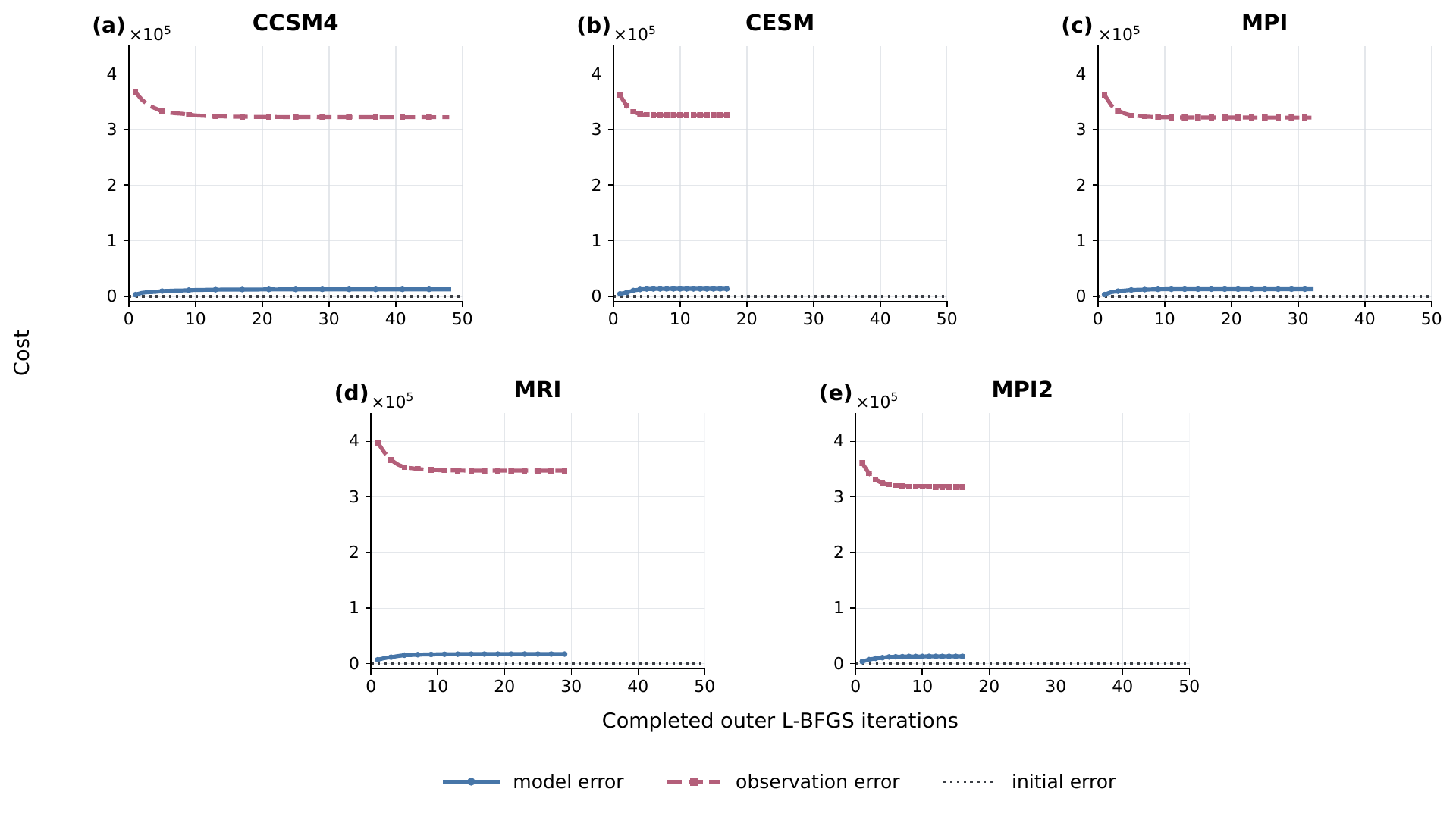}
\caption[Figure S2]{\textbf{Evolution of cost-function components during optimization of the P2k\_BH\_T12k experiment.} This experiment assimilates PAGES2k, borehole, and Temp12k observations. Panels show optimization histories for (a) CCSM4, (b) CESM-LME (labeled CESM), (c) MPI-ESM-P (labeled MPI), (d) MRI-ESM2-0 (labeled MRI), and (e) MPI-ESM1.2-LR (labeled MPI2). Blue solid curves with circles show the model-error penalty, pink dashed curves with squares show the proxy-observation misfit, and gray dotted curves show the initial-state penalty in Equation~\eqref{si:eq:si_final_cost_function_expanded}. Horizontal axes indicate the number of completed outer L-BFGS iterations, and vertical axes show the corresponding cost-function contributions. The final point of each curve gives the component value at the last completed iteration. Other proxy configurations show qualitatively similar optimization histories and are not shown.}
\label{si:fig:si_optimization_cost_history}
\end{figure}

\vspace{1em}
\begin{center}
{\large\bfseries List of Figures}
\end{center}
\noindent\hyperref[si:fig:si_model_ohc_changes]{Figure~\ref*{si:fig:si_model_ohc_changes}. Layer-resolved OHC changes in the model simulations}\dotfill\pageref{si:fig:si_model_ohc_changes}\\
\noindent\hyperref[si:fig:si_optimization_cost_history]{Figure~\ref*{si:fig:si_optimization_cost_history}. Optimization cost histories}\dotfill\pageref{si:fig:si_optimization_cost_history}\\
\noindent\hyperref[si:fig:si_borehole_error]{Figure~\ref*{si:fig:si_borehole_error}. Example construction of borehole observation error}\dotfill\pageref{si:fig:si_borehole_error}\\
\noindent\hyperref[si:fig:si_pseudoproxy_indices]{Figure~\ref*{si:fig:si_pseudoproxy_indices}. Pseudo-proxy comparison with the target CESM-LME simulation}\dotfill\pageref{si:fig:si_pseudoproxy_indices}\\
\noindent\hyperref[si:fig:si_pseudoproxy_boreholes]{Figure~\ref*{si:fig:si_pseudoproxy_boreholes}. Borehole contribution to deep-ocean heat-content recovery in the pseudo-proxy experiment}\dotfill\pageref{si:fig:si_pseudoproxy_boreholes}\\
\noindent\hyperref[si:fig:si_ohc_wu_corr]{Figure~\ref*{si:fig:si_ohc_wu_corr}. Spatial OHC correlations with Wu et al. (2025)}\dotfill\pageref{si:fig:si_ohc_wu_corr}\\
\noindent\hyperref[si:fig:si_ohc_wu_corr_per_model]{Figure~\ref*{si:fig:si_ohc_wu_corr_per_model}. Model-specific spatial OHC correlations with Wu et al. (2025)}\dotfill\pageref{si:fig:si_ohc_wu_corr_per_model}\\
\noindent\hyperref[si:fig:si_ohc_wu_trend_compare]{Figure~\ref*{si:fig:si_ohc_wu_trend_compare}. Comparison of 1880--2000 CE OHC trends with Wu et al. (2025)}\dotfill\pageref{si:fig:si_ohc_wu_trend_compare}\\
\noindent\hyperref[si:fig:si_ohc_wu_trend_per_model]{Figure~\ref*{si:fig:si_ohc_wu_trend_per_model}. Model-specific OHC trends compared with Wu et al. (2025)}\dotfill\pageref{si:fig:si_ohc_wu_trend_per_model}\\
\noindent\hyperref[si:fig:si_ohc_external_experiments]{Figure~\ref*{si:fig:si_ohc_external_experiments}. Global-mean OHC(300--2000 m) comparisons across experiments}\dotfill\pageref{si:fig:si_ohc_external_experiments}\\
\noindent\hyperref[si:fig:si_ohc_external]{Figure~\ref*{si:fig:si_ohc_external}. Layer-resolved OHC comparison with an external reconstruction}\dotfill\pageref{si:fig:si_ohc_external}\\
\noindent\hyperref[si:fig:si_ohc_mwp_lia_trend]{Figure~\ref*{si:fig:si_ohc_mwp_lia_trend}. OHC trends across the MWP-to-LIA transition}\dotfill\pageref{si:fig:si_ohc_mwp_lia_trend}\\
\noindent\hyperref[si:fig:figure14a]{Figure~\ref*{si:fig:figure14a}. Pairwise reconstruction comparisons, part A}\dotfill\pageref{si:fig:figure14a}\\
\noindent\hyperref[si:fig:figure14b]{Figure~\ref*{si:fig:figure14b}. Pairwise reconstruction comparisons, part B}\dotfill\pageref{si:fig:figure14b}\\
\noindent\hyperref[si:fig:si_ppe_ohc_trend_distributions]{Figure~\ref*{si:fig:si_ppe_ohc_trend_distributions}. Pseudo-proxy distributions of 130-year OHC trends}\dotfill\pageref{si:fig:si_ppe_ohc_trend_distributions}\\
\noindent\hyperref[si:fig:si_ohc_trend_nonoverlap]{Figure~\ref*{si:fig:si_ohc_trend_nonoverlap}. Non-overlapping historical OHC trend windows}\dotfill\pageref{si:fig:si_ohc_trend_nonoverlap}\\
\clearpage

\sisection{Abbreviations used throughout the manuscript}{si:sec:si_abbreviations}
Table~\ref{si:tab:si_abbreviations} summarizes the abbreviations used repeatedly in the main text and supporting information. After first definition in the main text, these abbreviations are used consistently throughout the manuscript.

\begin{table}[!htbp]
\centering
\caption[Table S1]{\textbf{Common abbreviations used in the manuscript.}}
\label{si:tab:si_abbreviations}
\begin{tabular}{ll}
\toprule
Abbreviation & Meaning \\
\midrule
LMR4D-Var & Last Millennium Reanalysis 4D-Var Framework\\
LIM & Linear Inverse Model \\
PSM & Proxy system model \\
CoE & Coefficient of Efficiency \\
CE & Common Era \\
GMT & Global mean temperature \\
OHC & Ocean heat content \\
OHC0--300 & Ocean heat content integrated over 0--300 m \\
OHC300--2000 & Ocean heat content integrated over 300--2000 m \\
OHC0--2000 & Ocean heat content integrated over 0--2000 m \\
PC & Principle Component \\
TAS & Near-surface air temperature \\
TOS & Sea-surface temperature \\
SIT & Northern Hemisphere sea-ice thickness \\
SIC & Northern Hemisphere sea-ice concentration \\
RWP & Roman Warm Period \\
MWP & Medieval Warm Period \\
LIA & Little Ice Age \\
\bottomrule
\end{tabular}
\end{table}

\sisection{Construction of the reduced-order LIM prior}{si:sec:si_lim_derivation}
The reduced-order dynamical prior is represented by a discrete Linear Inverse Model (LIM),
\begin{equation}
\state_{k+1} = \mathbf{G}\,\state_{k} + \mathbf{n}_{k},
\label{si:eq:si_discrete_lim}
\end{equation}
where $\state_k$ is the reduced climate state at seasonal analysis step $k$, $\mathbf{G}$ is the one-step propagator, and $\mathbf{n}_k$ is an additive model-error increment. Following standard LIM estimation, $\mathbf{G}$ is obtained from the lag-0 and lag-$\tau$ covariance matrices of the training trajectory \cite{penlandPredictionNinoSea1993}:
\begin{equation}
\mathbf{G}=\mathbf{C}(\tau)\mathbf{C}(0)^{-1}.
\label{si:eq:si_G_final}
\end{equation}
Assuming a stationary training process and model-error increments that are uncorrelated with the preceding state, the corresponding discrete model-error covariance is
\begin{equation}
\Qk = \mathbf{C}(0) - \mathbf{G}\mathbf{C}(0)\mathbf{G}^{T}.
\label{si:eq:si_Q_final}
\end{equation}
Thus $\mathbf{n}_k\sim\mathcal{N}(\mathbf{0},\Qk)$. The increments are assumed to be temporally uncorrelated between seasonal steps, while $\Qk$ retains the full covariance among reduced-state components within each step. Equations~\eqref{si:eq:si_discrete_lim}--\eqref{si:eq:si_Q_final} fully specify the discrete LIM prior used by LMR4D-Var; the construction of its reduced state and training data is described next.

\label{si:sec:si_lim_training}

The LIM is trained in a reduced EOF space built from coupled climate-model integrations. In the present study the state vector contains principal components for six variable groups: 2-m air temperature (TAS), sea-surface temperature (TOS), ocean heat content in the upper 300 m (OHC0--300), ocean heat content integrated over 300--2000 m (OHC300--2000), Northern Hemisphere sea-ice thickness (SIT), and Northern Hemisphere sea-ice concentration (SIC). We use the first 15 EOF modes for each variable except for OHC0--300 and OHC300--2000, for which we use the first 30 modes to capture the more complex spatial structure of upper-ocean heat content and low-frequency variability in the ocean interior. The total dimension of the reduced state is therefore 120. The reduced state is written schematically as
\begin{equation}
\state=
\left[
\mathbf{PC}_{\mathrm{TAS}}^{T},
\mathbf{PC}_{\mathrm{TOS}}^{T},
\mathbf{PC}_{\mathrm{OHC0\text{--}300}}^{T},
\mathbf{PC}_{\mathrm{OHC300\text{--}2000}}^{T},
\mathbf{PC}_{\mathrm{SIT}}^{T},
\mathbf{PC}_{\mathrm{SIC}}^{T}
\right]^{T}.
\label{si:eq:si_lim_state_vector}
\end{equation}
Upper-2000-m ocean heat content is then diagnosed as
\begin{equation}
\mathrm{OHC}_{0\text{--}2000} = \mathrm{OHC}_{0\text{--}300} + \mathrm{OHC}_{300\text{--}2000}.
\label{si:eq:si_ohc2000_sum}
\end{equation}
Here, $\mathrm{OHC}_{0\text{--}300}$ and $\mathrm{OHC}_{300\text{--}2000}$ are diagnosed from the respective EOF modes. This partition is useful because it allows the LIM to represent the fast upper-ocean component and the slower subsurface reservoir separately while still recovering the full upper-2000-m quantity analyzed in the main text.

Each variable is area weighted before EOF analysis. EOF truncation is applied separately to each variable class to limit the dimension of the coupled state while retaining the leading variance and covariance structures needed for prediction. In practice this truncation is a compromise: too few modes fail to capture important coupled variability, whereas too many modes can degrade skill and overfit noise. The resulting EOF-based state vectors are assembled from seasonal-mean fields and used to estimate lagged covariance statistics for each training ensemble.

Before the LIM is estimated, seasonal-mean anomalies are formed by subtracting the corresponding seasonal climatology after remapping and seasonal averaging. Each parent simulation is also linearly detrended over its training interval. This detrending removes slow unrealistically persistent adjustment associated with climate-model drift, or long equilibration of the ocean background state. If such behavior were left in the training data, the fitted LIM operator would absorb that nonstationary drift into its linear dynamics, which would in turn bias the propagator toward unstable or physically misleading low-frequency tendencies. The detrending step is therefore intended to isolate internally covarying seasonal-to-multidecadal variability rather than the model's particular centennial-scale drift history. The original and detrended OHC series are shown in Fig.~\ref{si:fig:si_model_ohc_changes}.

The LIMs are then trained from multiple coupled climate-model priors, including the CESM Last Millennium Ensemble and PMIP-class millennium simulations from MPI-ESM-P, MPI-ESM1.2-LR, MRI-ESM2-0, and CCSM4, after remapping to a common grid, forming seasonal means, and detrending the training trajectories. Training separate LIM priors from multiple parent simulations allows the reconstruction system to sample a broader range of coupled covariance structures and reduces dependence on any one model's drift history or initial condition. Table~\ref{si:tab:si_models} summarizes the training simulations. Table~\ref{si:tab:si_eof_variance} reports the retained EOF counts and cumulative explained variance for each variable and model, and Table~\ref{si:tab:si_experiments} summarizes the proxy composition and purpose of the reconstruction experiments.

\begin{table}[!tbp]
\centering
\caption[Table S2]{\textbf{Model simulations used to train the LIM priors.} Each model family is used to estimate lagged covariance statistics and train a separate reduced-order prior.}
\label{si:tab:si_models}
\small
\setlength{\tabcolsep}{4pt}
\begin{tabularx}{\textwidth}{lY Y lY}
\toprule
Model source & Model name & Experiment class & Simulation period & Role in LMR4D-Var \\
\midrule
CESM LME & CESM Last Millennium Ensemble & CESM millennium ensemble & 850--1850 CE & Multimember coupled prior \\
PMIP3 & MPI-ESM-P & Last millennium simulation & 850--1850 CE & Single-model LIM prior \\
PMIP4 & MPI-ESM1.2-LR & Last millennium simulation & 850--1850 CE & Single-model LIM prior \\
PMIP4 & MRI-ESM2-0 & Last millennium simulation & 0--1850 CE & Single-model LIM prior \\
PMIP3 & CCSM4 & Last millennium simulation & 850--1850 CE & Single-model LIM prior \\
\bottomrule
\end{tabularx}
\end{table}

\begin{table}[!tbp]
\centering
\caption[Table S3]{\textbf{Retained EOF modes and cumulative explained variance.} Entries in the five model columns give the cumulative percentage of variance explained by the retained EOFs for each variable and parent climate-model product. CESM-LME denotes the CESM Last Millennium Ensemble. SST corresponds to TOS in the reduced state vector, and NH denotes the Northern Hemisphere.}
\label{si:tab:si_eof_variance}
\small
\setlength{\tabcolsep}{4pt}
\begin{tabularx}{\textwidth}{Ycrrrrr}
\toprule
Variable & EOFs & CCSM4 & \shortstack{CESM-\\LME} & \shortstack{MPI-\\ESM-P} & \shortstack{MRI-\\ESM2-0} & \shortstack{MPI-\\ESM1.2-LR} \\
\midrule
Near-surface air temperature (TAS) & 15 & 74.83\% & 74.75\% & 65.41\% & 69.60\% & 65.22\% \\
Sea-surface temperature (SST) & 15 & 73.06\% & 75.59\% & 50.52\% & 65.13\% & 52.41\% \\
OHC(0--300 m) & 30 & 77.20\% & 81.24\% & 37.28\% & 67.65\% & 39.33\% \\
OHC(300--2000 m) & 30 & 66.56\% & 65.13\% & 36.66\% & 68.16\% & 40.23\% \\
NH sea-ice concentration (SIC) & 15 & 67.04\% & 64.83\% & 68.07\% & 67.83\% & 67.50\% \\
NH sea-ice thickness (SIT) & 15 & 73.16\% & 77.29\% & 85.42\% & 73.73\% & 88.67\% \\
\bottomrule
\end{tabularx}
 \end{table}

\begin{table}[!tbp]
\centering
\caption[Table S4]{\textbf{Summary of the five reconstruction experiments analyzed in the main text.} The experiment design isolates the effect of adding low-frequency and long-memory proxy information and includes a Temp12k-only experiment for Holocene-scale sensitivity tests.}
\label{si:tab:si_experiments}
\begin{tabularx}{\textwidth}{llllY}
\toprule
Experiment & PAGES2k & Temp12k & Boreholes & Purpose \\
\midrule
P2k & Yes & No & No & Baseline Common Era reconstruction \\
P2k\_BH & Yes & No & Yes & Isolate borehole long-memory constraints \\
P2k\_T12k & Yes & Yes & No & Isolate Temp12k low-frequency constraints \\
P2k\_BH\_T12k & Yes & Yes & Yes & Combined multiscale reconstruction \\
T12k & No & Yes & No & Temp12k-only Holocene and \(J_m\) sensitivity tests \\
\bottomrule
\end{tabularx}
\end{table}

\FloatBarrier

\sisection{Weak-constraint 4D-Var in control-variable form}{si:sec:si_wc4dvar}

For seasonal states $\state_0,\state_1,\ldots,\state_N$, we jointly optimize the initial state and additive model-error increments \cite{tremoletAccountingImperfectModel2006,fisherWeakConstraintLongWindow2011}. These unknowns form the control vector
\begin{equation}
\mathbf{v}=
\begin{pmatrix}
\state_0 \\
\mathbf{n}_0 \\
\mathbf{n}_1 \\
\vdots \\
\mathbf{n}_{N-1}
\end{pmatrix}.
\label{si:eq:si_control_vector}
\end{equation}
Applying Equation~\eqref{si:eq:si_discrete_lim} recursively gives the state at seasonal step $k$ as
\begin{equation}
\state_k
=
\mathbf{G}^{k}\state_0
+ \sum_{j=0}^{k-1}\mathbf{G}^{k-1-j}\mathbf{n}_j,
\qquad k=1,\ldots,N.
\label{si:eq:si_explicit_trajectory}
\end{equation}
Each increment modifies the next seasonal state and propagates through $\mathbf{G}$ to subsequent states. Stacking the seasonal states gives the linear control-to-trajectory mapping \cite{fisherKalmanSmoothing2005,fisherWeakConstraintLongWindow2011,evensen2022data}:
\begin{equation}
\state=\mathbf{U}\mathbf{v},
\label{si:eq:si_state_control_transform}
\end{equation}
where $\state=(\state_0,\state_1,\dots,\state_N)^T$ and $\mathbf{U}$ is the lower block-triangular matrix
\begin{equation}
\mathbf{U}=
\begin{pmatrix}
\mathbf{I} & \mathbf{0} & \mathbf{0} & \cdots & \mathbf{0} \\
\mathbf{G} & \mathbf{I} & \mathbf{0} & \cdots & \mathbf{0} \\
\mathbf{G}^2 & \mathbf{G} & \mathbf{I} & \cdots & \mathbf{0} \\
\vdots & \vdots & \vdots & \ddots & \vdots \\
\mathbf{G}^{N} & \mathbf{G}^{N-1} & \mathbf{G}^{N-2} & \cdots & \mathbf{I}
\end{pmatrix}.
\label{si:eq:si_U_matrix}
\end{equation}

The initial state and model-error increments are uncorrelated a priori, as are increments at different seasonal steps. Within each step, $\Qk$ represents spatial and cross-variable covariances among the reduced-state components. These assumptions give a block-diagonal control-space prior covariance \cite{tremoletAccountingImperfectModel2006,fisherWeakConstraintLongWindow2011,evensen2022data}:
\begin{equation}
\mathbf{B}_{v}=
\begin{pmatrix}
\mathbf{C}(0) & \mathbf{0} & \cdots & \mathbf{0} \\
\mathbf{0} & \Qk & \cdots & \mathbf{0} \\
\vdots & \vdots & \ddots & \vdots \\
\mathbf{0} & \mathbf{0} & \cdots & \Qk
\end{pmatrix},
\label{si:eq:si_Bv}
\end{equation}
Here $\mathbf{C}(0)$ is the stationary covariance of the initial reduced state, and $\Qk$ is defined by Equation~\eqref{si:eq:si_Q_final}.

Let $\mathcal{H}$ collect the proxy-specific observation operators, including their temporal averaging or memory, and let $\Rk$ denote the observation-error covariance. Under the assumed Gaussian prior and error models, the cost function is \cite{LeDimetTalagrand1986,fisherKalmanSmoothing2005,tremoletAccountingImperfectModel2006,evensen2022data}
\begin{equation*}
J(\mathbf{v})
=
\frac{1}{2}\mathbf{v}^{T}\mathbf{B}_{v}^{-1}\mathbf{v}
+ \frac{1}{2}\left[\obs-\mathcal{H}(\mathbf{U}\mathbf{v})\right]^{T}
\Rk^{-1}
\left[\obs-\mathcal{H}(\mathbf{U}\mathbf{v})\right].
\end{equation*}
Expanding the control-space prior term separates the initial-state and model-error contributions:
\begin{equation}
J(\mathbf{v})
=
\underbrace{\frac{1}{2}\state_0^{T}\mathbf{C}(0)^{-1}\state_0}_{\text{initial-state constraint}}
+ \underbrace{\frac{1}{2}\sum_{k=0}^{N-1}\mathbf{n}_k^{T}\Qk^{-1}\mathbf{n}_k}_{\text{model-error penalty}}
+ \underbrace{\frac{1}{2}\left[\obs-\mathcal{H}(\mathbf{U}\mathbf{v})\right]^T
\Rk^{-1}
\left[\obs-\mathcal{H}(\mathbf{U}\mathbf{v})\right]}_{\text{proxy observation misfit}}.
\label{si:eq:si_final_cost_function_expanded}
\end{equation}

The optimized control vector $\mathbf{v}^{*}$ gives the analyzed trajectory $\state_a=\mathbf{U}\mathbf{v}^{*}$. This formulation lets history-dependent observations constrain multiple analysis times without constructing a dense covariance for the full trajectory \cite{fisherKalmanSmoothing2005,fisherWeakConstraintLongWindow2011}.

Under the assumed Gaussian prior and error models, the cost function is the negative log posterior density of the control vector, up to an additive constant \cite{evensen2022data}. Minimizing it therefore yields a maximum a posteriori (MAP) estimate of the control vector and its associated climate trajectory, conditional on the prescribed emulator and error covariances. For linear dynamics and observation operators with Gaussian prior and error distributions, weak-constraint 4D-Var yields the same posterior-mean trajectory as the corresponding fixed-interval Kalman smoother \cite{fisherKalmanSmoothing2005}. This optimization does not itself sample the posterior distribution.

In a strong-constraint formulation, all model-error increments would be fixed to zero, giving $\state_k=\mathbf{G}^{k}\state_0$. For a stable, unforced LIM, initial anomalies decay, limiting the ability of this trajectory to fit proxy-constrained variability throughout a long reconstruction window.

Although temporally independent model-error increments can generate persistent climate variability through LIM dynamics, persistent model deficiencies may require coherent corrections over many steps. The independent-increment prior penalizes these corrections separately, potentially limiting the recovery of low-frequency variability. Temporally correlated model-error priors could provide a more flexible representation of such persistent deficiencies.

Figure~\ref{si:fig:si_optimization_cost_history} shows the three cost-function components across completed outer L-BFGS iterations for the P2k\_BH\_T12k experiment using the five LIM emulators. The proxy-observation misfit dominates the cost throughout optimization and decreases overall across all five emulators. The model-error penalty generally increases during the early iterations before approaching a plateau. The initial-state penalty remains negligible relative to the other two terms on the plotted scale.

\FloatBarrier

\subsection{Computational costs}

\begin{table}[!htbp]
\centering
\caption[Table S5]{\textbf{Computational resources and runtime.}}
\label{si:tab:si_computational_costs}
\small
\setlength{\tabcolsep}{4pt}
\renewcommand{\arraystretch}{1.08}
\begin{tabularx}{\textwidth}{@{}p{0.40\textwidth}>{\centering\arraybackslash}X>{\centering\arraybackslash}X@{}}
\toprule
Quantity & Common Era (500BCE -- 2000CE) & Holocene (10000BCE -- 2000CE) \\
\midrule
Experiment & P2k\_BH\_T12k & P2k\_BH\_T12k \\
GPU device & \multicolumn{2}{c}{1 NVIDIA A100 (80 GB)} \\
Average runtime & 29 min 9 s & 5 h 0 min 44 s \\
Peak host memory (GiB) & 31.71 & 80.00 \\
\bottomrule
\end{tabularx}
 \end{table}

\FloatBarrier

\sisection{PAGES2k proxy system models in the reduced LIM space}{si:sec:si_pages2k_psm}

The annually resolved baseline network uses the temperature-sensitive PAGES2k phase 2 proxy compilation \cite{PAGES2k2017global}. These records include tree rings, corals, ice cores, and other annually to seasonally resolved archives. In LMR4D-Var they are assimilated through simple temperature-sensitive proxy system models (PSMs) calibrated against instrumental target fields rather than against the full unreduced climate state.

Calibration is performed separately for terrestrial and marine records. Terrestrial proxies are calibrated against instrumental near-surface air temperature from GISTEMP v4 \cite{lenssen2019improvements}, whereas marine proxies are calibrated against instrumental sea-surface temperature from ERSST v5 \cite{huang2017extended}. Moisture variables are not included in the PSM calibration. This choice follows the temperature-centered design of the present reconstruction and avoids introducing additional predictor fields that are not represented uniformly across the reduced LIM state.

Because the LIM evolves a truncated EOF representation of the coupled climate state rather than the full local gridpoint climate field, the PSMs are calibrated against instrumental fields projected onto the same retained EOF subspace. Let $\mathbf{T}$ denote the original instrumental temperature field and $\widehat{\mathbf{T}}$ its EOF-truncated projection. Then
\begin{equation}
\widehat{\mathbf{T}}=\mathbf{E}\mathbf{E}^{T}\mathbf{T},
\label{si:eq:si_psm_projection}
\end{equation}
where $\mathbf{E}$ denotes the EOF basis matrix for the relevant climate field. The linear PSM for an individual record is written as
\begin{equation}
\obs = \mathbf{H}\widehat{\mathbf{T}} + \boldsymbol{\epsilon},
\label{si:eq:si_psm_linear}
\end{equation}
where $\mathbf{H}$ maps the truncated climate field to the proxy estimate and $\boldsymbol{\epsilon}$ is the residual error term. Calibrating in the EOF-truncated space partially absorbs representativeness error associated with unresolved local climate structure.

Observation-error statistics are represented with a diagonal covariance matrix, so that
\begin{equation}
\Rk = \langle \boldsymbol{\epsilon}\boldsymbol{\epsilon}^{T}\rangle
\end{equation}
is assumed to have negligible off-diagonal covariance among distinct PAGES2k records. This approximation is adopted because the calibration interval is too short and unevenly overlapping to estimate a stable full proxy-error covariance matrix, and exploratory tests indicate that the diagonal error variances dominate the off-diagonal terms.

Each record is calibrated by univariate linear regression against temperature at its location in the EOF-truncated instrumental field. Calibration uses the available proxy--instrumental overlap within 1850--2015 CE. Seasonality is selected by the best calibration fit among ten candidate averaging windows. These comprise DJF, MAM, JJA, and SON; December--May, March--August, and June--November; December--August and March--November; and December--November. For windows starting in December, December belongs to the preceding calendar year. Sub-seasonal coral records are first aggregated to seasonal resolution before calibration and assimilation.

Following LMR Seasonal \cite{meng2025coupled}, records are excluded if their PSM calibration correlation is below 0.05 or their calibration residuals have a one-year lag autocorrelation exceeding 0.90. The autocorrelation threshold excludes only strongly persistent residuals, so weaker residual autocorrelation can remain in the retained network. The short calibration interval also limits detection of longer-timescale residual persistence.

The full PAGES2k v2.0.0 compilation contains 692 temperature-sensitive records \cite{PAGES2k2017global}. The final assimilated PAGES2k network contains 506 records, leaving 186 records from the full compilation outside the retained network.

Table~\ref{si:tab:si_pages2k_selection} gives the archive-specific counts. The retained tree-ring records comprise 299 tree-ring-width (TRW) and 60 maximum-latewood-density (MXD) records. The table reports differences between the full compilation and the retained network, not rejection counts attributed to individual calibration criteria.

\begin{table}[!htbp]
\centering
\caption[Table S6]{\textbf{Archive composition of the full and retained PAGES2k networks.} Full-compilation counts follow \citeA{PAGES2k2017global}; retained counts correspond to the PAGES2k network shown in the main text. ``Not retained'' is the difference between these columns, not a rejection count for an individual screening criterion. Tree-ring records combine TRW and MXD. Other archives combine documentary records, sclerosponges, boreholes, bivalves, and hybrid tree/borehole records.}
\label{si:tab:si_pages2k_selection}
\begingroup
\small
\setlength{\tabcolsep}{6pt}
\begin{tabularx}{\linewidth}{@{}Yrrr@{}}
\toprule
Archive category & Full compilation & Retained & Not retained \\
\midrule
Tree rings & 415 & 359 & 56 \\
Corals & 96 & 91 & 5 \\
Ice cores & 49 & 30 & 19 \\
Lake sediments & 42 & 10 & 32 \\
Marine sediments & 58 & 9 & 49 \\
Speleothems & 4 & 3 & 1 \\
Other archives & 28 & 4 & 24 \\
\midrule
Total & 692 & 506 & 186 \\
\bottomrule
\end{tabularx}
\endgroup
 \end{table}

\FloatBarrier

\sisection{Borehole forward operator: from surface temperature history to subsurface profile}{si:sec:si_borehole_operator}

The borehole observation operator links a reconstructed ground-surface temperature history to the present-day temperature anomaly profile measured in the subsurface. Following standard borehole climatology theory \cite{huang2000temperature,cuestaValero2021longterm}, the continental subsurface is treated as a one-dimensional semi-infinite conductive medium with depth coordinate $z$ positive downward. The observed temperature profile is written as the sum of a long-term quasi-equilibrium profile and a transient perturbation,
\begin{equation}
T(z)=T_0+\Gamma_0 z + T_t(z),
\label{si:eq:si_borehole_profile_decomp}
\end{equation}
where $T_0$ is the long-term surface temperature, $\Gamma_0$ is the background geothermal gradient, and $T_t(z)$ is the transient anomaly induced by past changes in surface temperature.

For a homogeneous medium without internal heat sources, the transient temperature evolves according to the one-dimensional heat-diffusion equation
\begin{equation}
\frac{\partial T}{\partial t}=\kappa\frac{\partial^2 T}{\partial z^2},
\label{si:eq:si_heat_diffusion}
\end{equation}
where $\kappa$ is thermal diffusivity. An instantaneous surface step change propagates downward as a complementary-error-function profile, so a discretized surface-temperature history can be represented as a superposition of step responses. This solution also implies a characteristic diffusion timescale
\begin{equation*}
t \sim \frac{Z^2}{4\kappa},
\end{equation*}
obtained by setting the similarity variable $\eta=z/(2\sqrt{\kappa t})$ to order one. The relationship shows that the response time grows quadratically with depth: for $\kappa \sim 10^{-6}\ \mathrm{m^2\,s^{-1}}$, temperatures at 100 m depth primarily reflect variability on the order of many decades to about a century, with deeper levels integrating still longer timescales. If the surface-temperature history is sampled at discrete times and written as $\mathbf{T}_s=\left(T_s(t_1),T_s(t_2),\dots,T_s(t_{N_t})\right)^T$, then the transient anomaly at depth $z_i$ can be written as
\begin{equation}
T_t(z_i)=\sum_{j=1}^{N_t}T_s(t_j)
\left[
\operatorname{erfc}\!\left(\frac{z_i}{2\sqrt{\kappa t_j}}\right)
-
\operatorname{erfc}\!\left(\frac{z_i}{2\sqrt{\kappa t_{j-1}}}\right)
\right].
\label{si:eq:si_borehole_forward_sum}
\end{equation}
This is the forward model that maps a surface-temperature history to a borehole temperature anomaly profile.

In matrix form, the discrete borehole operator can be written as
\begin{equation}
\mathbf{y}_{\mathrm{bh}}=\mathbf{M}\mathbf{T}_s,
\label{si:eq:si_borehole_matrix_form}
\end{equation}
where $\mathbf{T}_s$ is the discretized surface-temperature history and $\mathbf{y}_{\mathrm{bh}}$ is the predicted borehole temperature anomaly profile at the observed depths. In other words, the borehole forward model in LMR4D-Var maps surface temperature through time directly to subsurface temperature through depth, $\mathbf{T}_s \mapsto \mathbf{y}_{\mathrm{bh}}$. The elements of $\mathbf{M}$ are
\begin{equation}
M_{ij}=
\operatorname{erfc}\!\left(\frac{z_i}{2\sqrt{\kappa t_j}}\right)
-
\operatorname{erfc}\!\left(\frac{z_i}{2\sqrt{\kappa t_{j-1}}}\right).
\label{si:eq:si_borehole_kernel}
\end{equation}
Within LMR4D-Var, $\mathbf{T}_s$ is supplied by the reconstructed near-surface temperature history at each borehole location, and the resulting profile is compared with the observed borehole anomaly profile in the observation-misfit term of the weak-constraint 4D-Var objective. In this way the borehole proxy constrains the low-frequency integrated thermal history of the land surface rather than the instantaneous climate state at a single analysis time. Because the purpose of the borehole constraint here is to isolate long-timescale information, only the part of each borehole profile below 100 m is used in the forward comparison and subsequent error construction.

\sisection{Borehole observation-error covariance}{si:sec:si_borehole_error_cov}

\begin{figure}[!tbp]
\centering
\includegraphics[width=\textwidth,max height=0.65\textheight,keepaspectratio]{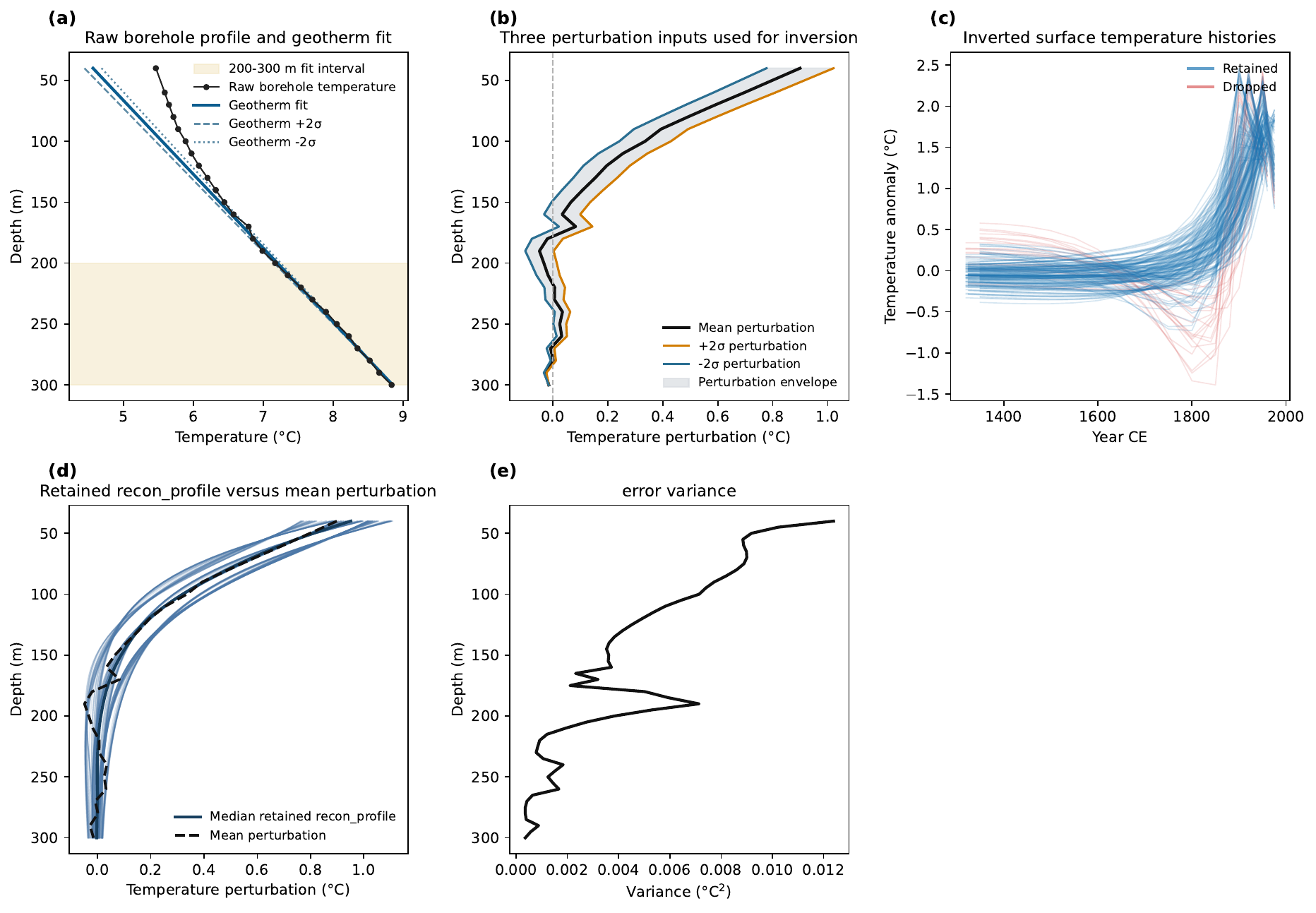}
\caption[Figure S3]{\textbf{Example construction of borehole error for station CA\_0011.} (a) Raw borehole temperature profile (black) as a function of depth, together with the mean geothermal fit estimated from the 200--300 m interval (blue) and the corresponding $\pm 2\sigma$ geothermal fits (dashed and dotted lines) derived from uncertainty in the fitted gradient. (b) Temperature perturbation profiles obtained by subtracting the mean and $\pm 2\sigma$ geothermal fits from the raw borehole temperature profile. These three perturbation estimates define the perturbation envelope associated with uncertainty in the background geothermal trend. (c) Surface-temperature histories inferred from the perturbation profiles using the same inverse geothermal reconstruction method as in Section~\ref{si:sec:si_borehole_error_cov}. (d) Perturbation profiles reconstructed by forward-projecting the retained temperature histories back into depth space. (e) Depth-dependent variance of the residual between the forward-reconstructed perturbation profiles and the original perturbation profile. This residual variance is used to approximate the borehole observation-error covariance, with the diagonal borehole error matrix constructed from the depthwise mean squared residuals.}
\label{si:fig:si_borehole_error}
\end{figure}

The borehole observation error used in the present study is intended as an effective error model rather than as a pure estimate of measurement noise alone. In practice, borehole profiles can be affected by site-specific environmental disturbances and observational complications, including groundwater flow, snow and surface-condition effects, and other departures from ideal one-dimensional conductive behavior. A second source of uncertainty arises from the forward model itself: LMR4D-Var uses a simplified conductive diffusion model to simulate borehole temperatures, and any variance not captured by that model must also be absorbed into the observation-error description. Rather than assigning a depth-independent error or a single site-wide scalar uncertainty, we therefore estimate a depth-dependent residual variance for each borehole profile and use that quantity to define the diagonal borehole error covariance. Fig.~\ref{si:fig:si_borehole_error} summarizes this procedure panel by panel.

The construction begins from the raw borehole temperature profile $T_{\mathrm{raw}}(z)$ shown in panel A. Following the geothermal treatment used in borehole climatology studies \cite{cuestaValero2021longterm}, we estimate the quasi-equilibrium geothermal profile by fitting a straight line over the 200--300 m interval,
\begin{equation*}
T_{\mathrm{geo}}(z)=a+bz,
\end{equation*}
where the fitted slope $b$ defines the local geothermal gradient. Uncertainty in the fitted slope is propagated to construct a mean geothermal fit together with upper and lower $\pm 2\sigma$ fits, also shown in panel A. Subtracting these fits from the raw profile yields the perturbation profiles shown in panel B,
\begin{equation*}
T_{\mathrm{pert}}(z)=T_{\mathrm{raw}}(z)-T_{\mathrm{geo}}(z),
\end{equation*}
including the mean perturbation profile and the perturbation envelope associated with uncertainty in the background geothermal trend.

Each perturbation profile is then inverted to obtain a candidate ground-surface temperature history, as illustrated in panel C, using the same inverse geothermal reconstruction framework as in \citeA{cuestaValero2021longterm}. In that framework the perturbation profile is written in matrix form as
\begin{equation*}
\mathbf{y}_{\mathrm{bh}}=\mathbf{M}\mathbf{T}_s,
\end{equation*}
and the inverse problem is solved with singular value decomposition (SVD). If
\begin{equation*}
\mathbf{M}=\mathbf{U}\mathbf{S}\mathbf{V}^{T},
\end{equation*}
then the candidate surface-temperature history is obtained from the truncated pseudo-inverse,
\begin{equation*}
\widehat{\mathbf{T}}_{s}^{(k)}=\mathbf{V}\mathbf{S}_{k}^{-1}\mathbf{U}^{T}\mathbf{y}_{\mathrm{bh}},
\end{equation*}
where $\mathbf{S}_{k}^{-1}$ retains only the leading $k$ singular values and sets the smaller ones to zero. Writing the inverse explicitly in this way is useful because the truncation level is one of the ensemble dimensions used to characterize borehole uncertainty.

Following \citeA{cuestaValero2021longterm}, we consider step-change models with 25-, 40-, and 50-year temporal resolution, corresponding to 16, 10, and 8 time steps over the reconstructed interval. We also allow the thermal properties entering the conductive kernel to vary across a broad but physically plausible range. Specifically, the inversion ensemble spans thermal conductivities of 2.5, 3.0, and 3.5 W\,m$^{-1}$\,K$^{-1}$ and volumetric heat capacities of $2.5$, $3.0$, and $3.5\times10^{6}$ J\,m$^{-3}$\,K$^{-1}$, which together yield nine thermal diffusivities ranging from approximately $0.7\times10^{-6}$ to $1.4\times10^{-6}$ m$^{2}$\,s$^{-1}$. We use the same SVD truncation strategy: the 25-year model retains the leading 3, 4, or 5 singular values; the 40-year model retains the leading 2, 3, or 4 singular values; and the 50-year model also retains the leading 2, 3, or 4 singular values. These alternative truncations suppress unrealistically amplified small-eigenvalue solutions while still allowing a broad family of admissible inverse histories.

The total number of candidate histories generated from a single original borehole profile is therefore
\begin{equation*}
N_{\mathrm{cand}} = 3 \times 9 \times 3 \times 3 = 243,
\end{equation*}
where the four factors correspond to the three perturbation profiles from panel B, the nine thermal-property combinations, the three step-change models, and the three SVD truncation choices available within each step-change model. The borehole inversion in panel C should therefore be interpreted not as a single reconstruction but as a 243-member inverse ensemble for each site.

In the present implementation, the candidate histories obtained from this large inversion ensemble are then filtered, as shown in panel C, to remove solutions with implausibly large temporal gradients exceeding 4 K per decade. This step is intended to discard unstable inverse solutions that fit the perturbation profile only by introducing unrealistically abrupt or high-magnitude temperature changes. The retained ensemble therefore represents the subset of inverse solutions that remain physically plausible while still reproducing the observed perturbation structure.

The retained temperature histories are then projected forward again through the borehole operator to reconstruct perturbation profiles in depth space, producing the curves shown in panel D:
\begin{equation*}
\widehat{\mathbf{y}}_{\mathrm{bh}}=\mathbf{M}\widehat{\mathbf{T}}_{s},
\end{equation*}
where $\widehat{\mathbf{T}}_{s}$ denotes a retained inverse temperature history. These forward-reconstructed perturbation profiles are compared with the original perturbation profile from panel B to form a depth-dependent residual ensemble. Let $r_i^{(m)}$ denote the residual at depth $z_i$ for retained solution $m$,
\begin{equation*}
r_i^{(m)} = \widehat{y}_{\mathrm{bh},i}^{(m)} - y_{\mathrm{bh},i}^{\mathrm{orig}}.
\end{equation*}
Panel E summarizes the resulting depth-dependent residual variance, approximated by the mean squared residual across the retained ensemble,
\begin{equation}
\sigma_{\mathrm{bh}}^2(z_i)=\frac{1}{N_m}\sum_{m=1}^{N_m}\left(r_i^{(m)}\right)^2,
\label{si:eq:si_borehole_variance}
\end{equation}
which is then used to define the site-level diagonal borehole observation-error covariance,
\begin{equation}
\mathbf{R}_{\mathrm{bh}}=\operatorname{diag}\!\left(\sigma_{\mathrm{bh}}^2(z_1),\sigma_{\mathrm{bh}}^2(z_2),\ldots,\sigma_{\mathrm{bh}}^2(z_{N_z})\right).
\label{si:eq:si_borehole_R}
\end{equation}
This construction allows the borehole uncertainty to vary with depth, so that levels where the inverse-plus-forward reconstruction is less stable carry larger observational uncertainty than levels where the reconstruction is more tightly constrained.

\FloatBarrier

\sisection{Significance testing of spatial OHC trends}{si:sec:si_trend_sig}
This section describes the significance test used for the spatial trend maps in Figs.~\ref{si:fig:si_ohc_wu_trend_compare} and \ref{si:fig:si_ohc_mwp_lia_trend}. At each grid point, we estimate the linear trend in annual OHC anomalies for each depth layer with an ordinary least squares regression, following standard practice for climate-trend estimation \cite{santer2000significance,wilks2011statistical},
\[
y(t)=a+bt+\varepsilon(t),
\]
where \(y(t)\) is the annual OHC anomaly, \(a\) is the intercept, \(b\) is the linear trend, and \(\varepsilon(t)\) is the residual. The plotted trend is the slope estimate \(\hat{b}\), reported in \(\mathrm{J\,m^{-2}\,century^{-1}}\).

We test the null hypothesis \(H_0:b=0\) against the two-sided alternative \(H_1:b\neq 0\). Because annual OHC anomalies are serially correlated, the nominal sample size overstates the effective degrees of freedom. We therefore estimate the lag-1 autocorrelation \(r_1\) of the regression residuals and use it to define the effective sample size
\[
n_{\mathrm{eff}} = n \frac{1-r_1}{1+r_1},
\]
where \(n\) is the number of annual observations. This lag-1 adjustment is a standard approximation for climate time series with red-noise persistence \cite{santer2000significance,wilks2011statistical}. The effective degrees of freedom are then taken as \(\mathrm{df}=n_{\mathrm{eff}}-2\).

Using this adjusted sample size, we estimate the residual variance and the standard error of the slope in the usual way,
\[
\mathrm{MSE}=\frac{\sum_{i=1}^{n}\varepsilon_i^2}{\mathrm{df}},
\qquad
\mathrm{SE}(\hat{b})=\sqrt{\frac{\mathrm{MSE}}{\sum_{i=1}^{n}(t_i-\bar{t})^2}},
\]
and compute the corresponding \(t\)-statistic,
\[
t=\frac{\hat{b}}{\mathrm{SE}(\hat{b})}.
\]
The two-sided \(p\)-value is obtained from the Student's \(t\) distribution with \(\mathrm{df}\) degrees of freedom. We account for simultaneous testing across each spatial field using the Benjamini-Hochberg procedure, with the false discovery rate (FDR) controlled at \(q=0.05\) \cite{benjamini1995controlling,wilks2016stippling}. For each panel, the \(m\) valid ocean-grid-cell \(p\)-values are ordered as
\[
p_{(1)}\le p_{(2)}\le\cdots\le p_{(m)}.
\]
We identify the largest rank \(k\) satisfying
\[
p_{(k)}\le \frac{k}{m}q
\]
and define the panel-specific threshold as \(p_{\mathrm{FDR}}^{*}=p_{(k)}\). Grid cells with \(p_i\le p_{\mathrm{FDR}}^{*}\) are classified as FDR significant. Thus, \(q=0.05\) controls the expected proportion of false discoveries among the grid cells classified as significant, rather than the probability that any individual grid cell is a false positive.

The correction is applied separately to all valid ocean grid cells in each trend-map panel in Figs.~\ref{si:fig:si_ohc_wu_trend_compare}, \ref{si:fig:si_ohc_mwp_lia_trend}, and \ref{si:fig:si_ohc_wu_trend_per_model}. Land and invalid grid cells are excluded and remain missing in the trend, raw-\(p\), FDR-adjusted \(q\), and significance-mask fields. The effective-sample-size adjustment addresses temporal autocorrelation, whereas the panel-wise BH-FDR procedure addresses spatial multiple testing. In the figures, unmasked colored regions pass the FDR-controlled test, semitransparent gray overlays identify regions that do not pass, and land or invalid grid cells remain blank.

\FloatBarrier

\sisection{Pseudo-proxy experiment}{si:sec:si_ppe}

We use CESM-LME member 001 as the known target and evaluate recovery of GMT and layer-resolved OHC over 1000--2000 CE. The main experiment uses the P2k\_BH\_T12k proxy configuration. We adopt a leave-one-model-family-out design, excluding all CESM-LME simulations from construction of the EOF basis and LIM priors. The reconstruction uses the four remaining model-specific LIMs, so the target model family supplies no dynamical information.

Synthetic observations are generated from the target simulation using the same proxy locations and observation operators as in the Common Era reconstruction. For each Temp12k record, averaging intervals are bounded by the midpoints between consecutive observation times. Target-simulation temperatures within each interval are averaged to generate the noiseless pseudo-proxy, after which independent Gaussian observation errors are added. This construction represents time-integrated observations and matches the temporal-averaging operator used in assimilation. Block averaging does not eliminate aliasing and provides a simplified representation of archive-specific smoothing and sampling. The pseudo-proxy experiment assumes independent Gaussian observation errors, consistent with the diagonal observation-error covariance used in assimilation. Temporally correlated proxy errors can retain more noise at multidecadal to centennial timescales, so reconstruction skill at these timescales may be optimistic under the white-noise assumption.

\begin{figure}[!htbp]
\centering
\includegraphics[width=\textwidth,max height=0.65\textheight,keepaspectratio]{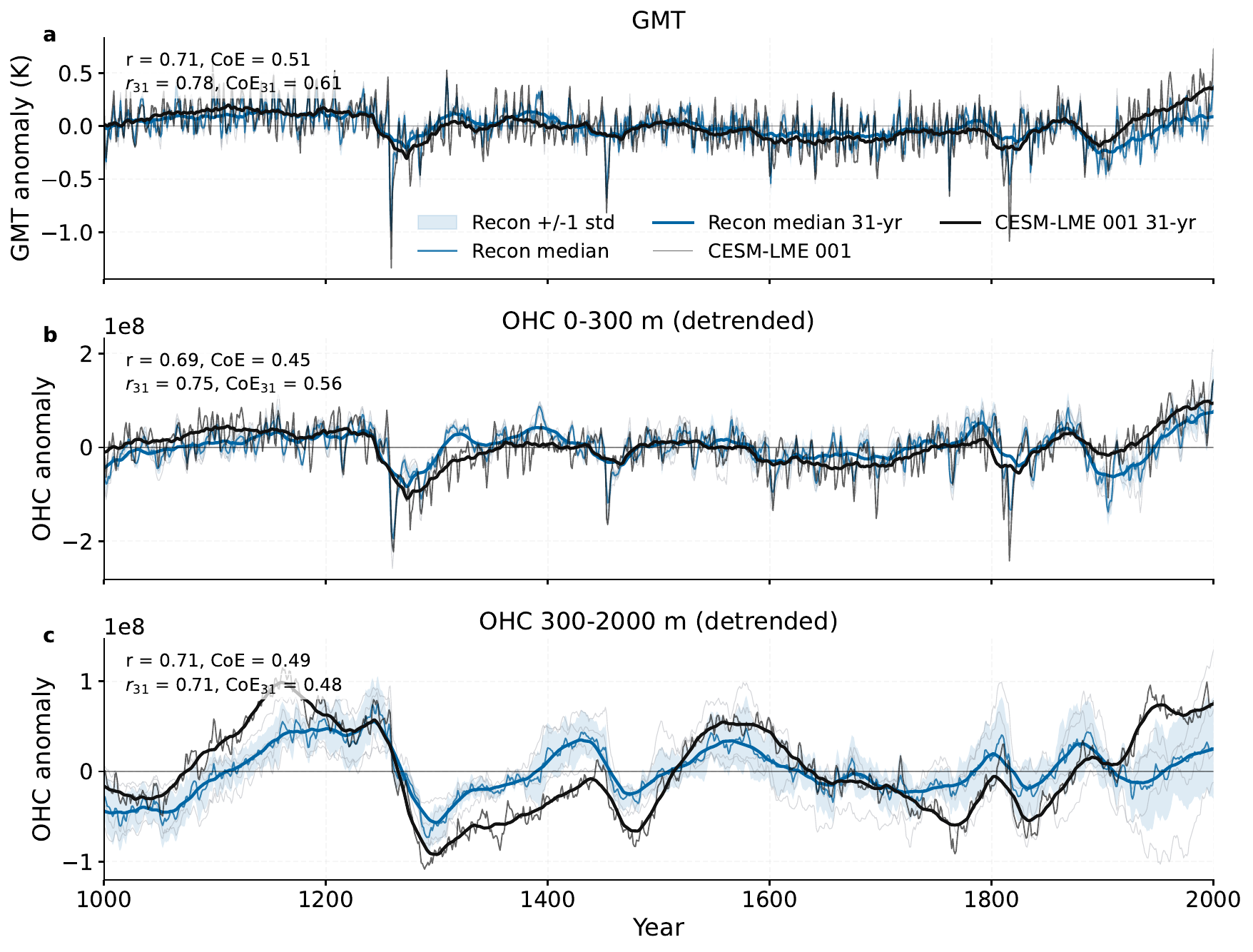}
\caption[Figure S4]{\textbf{Recovery of GMT and layer-resolved OHC in the P2k\_BH\_T12k pseudo-proxy experiment.} Panels compare reconstructed anomalies with CESM-LME member 001 over 1000--2000 CE for (a) global mean temperature (GMT), (b) OHC(0--300 m), and (c) OHC(300--2000 m). All CESM-LME simulations are excluded from construction of the EOF basis and LIM emulators. Blue curves show the median across the four remaining model-emulator-specific reconstructions, and black curves show the target simulation. Thin curves represent annual values, and thick curves show 31-year running medians. Light blue shading indicates $\pm 1$ standard deviation across the four reconstructions. OHC series are linearly detrended before anomaly calculation. Panel annotations report Pearson correlation ($r$) and coefficient of efficiency (CoE) for the annual series; $r_{31}$ and $\mathrm{CoE}_{31}$ denote the corresponding metrics for the 31-year running medians. GMT anomalies are in K, and OHC anomalies are in J\,m$^{-2}$.}
\label{si:fig:si_pseudoproxy_indices}
\end{figure}

\begin{figure}[!htbp]
\centering
\includegraphics[width=\textwidth,max height=0.65\textheight,keepaspectratio]{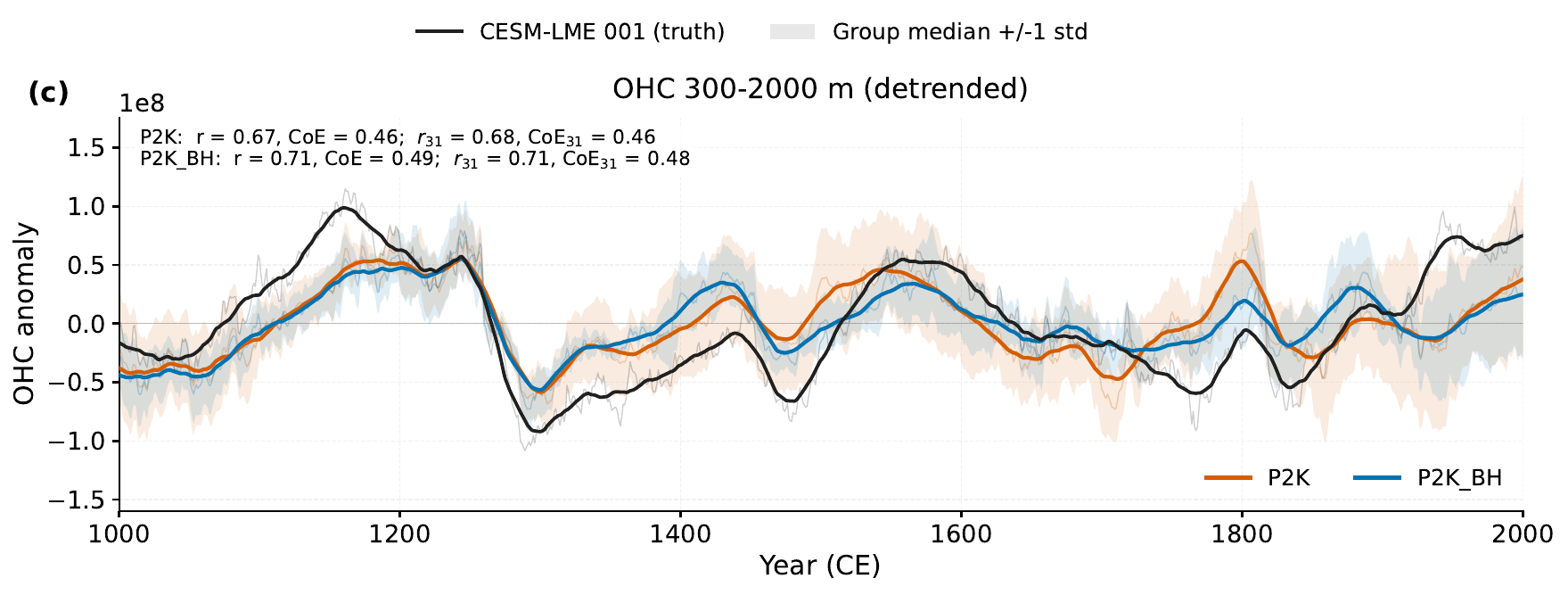}
\caption[Figure S5]{\textbf{Borehole contribution to reconstructed deep-ocean heat content in the pseudo-proxy experiment.} Detrended global-mean OHC(300--2000 m) anomalies from P2k (orange) and P2k\_BH (blue) are compared with CESM-LME member 001 (black) over 1000--2000 CE. Each experiment uses the four model-specific emulators trained without CESM-LME simulations. Thin curves show annual values, and thick curves show 31-year running medians. Colored shading indicates $\pm 1$ standard deviation across the four model-emulator-specific reconstructions within each configuration, rather than a posterior confidence interval. Annotations report annual Pearson correlation ($r$) and coefficient of efficiency (CoE), together with their 31-year-smoothed counterparts ($r_{31}$ and $\mathrm{CoE}_{31}$). OHC anomalies are in J\,m$^{-2}$.}
\label{si:fig:si_pseudoproxy_boreholes}
\end{figure}

\begin{figure}[p]
\centering
\includegraphics[width=\textwidth,max height=0.65\textheight,keepaspectratio]{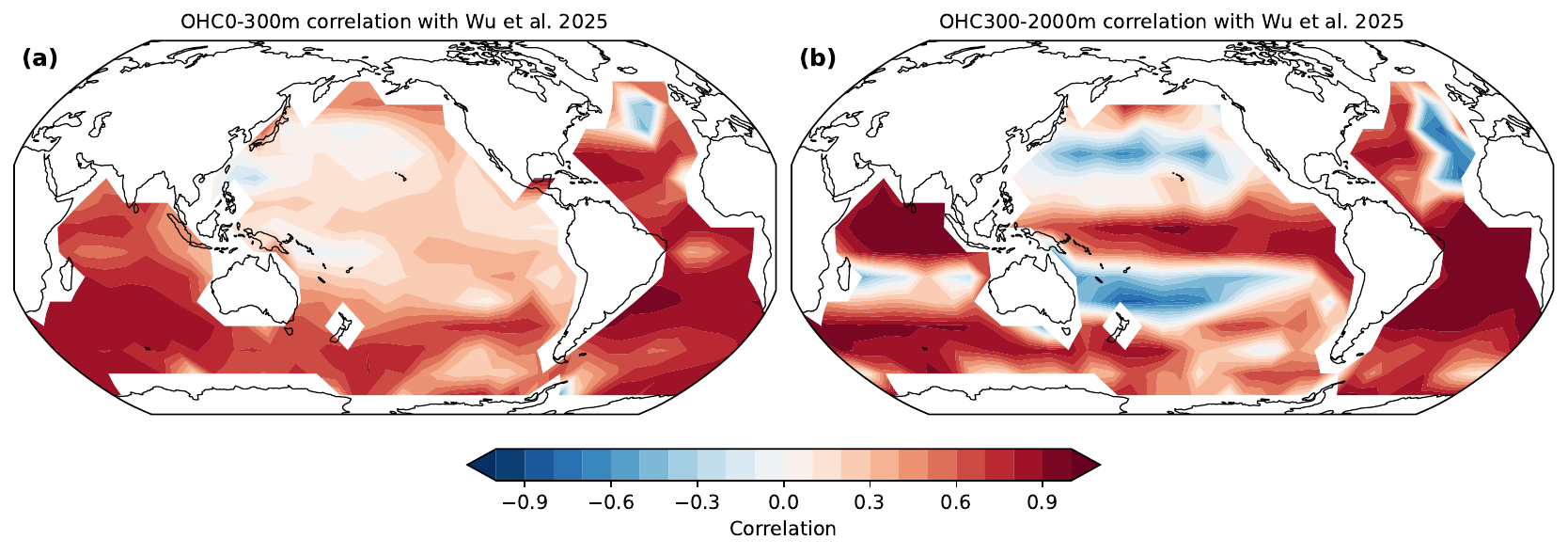}
\caption[Figure S6]{\textbf{Spatial correlation between reconstructed ocean heat content and \citeA{wu2025energybudget}.} Maps show gridpoint correlations between the ocean heat content reconstruction in this study and the corresponding reconstruction from \citeA{wu2025energybudget} from 1880 to 2000 for 0--300 m and 300--2000 m.}
\label{si:fig:si_ohc_wu_corr}
\end{figure}

Figure~\ref{si:fig:si_pseudoproxy_indices} compares the P2k\_BH\_T12k reconstruction with the known target. OHC series are linearly detrended before anomaly calculation to emphasize reconstructed variability rather than differences in background model drift. We compare P2k and P2k\_BH to assess the effect of borehole constraints on recovery of the known target (Fig.~\ref{si:fig:si_pseudoproxy_boreholes}). For annual detrended OHC(300--2000 m), adding boreholes increases $r$ from 0.67 to 0.71 and CoE from 0.46 to 0.49. For the 31-year running medians, $r_{31}$ increases from 0.68 to 0.71 and $\mathrm{CoE}_{31}$ from 0.46 to 0.48. These changes indicate a modest improvement in recovering deep-ocean variability in this model-world experiment, although some multicentennial excursions remain underestimated.

\FloatBarrier

\FloatBarrier

\begin{figure}[!p]
\centering
\includegraphics[width=\textwidth,max height=0.65\textheight,keepaspectratio]{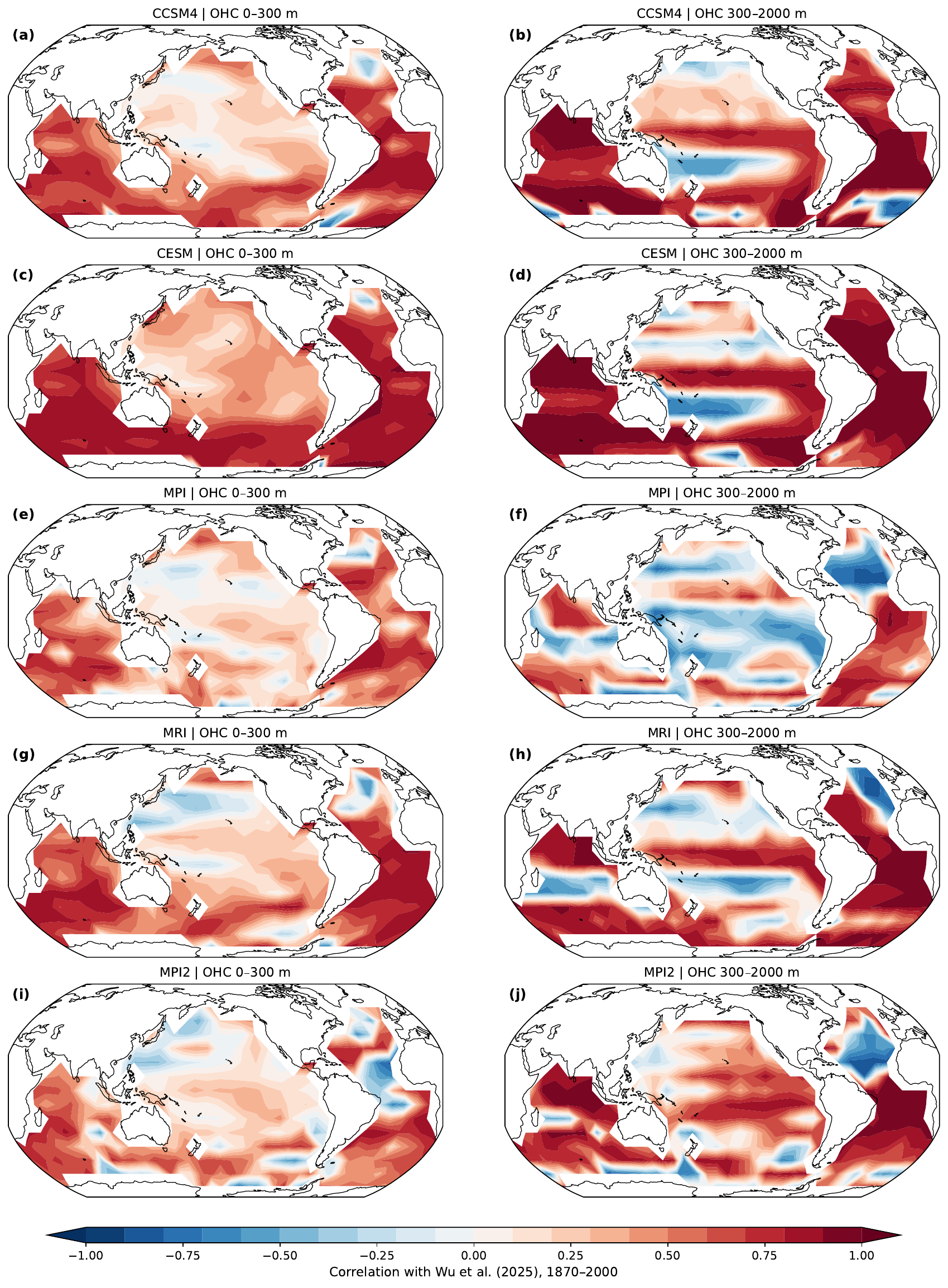}
\caption[Figure S7]{\textbf{Model-specific spatial correlations between reconstructed OHC and \citeA{wu2025energybudget}.} Gridpoint correlations are calculated over 1870--2000 CE. Left and right columns show OHC(0--300 m) and OHC(300--2000 m), respectively. Rows show reconstructions using (a and b) CCSM4, (c and d) CESM-LME (labeled CESM), (e and f) MPI-ESM-P (MPI), (g and h) MRI-ESM2-0 (MRI), and (i and j) MPI-ESM1.2-LR (MPI2). All panels share the same correlation color scale.}
\label{si:fig:si_ohc_wu_corr_per_model}
\end{figure}

\begin{figure}[p]
\centering
\includegraphics[width=\textwidth,max height=0.65\textheight,keepaspectratio]{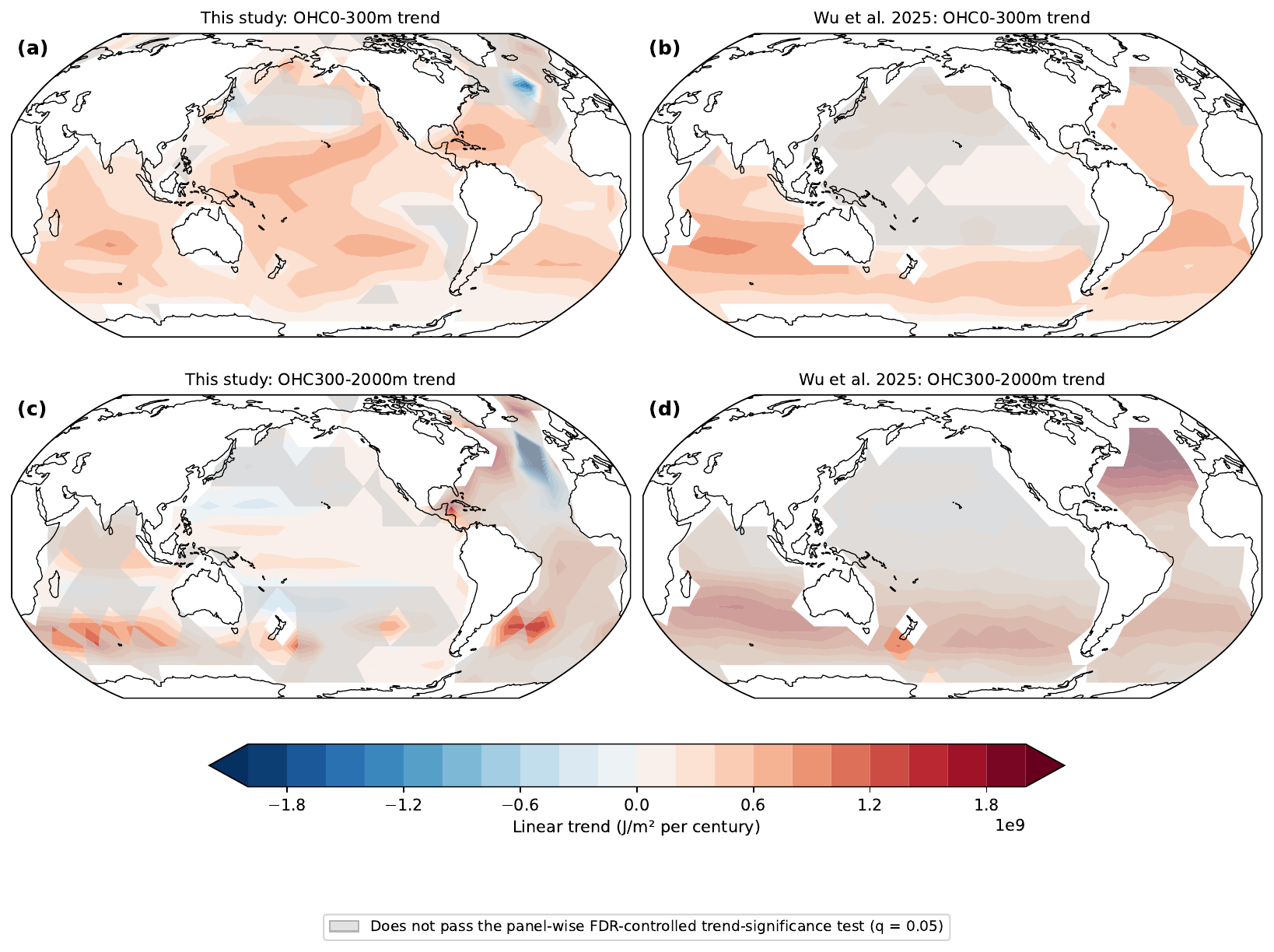}
\caption[Figure S8]{\textbf{Comparison of 1880--2000 CE ocean heat content trends in this study and \citeA{wu2025energybudget}.} Panels compare trends from this study with the corresponding trends from \citeA{wu2025energybudget} for (a and b) OHC(0--300 m) and (c and d) OHC(300--2000 m). Semitransparent gray overlays identify grid cells that do not pass the panel-wise FDR-controlled trend-significance test at \(q=0.05\); unmasked colored regions pass the test described in Supplementary Section~\ref{si:sec:si_trend_sig}. The FDR-significant regions represent 73.4\%, 59.2\%, 23.1\%, and 0.4\% of the ocean area in (a) to (d), respectively. Only two isolated grid cells pass the test in (d), providing no evidence for a large-scale significant trend pattern in the \citeA{wu2025energybudget} 300--2000-m field.}
\label{si:fig:si_ohc_wu_trend_compare}
\end{figure}

\begin{figure}[!p]
\centering
\includegraphics[width=\textwidth,max height=0.65\textheight,keepaspectratio]{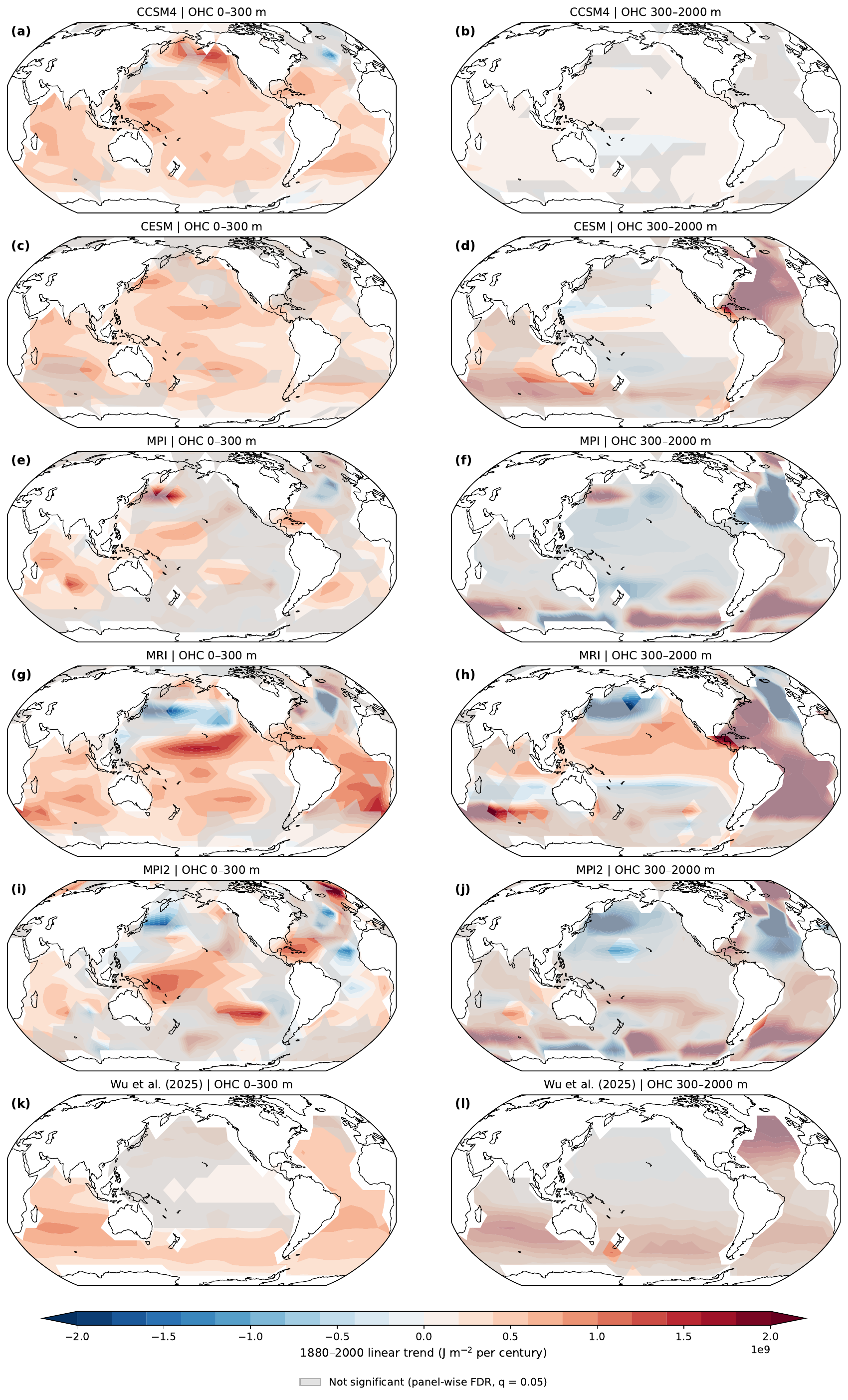}
\caption[Figure S9]{\textbf{Model-specific 1880--2000 CE OHC trends compared with \citeA{wu2025energybudget}.} Left and right columns show OHC(0--300 m) and OHC(300--2000 m), respectively. Rows show (a and b) CCSM4, (c and d) CESM-LME (CESM), (e and f) MPI-ESM-P (MPI), (g and h) MRI-ESM2-0 (MRI), (i and j) MPI-ESM1.2-LR (MPI2), and (k and l) \citeA{wu2025energybudget}. Gray overlays identify grid cells that do not pass the panel-wise FDR test at \(q=0.05\) (Section~\ref{si:sec:si_trend_sig}). Trends are in J\,m$^{-2}$\,century$^{-1}$, with a \(10^9\) scale factor on the shared color bar.}
\label{si:fig:si_ohc_wu_trend_per_model}
\end{figure}

\begin{figure}[p]
\centering
\includegraphics[width=\textwidth,max height=0.65\textheight,keepaspectratio]{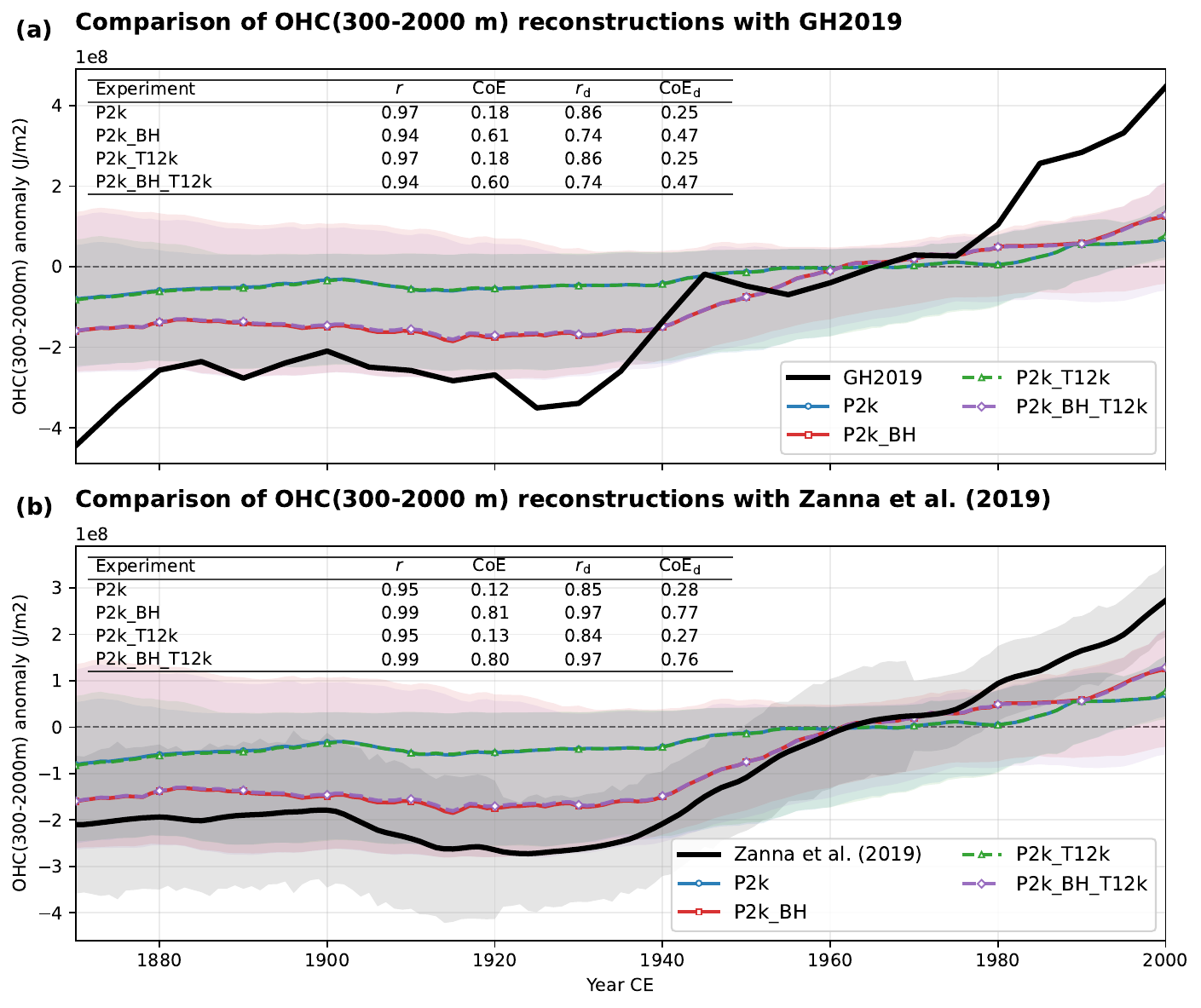}
\caption[Figure S10]{\textbf{Comparison of reconstructed global-mean OHC(300--2000 m) anomalies with external estimates.} (a) Comparison of the P2k, P2k\_BH, P2k\_T12k, and P2k\_BH\_T12k reconstructions with \citeA{gebbie2019little} (GH2019). (b) Corresponding comparison with \citeA{zannaGlobalReconstructionHistorical2019a}. Colored lines show the mean across the five model-emulator-specific reconstructions for each experiment. Colored shading denotes the pointwise 5th--95th percentile range across the 100 reconstruction trajectories for each experiment. Black lines show the external estimates; gray shading in (b) denotes the reported uncertainty. In both panels, P2k closely overlaps P2k\_T12k, while P2k\_BH closely overlaps P2k\_BH\_T12k. All anomalies are referenced to the 1950--1980 CE climatology. Inset tables report Pearson correlation coefficients ($r$), coefficients of efficiency (CoE), and their detrended counterparts ($r_{\mathrm d}$ and $\mathrm{CoE}_{\mathrm d}$). Detrended metrics are calculated after linearly detrending each reconstruction-reference pair over its common interval.}
\label{si:fig:si_ohc_external_experiments}
\end{figure}

\begin{figure}[p]
\centering
\includegraphics[width=\textwidth,max height=0.65\textheight,keepaspectratio]{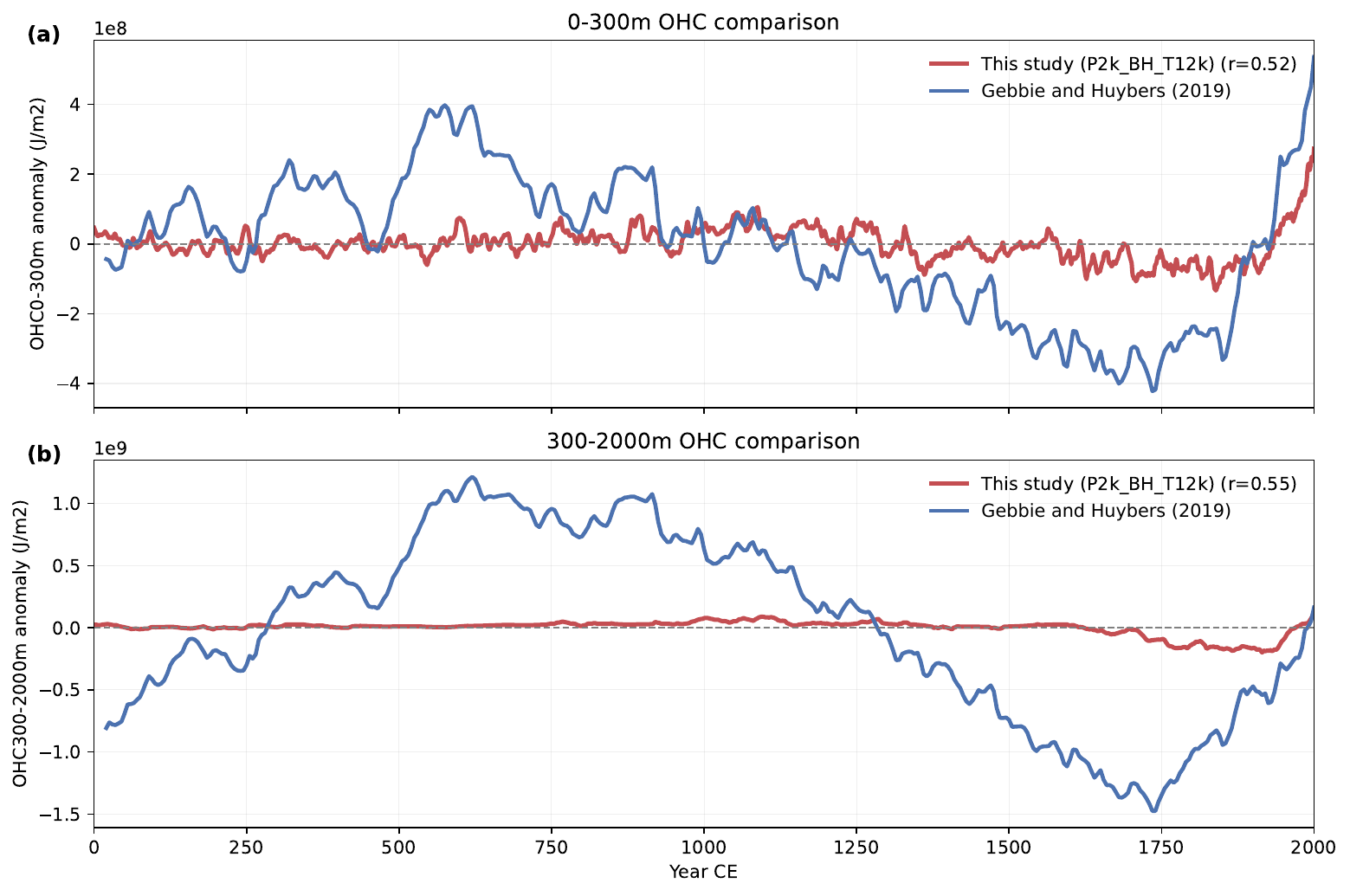}
\caption[Figure S11]{\textbf{Comparison of reconstructed upper-300 m and 300--2000 m ocean heat content with an external reconstruction.} The ocean heat content anomaly from the P2k\_BH\_T12k reconstruction is compared with the product derived from \citeA{gebbie2019little}.}
\label{si:fig:si_ohc_external}
\end{figure}

\begin{figure}[p]
\centering
\includegraphics[width=\textwidth,max height=0.65\textheight,keepaspectratio]{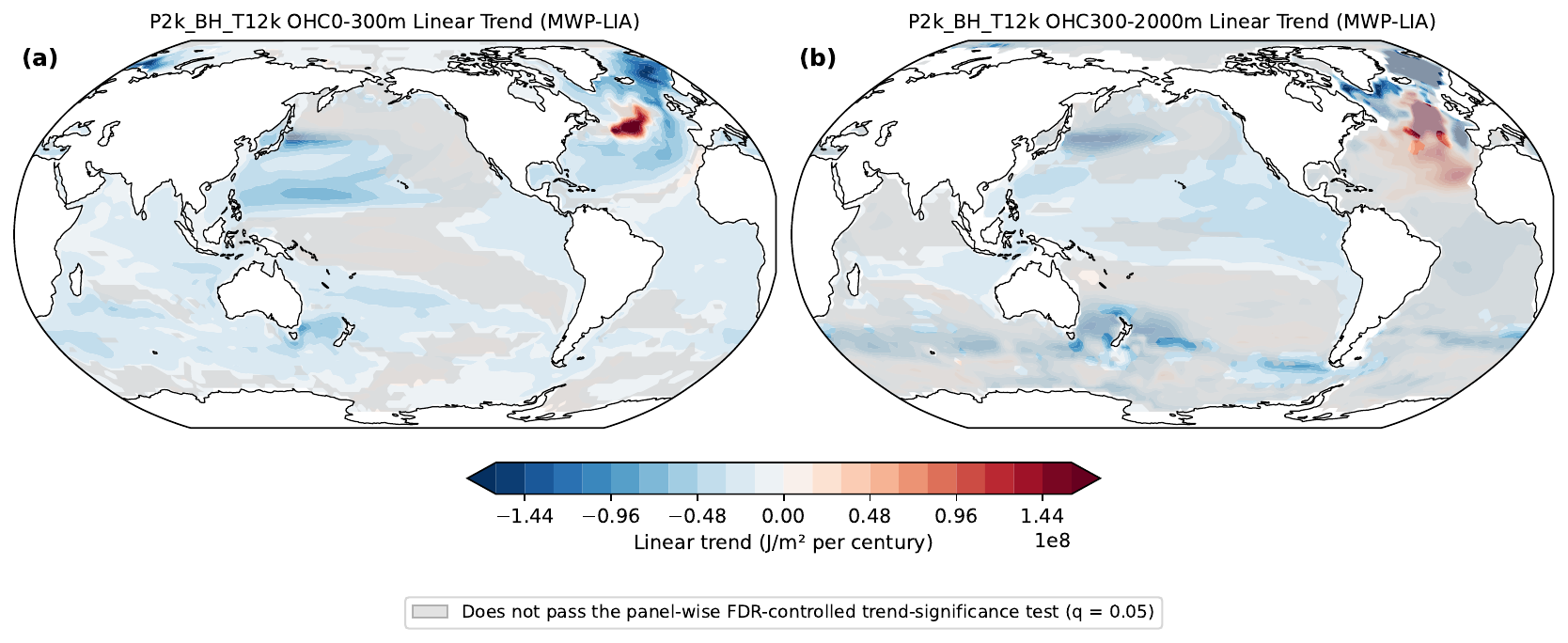}
\caption[Figure S12]{\textbf{Linear trends in ocean heat content across the Medieval Warm Period to Little Ice Age transition.} Spatial patterns show trends from the P2k\_BH\_T12k reconstruction for (a) OHC(0--300 m) and (b) OHC(300--2000 m) over the interval spanning the Medieval Warm Period to the Little Ice Age. Semitransparent gray overlays identify grid cells that do not pass the panel-wise FDR-controlled trend-significance test at \(q=0.05\); unmasked colored regions pass the test described in Supplementary Section~\ref{si:sec:si_trend_sig}. The FDR-significant regions represent 64.4\% and 26.3\% of the ocean area in (a) and (b), respectively.}
\label{si:fig:si_ohc_mwp_lia_trend}
\end{figure}

\begin{figure}[!p]
\centering
\includegraphics[width=\textwidth,max height=0.65\textheight,keepaspectratio]{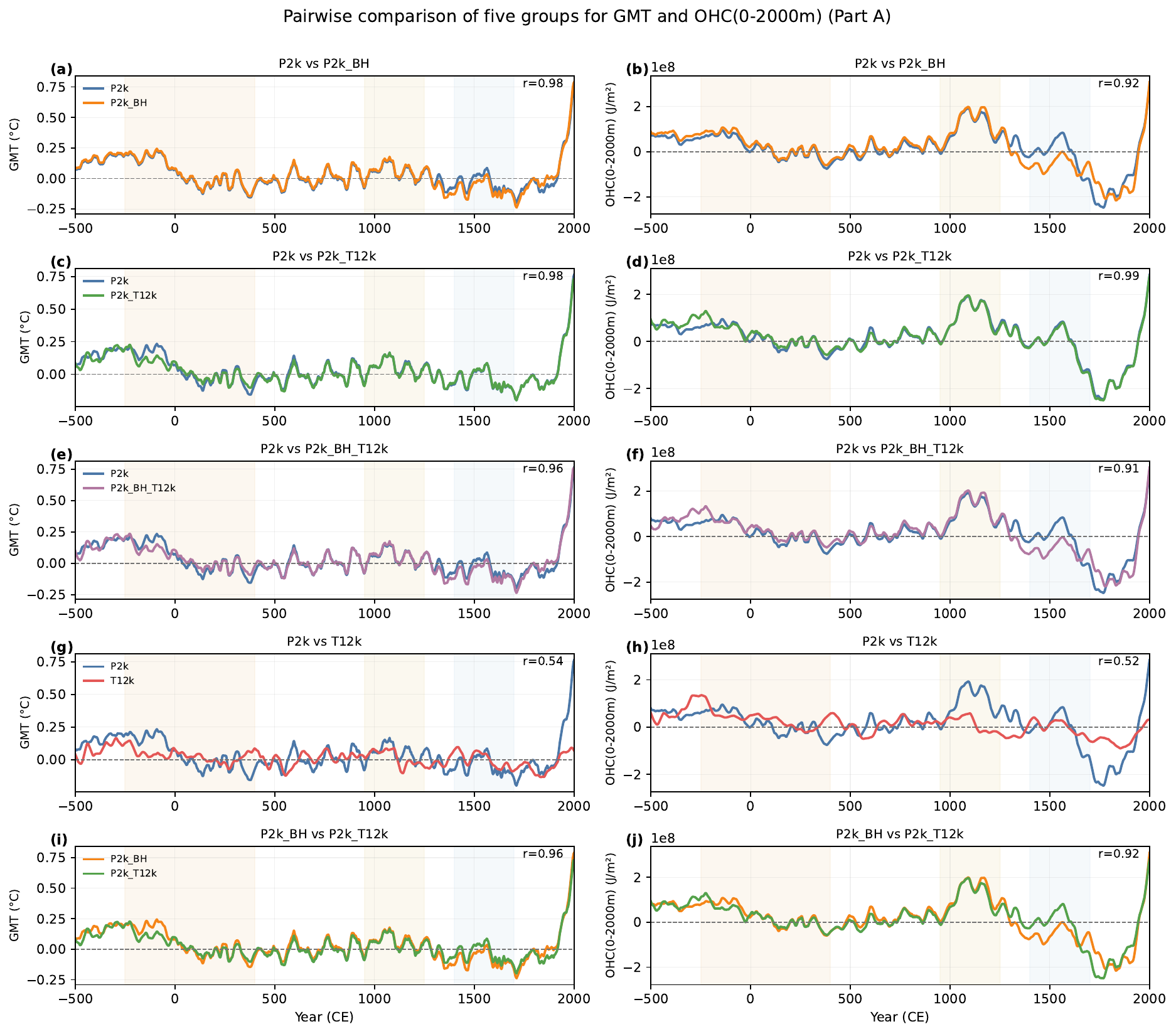}
\caption[Figure S13]{\textbf{Pairwise comparisons of global mean temperature and upper-2000-m ocean heat content across reconstruction experiments, part A.} Each row shows one pairwise comparison between two experiments. Left panels show 30-year low-pass-filtered global mean temperature (GMT) anomalies, and right panels show 30-year low-pass-filtered upper-2000-m ocean heat content anomalies, computed as OHC0--300 + OHC300--2000. The five experiments are P2k = PAGES2k only, P2k\_BH = PAGES2k + boreholes, P2k\_T12k = PAGES2k + Temp12k, P2k\_BH\_T12k = PAGES2k + boreholes + Temp12k, and T12k = Temp12k only (\(J_m=1\)). Correlation coefficients for each pair are shown in the panel corners. All anomalies are referenced to the 0--2000 CE climatology, and the Roman Warm Period (RWP), Medieval Warm Period (MWP), and Little Ice Age (LIA) are indicated.}
\label{si:fig:figure14a}
\end{figure}

\begin{figure}[!p]
\centering
\includegraphics[width=\textwidth,max height=0.65\textheight,keepaspectratio]{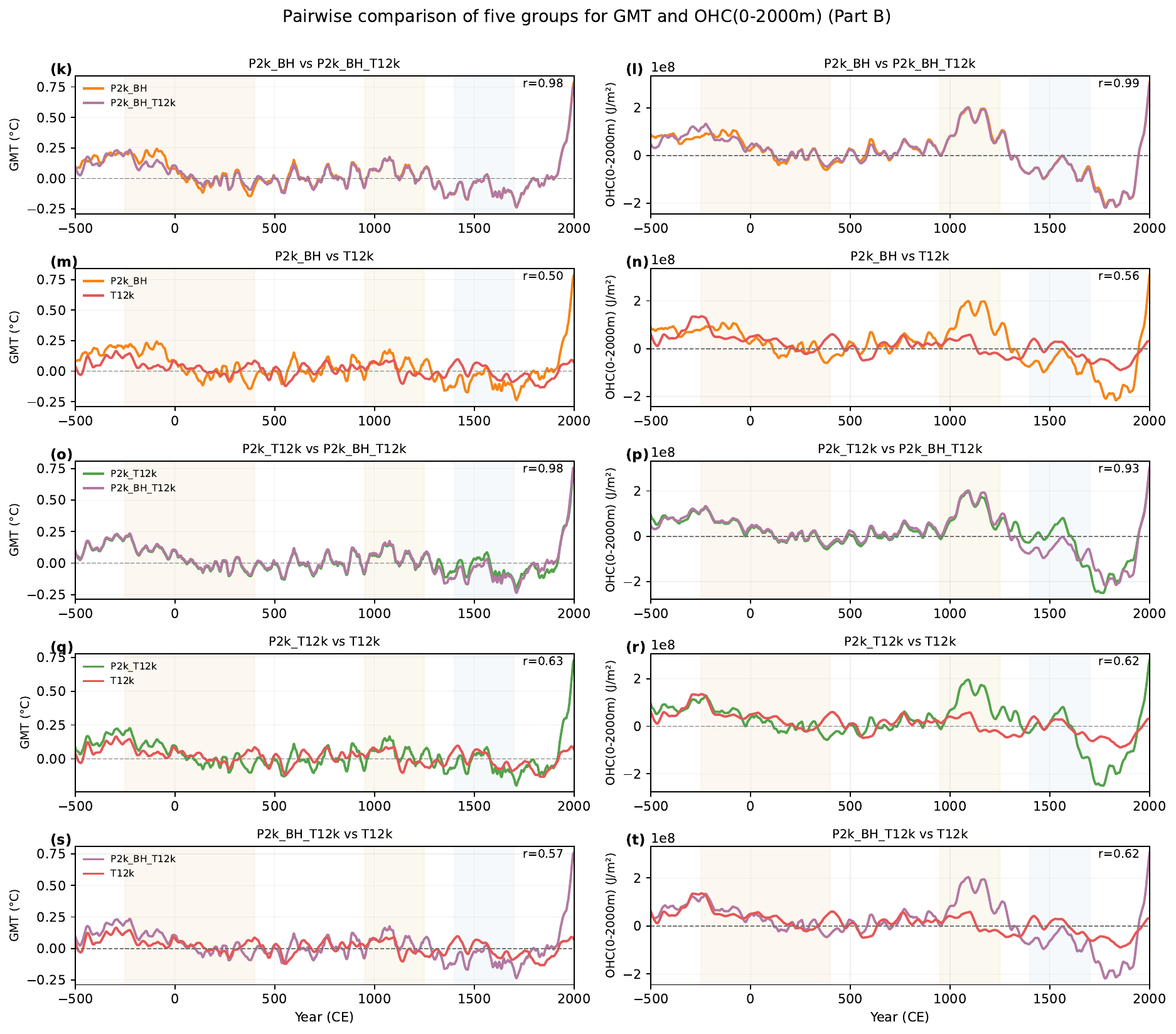}
\caption[Figure S14]{\textbf{Pairwise comparisons of global mean temperature and upper-2000-m ocean heat content across reconstruction experiments, part B.} Continued from Fig.~\ref{si:fig:figure14a}.}
\label{si:fig:figure14b}
\end{figure}

\sisection{Additional Common Era reconstruction and validation diagnostics}{si:sec:si_ohc_diagnostics}
The following figures provide spatial and layer-resolved comparisons of reconstructed OHC with external estimates, compare historical OHC changes across selected intervals, and show pairwise comparisons among the proxy-assimilation experiments. Model-specific correlation and trend maps show substantial differences in regional structure and amplitude (Figs.~\ref{si:fig:si_ohc_wu_corr_per_model} and \ref{si:fig:si_ohc_wu_trend_per_model}).

\FloatBarrier

\FloatBarrier

\sisection{Sensitivity diagnostics for 130-year OHC trends}{si:sec:si_ohc_trend_diagnostics}

\begin{figure}[!htbp]
\centering
\includegraphics[width=\textwidth,max height=0.65\textheight,keepaspectratio]{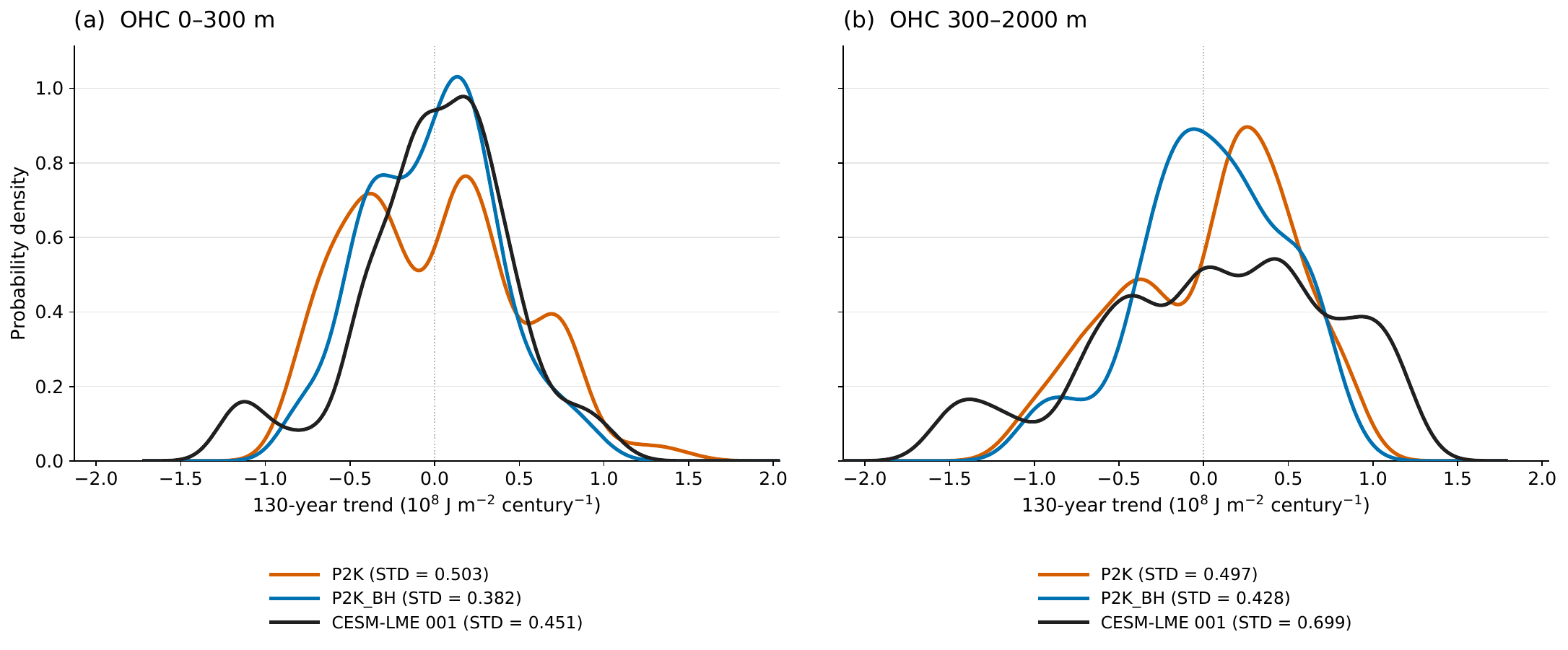}
\caption[Figure S15]{\textbf{Distributions of 130-year OHC trends in the pseudo-proxy experiment.} Probability-density curves compare P2k (orange), P2k\_BH (blue), and the CESM-LME member 001 target (black) for (a) OHC(0--300 m) and (b) OHC(300--2000 m) over 1000--2000 CE. Reconstruction trends are calculated from individual trajectories rather than a multi-emulator median. Legend values give the standard deviations of the trend distributions: 0.503, 0.382, and 0.451 in (a), and 0.497, 0.428, and 0.699 in (b), respectively. Trend units are \(10^{8}\) J m\(^{-2}\) century\(^{-1}\).}
\label{si:fig:si_ppe_ohc_trend_distributions}
\end{figure}

\textbf{Pseudo-proxy trend distributions.} We compare 130-year OHC trend distributions from the P2k and P2k\_BH pseudo-proxy reconstructions with CESM-LME member 001 over 1000--2000 CE (Fig.~\ref{si:fig:si_ppe_ohc_trend_distributions}). Trends are calculated over complete windows containing 131 annual values. The reconstruction distributions use individual trajectories rather than a multi-emulator median. The standard deviations therefore characterize the spread of 130-year trends, not annual OHC variability. Upper-layer trend variability is broadly retained, whereas the deeper-layer distributions remain narrower than the target distribution. This attenuation is present in individual reconstruction trajectories, indicating that taking a multi-emulator median is not its sole explanation.

\FloatBarrier

\begin{figure}[!p]
\centering
\includegraphics[width=0.88\textwidth,max height=0.65\textheight,keepaspectratio]{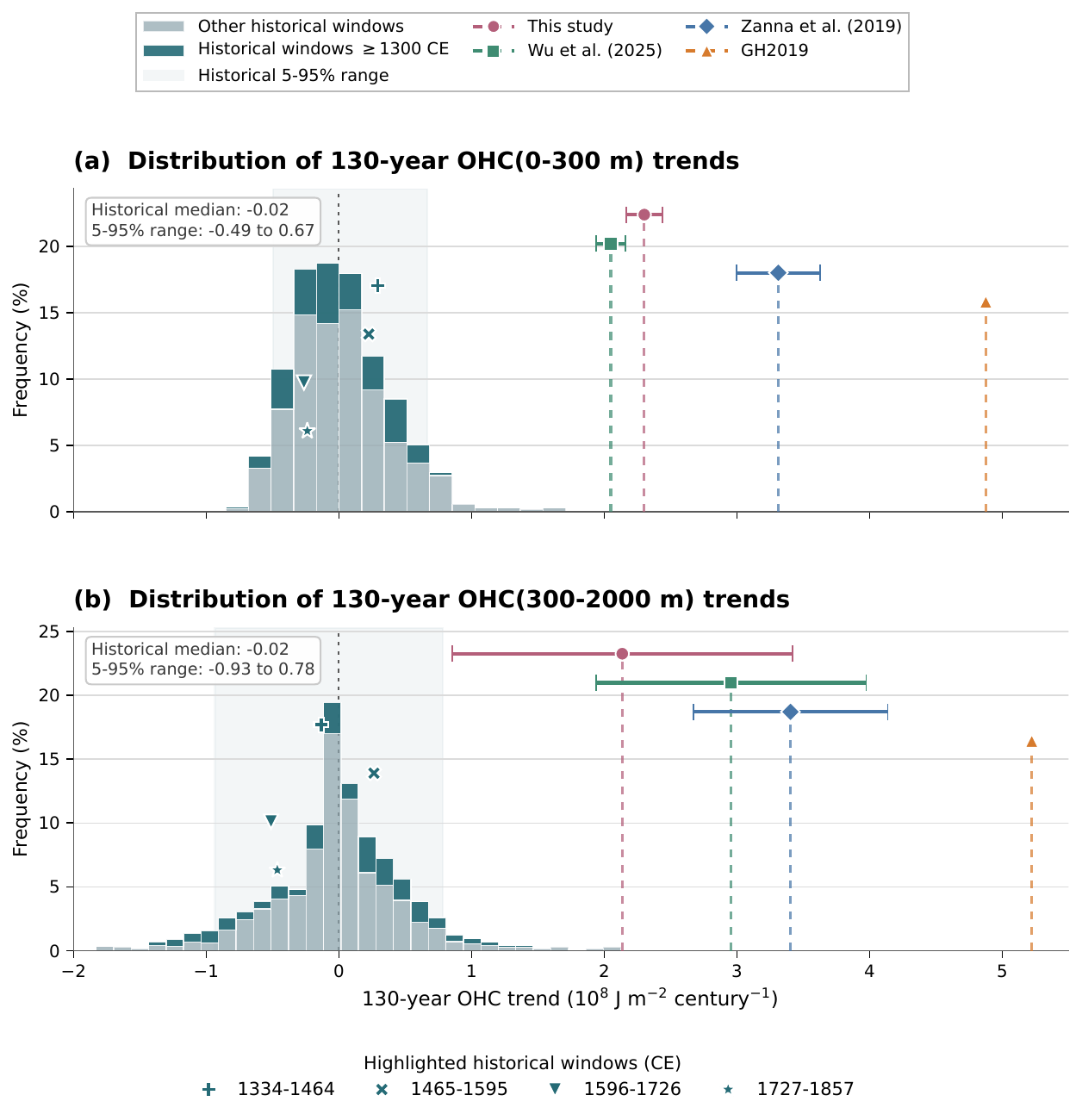}
\caption[Figure S16]{\textbf{Non-overlapping historical windows in the OHC trend comparison.} Histograms pool trajectory-specific 130-year trends for (a) OHC(0--300 m) and (b) OHC(300--2000 m). Each of the 100 reconstruction trajectories contributes 18 non-overlapping windows within 500 BCE--1869 CE. Dark bars and symbols highlight four historical intervals beginning after 1300 CE: 1334--1464, 1465--1595, 1596--1726, and 1727--1857 CE. Light bars show the other historical windows. Gray shading denotes the pooled historical 5th--95th percentile range. Colored markers show the 1870--2000 CE estimates identified in the legend. This study's error bars show the 5th--95th percentiles across the 100 modern trajectory-specific trends; external error bars show reported uncertainty where available. Trend units are \(10^{8}\) J m\(^{-2}\) century\(^{-1}\).}
\label{si:fig:si_ohc_trend_nonoverlap}
\end{figure}

\textbf{Non-overlapping-window sensitivity.} We repeat the historical comparison using consecutive, non-overlapping windows beginning at 500 BCE, each containing 131 annual values. Advancing by 131 values gives 18 windows per trajectory, with the final complete interval ending in 1857 CE. Figure~\ref{si:fig:si_ohc_trend_nonoverlap} highlights four intervals beginning after 1300 CE: 1334--1464, 1465--1595, 1596--1726, and 1727--1857 CE. Their trends remain below the modern estimates in the plotted comparison. These diagnostics assess sensitivity to window selection; neither overlapping windows nor reconstruction trajectories are treated as independent realizations of climate history.

\FloatBarrier

\clearpage

\clearpage
\bibliographystyle{ametsocV6}
\bibliography{references}

\begin{thebibliography}{84}
\providecommand{\natexlab}[1]{#1}
\providecommand{\url}[1]{\texttt{#1}}
\renewcommand{\UrlFont}{\rmfamily}
\providecommand{\urlprefix}{URL }
\expandafter\ifx\csname urlstyle\endcsname\relax
  \providecommand{\doi}[1]{https://doi.org/\discretionary{}{}{}#1}\else
  \providecommand{\doi}{https://doi.org/\discretionary{}{}{}\begingroup
  \urlstyle{rm}\Url}\fi
\providecommand{\eprint}[2][]{\url{#2}}

\bibitem[{Amrhein et~al.(2020)Amrhein, Hakim,, and
  Parsons}]{amrhein2020quantifying}
Amrhein, D.~E., G.~J. Hakim, and L.~A. Parsons, 2020: Quantifying structural
  uncertainty in paleoclimate data assimilation with an application to the last
  millennium. \textit{Geophysical Research Letters}, \textbf{47~(22)},
  e2020GL090\,485.

\bibitem[{Anderson(2012)}]{anderson2012localization}
Anderson, J.~L., 2012: {Localization and sampling error correction in ensemble
  Kalman filter data assimilation}. \textit{Monthly Weather Review},
  \textbf{140~(7)}, 2359--2371.

\bibitem[{Benjamini and Hochberg(1995)Benjamini, and
  Hochberg}]{benjamini1995controlling}
Benjamini, Y., and Y.~Hochberg, 1995: Controlling the false discovery rate: A
  practical and powerful approach to multiple testing. \textit{Journal of the
  Royal Statistical Society: Series B (Methodological)}, \textbf{57~(1)},
  289--300, \doi{10.1111/j.2517-6161.1995.tb02031.x}.

\bibitem[{Bradley(1999)}]{bradley1999paleoclimatology}
Bradley, R.~S., 1999: \textit{Paleoclimatology: reconstructing climates of the
  Quaternary}, Vol.~68. Elsevier.

\bibitem[{Christiansen and Ljungqvist(2017)Christiansen, and
  Ljungqvist}]{christiansen2017challenges}
Christiansen, B., and F.~C. Ljungqvist, 2017: {Challenges and Perspectives for
  Large-Scale Temperature Reconstructions of the Past Two Millennia}.
  \textit{Reviews of Geophysics}, \textbf{55~(1)}, 40--96.

\bibitem[{Council et~al.(2007)Council, on~Earth, Studies,
  on~Atmospheric~Sciences,, and Committee on Surface Temperature
  Reconstructions for~the Last~2}]{national2007surface}
Council, N.~R., D.~on~Earth, L.~Studies, B.~on~Atmospheric~Sciences, and .~Y.
  Committee on Surface Temperature Reconstructions for~the Last~2, 2007:
  \textit{Surface temperature reconstructions for the last 2,000 years}.
  National Academies Press.

\bibitem[{Courtier et~al.(1994)Courtier, Th{\'e}paut,, and
  Hollingsworth}]{Courtier1994}
Courtier, P., J.-N. Th{\'e}paut, and A.~Hollingsworth, 1994: A strategy for
  operational implementation of 4d-var, using an incremental approach.
  \textit{Quarterly Journal of the Royal Meteorological Society},
  \textbf{120~(519)}, 1367--1387, \doi{10.1002/qj.49712051912}.

\bibitem[{Cuesta-Valero et~al.(2021)Cuesta-Valero, Garc{\'\i}a-Garc{\'\i}a,
  Beltrami, Gonz{\'a}lez-Rouco, Garc{\'\i}a-Bustamante,, and
  Mont{\'a}vez}]{cuestaValero2021longterm}
Cuesta-Valero, F.~J., A.~Garc{\'\i}a-Garc{\'\i}a, H.~Beltrami, J.~F.
  Gonz{\'a}lez-Rouco, E.~Garc{\'\i}a-Bustamante, and J.~P. Mont{\'a}vez, 2021:
  {Long-term global ground heat flux and continental heat storage from
  geothermal data}. \textit{Climate of the Past}, \textbf{17~(1)}, 451--468,
  \doi{10.5194/cp-17-451-2021}.

\bibitem[{Dolman and Laepple(2018)Dolman, and Laepple}]{dolman2018sedproxy}
Dolman, A.~M., and T.~Laepple, 2018: Sedproxy: a forward model for
  sediment-archived climate proxies. \textit{Climate of the Past},
  \textbf{14~(12)}, 1851--1868.

\bibitem[{Emile-Geay et~al.(2025)Emile-Geay, Hakim, Viens, Zhu,, and
  Amrhein}]{Emile-Geay:2025}
Emile-Geay, J., G.~J. Hakim, F.~Viens, F.~Zhu, and D.~E. Amrhein, 2025:
  Temporal comparisons involving paleoclimate data assimilation: Challenges and
  remedies. \textit{Journal of Climate}, \textbf{38~(5)}, 1365--1385,
  \doi{10.1175/JCLI-D-24-0101.1}.

\bibitem[{Erb et~al.(2022)Erb, McKay, Steiger, Dee, Hancock, Ivanovic,
  Gregoire,, and Valdes}]{erb2022reconstructing}
Erb, M.~P., N.~P. McKay, N.~Steiger, S.~Dee, C.~Hancock, R.~F. Ivanovic, L.~J.
  Gregoire, and P.~Valdes, 2022: Reconstructing holocene temperatures in time
  and space using paleoclimate data assimilation. \textit{Climate of the Past},
  \textbf{18~(12)}, 2599--2629.

\bibitem[{Evans et~al.(2013)Evans, Tolwinski-Ward, Thompson,, and
  Anchukaitis}]{Evans_QSR13}
Evans, M.~N., S.~E. Tolwinski-Ward, D.~M. Thompson, and K.~J. Anchukaitis,
  2013: Applications of proxy system modeling in high resolution
  paleoclimatology. \textit{Quaternary Science Reviews}, \textbf{76~(0)},
  16--28, \doi{10.1016/j.quascirev.2013.05.024}.

\bibitem[{Evensen et~al.(2022)Evensen, Vossepoel,, and van
  Leeuwen}]{evensen2022data}
Evensen, G., F.~C. Vossepoel, and P.~J. van Leeuwen, 2022: \textit{Data
  Assimilation Fundamentals: A Unified Formulation of the State and Parameter
  Estimation Problem}. Springer Textbooks in Earth Sciences, Geography and
  Environment, Springer, Cham, \doi{10.1007/978-3-030-96709-3}.

\bibitem[{Fisher et~al.(2005)Fisher, Leutbecher,, and
  Kelly}]{fisherKalmanSmoothing2005}
Fisher, M., M.~Leutbecher, and G.~A. Kelly, 2005: On the equivalence between
  kalman smoothing and weak-constraint four-dimensional variational data
  assimilation. \textit{Quarterly Journal of the Royal Meteorological Society},
  \textbf{131~(613)}, 3235--3246, \doi{10.1256/qj.04.142}.

\bibitem[{Fisher et~al.(2011)Fisher, Tr{\'e}molet, Auvinen, Tan,, and
  Poli}]{fisherWeakConstraintLongWindow2011}
Fisher, M., Y.~Tr{\'e}molet, H.~Auvinen, D.~Tan, and P.~Poli, 2011:
  Weak-constraint and long-window {4D-Var}. ECMWF Technical Memorandum 655,
  European Centre for Medium-Range Weather Forecasts. \doi{10.21957/9ii4d4dsq}.

\bibitem[{Franke et~al.(2017)Franke, Br{\"o}nnimann, Bhend,, and
  Brugnara}]{franke2017monthly}
Franke, J., S.~Br{\"o}nnimann, J.~Bhend, and Y.~Brugnara, 2017: {A monthly
  global paleo-reanalysis of the atmosphere from 1600 to 2005 for studying past
  climatic variations}. \textit{Scientific data}, \textbf{4~(1)}, 1--19.

\bibitem[{Gebbie and Huybers(2019)Gebbie, and Huybers}]{gebbie2019little}
Gebbie, G., and P.~Huybers, 2019: {The little ice age and 20th-century deep
  Pacific cooling}. \textit{Science}, \textbf{363~(6422)}, 70--74.

\bibitem[{Giorgetta et~al.(2013)}]{giorgetta2013mpi}
Giorgetta, M.~A., and Coauthors, 2013: Climate and carbon cycle changes from
  1850 to 2100 in {MPI-ESM} simulations for the coupled model intercomparison
  project phase 5. \textit{Journal of Advances in Modeling Earth Systems},
  \textbf{5~(3)}, 572--597, \doi{10.1002/jame.20038}.

\bibitem[{Gleckler et~al.(2012)Gleckler, Santer, Domingues
  et~al.}]{gleckler2012human}
Gleckler, P.~J., B.~D. Santer, C.~M. Domingues, and Coauthors, 2012:
  Human-induced global ocean warming on multidecadal timescales. \textit{Nature
  Climate Change}, \textbf{2}, 524--529, \doi{10.1038/nclimate1553}.

\bibitem[{Hakim et~al.(2016)Hakim, Emile-Geay, Steig, Noone, Anderson, Tardif,
  Steiger,, and Perkins}]{hakim2016last}
Hakim, G.~J., J.~Emile-Geay, E.~J. Steig, D.~Noone, D.~M. Anderson, R.~Tardif,
  N.~Steiger, and W.~A. Perkins, 2016: {The last millennium climate reanalysis
  project: Framework and first results}. \textit{Journal of Geophysical
  Research: Atmospheres}, \textbf{121~(12)}, 6745--6764.

\bibitem[{Huang et~al.(2017)}]{huang2017extended}
Huang, B., and Coauthors, 2017: {Extended reconstructed sea surface
  temperature, version 5 (ERSSTv5): upgrades, validations, and
  intercomparisons}. \textit{Journal of Climate}, \textbf{30~(20)}, 8179--8205.

\bibitem[{Huang et~al.(2000)Huang, Pollack,, and Shen}]{huang2000temperature}
Huang, S., H.~N. Pollack, and P.-Y. Shen, 2000: Temperature trends over the
  past five centuries reconstructed from borehole temperatures.
  \textit{Nature}, \textbf{403~(6771)}, 756--758.

\bibitem[{{IPCC}(2021)}]{IPCC2021WGI}
{IPCC}, 2021: \textit{Climate Change 2021: The Physical Science Basis}.
  Cambridge University Press, \doi{10.1017/9781009157896}.

\bibitem[{{IPCC}(2023)}]{IPCC_AR6_SYR_SPM_2023}
{IPCC}, 2023: Climate change 2023: Synthesis report, summary for policymakers.
  Intergovernmental Panel on Climate Change.

\bibitem[{Ishii et~al.(2017)Ishii, Fukuda, Hirahara, Yasui, Suzuki,, and
  Sato}]{ishii2017accuracy}
Ishii, M., Y.~Fukuda, S.~Hirahara, S.~Yasui, T.~Suzuki, and K.~Sato, 2017:
  Accuracy of global upper ocean heat content estimation expected from present
  observational data sets. \textit{Sola}, \textbf{13}, 163--167.

\bibitem[{Jones et~al.(2009)}]{Jones_Holocene09}
Jones, P., and Coauthors, 2009: {High-resolution palaeoclimatology of the last
  millennium: a review of current status and future prospects}. \textit{The
  Holocene}, \textbf{19~(1)}, 3--49, \doi{10.1177/0959683608098952}.

\bibitem[{Kaufman et~al.(2020{\natexlab{a}})Kaufman, McKay, Routson, Erb,
  D{\"a}twyler, Sommer, Heiri,, and Davis}]{kaufman2020holocene}
Kaufman, D., N.~McKay, C.~Routson, M.~Erb, C.~D{\"a}twyler, P.~S. Sommer,
  O.~Heiri, and B.~Davis, 2020{\natexlab{a}}: {Holocene global mean surface
  temperature, a multi-method reconstruction approach}. \textit{Scientific
  Data}, \textbf{7}, 201, \doi{10.1038/s41597-020-0530-7}.

\bibitem[{Kaufman et~al.(2020{\natexlab{b}})}]{kaufman2020global}
Kaufman, D., and Coauthors, 2020{\natexlab{b}}: A global database of holocene
  paleotemperature records. \textit{Scientific data}, \textbf{7~(1)}, 115.

\bibitem[{Landrum et~al.(2013)Landrum, Otto-Bliesner, Wahl, Conley, Lawrence,
  Rosenbloom,, and Teng}]{landrum2013last}
Landrum, L., B.~L. Otto-Bliesner, E.~R. Wahl, A.~Conley, P.~J. Lawrence,
  N.~Rosenbloom, and H.~Teng, 2013: Last millennium climate and its variability
  in {CCSM4}. \textit{Journal of Climate}, \textbf{26~(4)}, 1085--1111,
  \doi{10.1175/JCLI-D-11-00326.1}.

\bibitem[{Le~Dimet and Talagrand(1986)Le~Dimet, and
  Talagrand}]{LeDimetTalagrand1986}
Le~Dimet, F.-X., and O.~Talagrand, 1986: Variational algorithms for analysis
  and assimilation of meteorological observations: Theoretical aspects.
  \textit{Tellus A}, \textbf{38A~(2)}, 97--110,
  \doi{10.3402/tellusa.v38i2.11706}.

\bibitem[{Lenssen et~al.(2019)Lenssen, Schmidt, Hansen, Menne, Persin, Ruedy,,
  and Zyss}]{lenssen2019improvements}
Lenssen, N.~J., G.~A. Schmidt, J.~E. Hansen, M.~J. Menne, A.~Persin, R.~Ruedy,
  and D.~Zyss, 2019: {Improvements in the {GISTEMP} uncertainty model}.
  \textit{Journal of Geophysical Research: Atmospheres}, \textbf{124~(12)},
  6307--6326.

\bibitem[{Liu and Nocedal(1989)Liu, and Nocedal}]{liu1989limited}
Liu, D.~C., and J.~Nocedal, 1989: On the limited memory bfgs method for large
  scale optimization. \textit{Mathematical programming}, \textbf{45~(1)},
  503--528.

\bibitem[{Lorenc(2003)}]{Lorenc2003}
Lorenc, A.~C., 2003: The potential of the ensemble kalman filter for nwp---a
  comparison with 4d-var. \textit{Quarterly Journal of the Royal Meteorological
  Society}, \textbf{129~(595)}, 3183--3203, \doi{10.1256/qj.02.132}.

\bibitem[{Mann et~al.(2009)}]{mann2009global}
Mann, M.~E., and Coauthors, 2009: {Global Signatures and Dynamical Origins of
  the Little Ice Age and Medieval Climate Anomaly}. \textit{Science},
  \textbf{326~(5957)}, 1256--1260.

\bibitem[{Mauritsen et~al.(2019)}]{mauritsen2019mpi}
Mauritsen, T., and Coauthors, 2019: Developments in the {MPI-M} earth system
  model version 1.2 ({MPI-ESM1.2}) and its response to increasing {CO2}.
  \textit{Journal of Advances in Modeling Earth Systems}, \textbf{11~(4)},
  998--1038, \doi{10.1029/2018MS001400}.

\bibitem[{Meng and Hakim(2024)Meng, and Hakim}]{meng2024reconstructing}
Meng, Z., and G.~J. Hakim, 2024: {Reconstructing the tropical Pacific upper
  ocean using online data assimilation with a deep learning model}.
  \textit{Journal of Advances in Modeling Earth Systems}, \textbf{16~(11)},
  e2024MS004\,422.

\bibitem[{Meng et~al.(2025)Meng, Hakim,, and Steig}]{meng2025coupled}
Meng, Z., G.~J. Hakim, and E.~J. Steig, 2025: Coupled seasonal data
  assimilation of sea ice, ocean, and atmospheric dynamics over the last
  millennium. \textit{Journal of Climate}, \textbf{38~(23)}, 7229--7247.

\bibitem[{Meng and Li(2024)Meng, and Li}]{meng2024pacific}
Meng, Z., and T.~Li, 2024: {Why is the Pacific meridional mode most pronounced
  in boreal spring?} \textit{Climate Dynamics}, \textbf{62~(1)}, 459--471.

\bibitem[{Meng and Polvani(2026)Meng, and
  Polvani}]{mengNoEvidenceStatistically2026}
Meng, Z., and L.~M. Polvani, 2026: No evidence of statistically significant
  eurasian winter warming or cooling caused by large eruptions over the last
  millennium. \textit{npj Climate and Atmospheric Science},
  \doi{10.1038/s41612-026-01535-0}.

\bibitem[{Meng et~al.(2023)Meng, Zhu,, and Hakim}]{meng2023sacpy}
Meng, Z., F.~Zhu, and G.~J. Hakim, 2023: {Sacpy--A Python Package for
  Statistical Analysis of Climate}. \textit{AGU23}.

\bibitem[{Nash and Sutcliffe(1970)Nash, and Sutcliffe}]{nash1970river}
Nash, J.~E., and J.~V. Sutcliffe, 1970: {River flow forecasting through
  conceptual models part I—A discussion of principles}. \textit{Journal of
  hydrology}, \textbf{10~(3)}, 282--290.

\bibitem[{Ning et~al.(2026)}]{ning2026progress}
Ning, L., and Coauthors, 2026: Progress and prospects of paleoclimate data
  assimilation. \textit{Science China Earth Sciences}, \textbf{69~(5)},
  1605--1617.

\bibitem[{Okazaki et~al.(2021)Okazaki, Miyoshi, Yoshimura, Greybush,, and
  Zhang}]{okazaki2021revisiting}
Okazaki, A., T.~Miyoshi, K.~Yoshimura, S.~J. Greybush, and F.~Zhang, 2021:
  {Revisiting online and offline data assimilation comparison for paleoclimate
  reconstruction: An idealized OSSE study}. \textit{Journal of Geophysical
  Research: Atmospheres}, \textbf{126~(16)}, e2020JD034\,214.

\bibitem[{O'Neill et~al.(2016)}]{o2016scenario}
O'Neill, B.~C., and Coauthors, 2016: The scenario model intercomparison project
  (scenariomip) for cmip6. \textit{Geoscientific Model Development},
  \textbf{9~(9)}, 3461--3482.

\bibitem[{Orsi et~al.(2012)Orsi, Cornuelle,, and Severinghaus}]{orsi2012little}
Orsi, A.~J., B.~D. Cornuelle, and J.~P. Severinghaus, 2012: {Little Ice Age
  cold interval in West Antarctica: evidence from borehole temperature at the
  West Antarctic Ice Sheet (WAIS) divide}. \textit{Geophysical Research
  Letters}, \textbf{39~(9)}.

\bibitem[{Osman et~al.(2021)Osman, Tierney, Zhu, Tardif, Hakim, King,, and
  Poulsen}]{osman2021globally}
Osman, M.~B., J.~E. Tierney, J.~Zhu, R.~Tardif, G.~J. Hakim, J.~King, and C.~J.
  Poulsen, 2021: Globally resolved surface temperatures since the last glacial
  maximum. \textit{Nature}, \textbf{599~(7884)}, 239--244.

\bibitem[{Otto-Bliesner et~al.(2016)}]{OttoBliesner_etal2016}
Otto-Bliesner, B.~L., and Coauthors, 2016: Climate variability and change since
  850 {CE}: An ensemble approach with the {Community Earth System Model}.
  \textit{Bulletin of the American Meteorological Society}, \textbf{97~(5)},
  735--754, \doi{10.1175/BAMS-D-14-00233.1}.

\bibitem[{{PAGES2k Consortium and others}(2017)}]{PAGES2k2017global}
{PAGES2k Consortium and others}, 2017: {A global multiproxy database for
  temperature reconstructions of the Common Era}. \textit{Scientific data},
  \textbf{4}.

\bibitem[{Parsons et~al.(2021)Parsons, Amrhein, Sanchez, Tardif, Brennan,, and
  Hakim}]{parsons2021multi}
Parsons, L.~A., D.~E. Amrhein, S.~C. Sanchez, R.~Tardif, M.~K. Brennan, and
  G.~J. Hakim, 2021: Do multi-model ensembles improve reconstruction skill in
  paleoclimate data assimilation? \textit{Earth and Space Science},
  \textbf{8~(4)}, e2020EA001\,467.

\bibitem[{Paszke et~al.(2019)}]{paszke2019pytorch}
Paszke, A., and Coauthors, 2019: Pytorch: An imperative style, high-performance
  deep learning library. \textit{Advances in neural information processing
  systems}, \textbf{32}.

\bibitem[{Penland and Magorian(1993{\natexlab{a}})Penland, and
  Magorian}]{penland1993prediction}
Penland, C., and T.~Magorian, 1993{\natexlab{a}}: {Prediction of Ni{\~n}o 3 sea
  surface temperatures using linear inverse modeling}. \textit{Journal of
  Climate}, \textbf{6~(6)}, 1067--1076.

\bibitem[{Penland and Magorian(1993{\natexlab{b}})Penland, and
  Magorian}]{penlandPredictionNinoSea1993}
Penland, C., and T.~Magorian, 1993{\natexlab{b}}: {Prediction of Niño 3 Sea
  Surface Temperatures Using Linear Inverse Modeling}. \textit{Journal of
  Climate}, \textbf{6~(6)}, 1067--1076,
  \doi{10.1175/1520-0442(1993)006<1067:PONSST>2.0.CO;2}.

\bibitem[{Penland and Matrosova(1994)Penland, and
  Matrosova}]{penland1994balance}
Penland, C., and L.~Matrosova, 1994: {A balance condition for stochastic
  numerical models with application to the El Ni{\~n}o-Southern Oscillation}.
  \textit{Journal of climate}, \textbf{7~(9)}, 1352--1372.

\bibitem[{Perkins and Hakim(2021)Perkins, and Hakim}]{perkins2021coupled}
Perkins, W., and G.~Hakim, 2021: {Coupled atmosphere--ocean reconstruction of
  the last millennium using online data assimilation}. \textit{Paleoceanography
  and Paleoclimatology}, \textbf{36~(5)}, e2020PA003\,959.

\bibitem[{Perkins and Hakim(2020)Perkins, and
  Hakim}]{perkinsLinearInverseModeling2020}
Perkins, W.~A., and G.~Hakim, 2020: {Linear Inverse Modeling for Coupled
  Atmosphere-Ocean Ensemble Climate Prediction}. \textit{Journal of Advances in
  Modeling Earth Systems}, \textbf{12~(1)}, e2019MS001\,778.

\bibitem[{Perkins and Hakim(2017)Perkins, and
  Hakim}]{perkins2017reconstructing}
Perkins, W.~A., and G.~J. Hakim, 2017: {Reconstructing paleoclimate fields
  using online data assimilation with a linear inverse model}. \textit{Climate
  of the Past}, \textbf{13~(5)}, 421--436.

\bibitem[{Pollack and Huang(2000)Pollack, and Huang}]{PollackHuang:2000}
Pollack, H.~N., and S.~Huang, 2000: Climate reconstruction from subsurface
  temperatures. \textit{Annual Review of Earth and Planetary Sciences},
  \textbf{28~(1)}, 339--365, \doi{10.1146/annurev.earth.28.1.339}.

\bibitem[{Rohde and Hausfather(2020)Rohde, and Hausfather}]{rohde2020berkeley}
Rohde, R.~A., and Z.~Hausfather, 2020: {The Berkeley Earth land/ocean
  temperature record}. \textit{Earth System Science Data}, \textbf{12~(4)},
  3469--3479.

\bibitem[{Santer et~al.(2000)Santer, Wigley, Boyle, Gaffen, Hnilo, Nychka,
  Parker,, and Taylor}]{santer2000significance}
Santer, B.~D., T.~M.~L. Wigley, J.~S. Boyle, D.~J. Gaffen, J.~J. Hnilo,
  D.~Nychka, D.~E. Parker, and K.~E. Taylor, 2000: Statistical significance of
  trends and trend differences in layer-average atmospheric temperature time
  series. \textit{Journal of Geophysical Research: Atmospheres},
  \textbf{105~(D6)}, 7337--7356, \doi{10.1029/1999JD901105}.

\bibitem[{Sasaki(1970)}]{Sasaki1970}
Sasaki, Y., 1970: Some basic formalisms in numerical variational analysis.
  \textit{Monthly Weather Review}, \textbf{98~(12)}, 875--883,
  \doi{10.1175/1520-0493(1970)098<0875:SBFINV>2.3.CO;2}.

\bibitem[{Steig and Orsi(2013)Steig, and Orsi}]{steigorsi2013}
Steig, E.~J., and A.~J. Orsi, 2013: The heat is on in antarctica.
  \textit{Nature Geoscience}, \textbf{6~(2)}, 87--88, \doi{10.1038/ngeo1717}.

\bibitem[{Steiger and Hakim(2016)Steiger, and
  Hakim}]{steiger2016multitimescale}
Steiger, N.~J., and G.~J. Hakim, 2016: {Multi-timescale data assimilation for
  atmosphere--ocean state estimates}. \textit{Climate of the Past},
  \textbf{12~(6)}, 1375--1388, \doi{10.5194/cp-12-1375-2016}.

\bibitem[{Steiger et~al.(2018)Steiger, Smerdon, Cook,, and
  Cook}]{steiger2018reconstruction}
Steiger, N.~J., J.~E. Smerdon, E.~R. Cook, and B.~I. Cook, 2018: {A
  reconstruction of global hydroclimate and dynamical variables over the Common
  Era}. \textit{Scientific data}, \textbf{5~(1)}, 1--15.

\bibitem[{Stiller and Hakim(2026)Stiller, and Hakim}]{stiller2026top}
Stiller, D., and G.~J. Hakim, 2026: Top-of-atmosphere radiation over the last
  millennium reconstructed from proxies. \textit{Journal of Climate}, e250568.

\bibitem[{Sun et~al.(2022)Sun, Lei, Liu, Ning,, and Tan}]{sun2022analog}
Sun, H., L.~Lei, Z.~Liu, L.~Ning, and Z.-M. Tan, 2022: {An analog offline EnKF
  for paleoclimate data assimilation}. \textit{Journal of Advances in Modeling
  Earth Systems}, \textbf{14~(5)}, e2021MS002\,674.

\bibitem[{Sun et~al.(2024)Sun, Lei, Liu, Ning,, and Tan}]{sun2024hybrid}
Sun, H., L.~Lei, Z.~Liu, L.~Ning, and Z.-M. Tan, 2024: A hybrid gain analog
  offline enkf for paleoclimate data assimilation. \textit{Journal of Advances
  in Modeling Earth Systems}, \textbf{16~(1)}, e2022MS003\,414.

\bibitem[{Sun et~al.(2025)Sun, Lei, Liu, Ning,, and Tan}]{sun2025online}
Sun, H., L.~Lei, Z.~Liu, L.~Ning, and Z.-M. Tan, 2025: An online paleoclimate
  data assimilation with a deep learning-based network. \textit{Journal of
  Advances in Modeling Earth Systems}, \textbf{17~(6)}, e2024MS004\,675.

\bibitem[{Sun et~al.(2026)Sun, Lei, Tan, Ning,, and Liu}]{sun2026ensemble}
Sun, H., L.~Lei, Z.-M. Tan, L.~Ning, and Z.~Liu, 2026: {An Ensemble Kalman
  Smoother for Online Paleoclimate Data Assimilation}. \textit{Journal of
  Advances in Modeling Earth Systems}, \textbf{18~(10)}, e2025MS005\,730,
  \doi{10.1029/2025MS005730}.

\bibitem[{Talagrand and Courtier(1987)Talagrand, and
  Courtier}]{TalagrandCourtier1987}
Talagrand, O., and P.~Courtier, 1987: Variational assimilation of
  meteorological observations with the adjoint vorticity equation. i: Theory.
  \textit{Quarterly Journal of the Royal Meteorological Society},
  \textbf{113~(478)}, 1311--1328, \doi{10.1002/qj.49711347812}.

\bibitem[{Tardif et~al.(2019)}]{tardif2019last}
Tardif, R., and Coauthors, 2019: {Last Millennium Reanalysis with an expanded
  proxy database and seasonal proxy modeling}. \textit{Climate of the Past},
  \textbf{15~(4)}, 1251--1273.

\bibitem[{Tierney et~al.(2025)Tierney, Judd, Osman, King, Truax, Steiger,
  Amrhein,, and Anchukaitis}]{tierney2025advances}
Tierney, J.~E., E.~J. Judd, M.~B. Osman, J.~M. King, O.~J. Truax, N.~J.
  Steiger, D.~E. Amrhein, and K.~J. Anchukaitis, 2025: {Advances in
  Paleoclimate Data Assimilation}. \textit{Annual Review of Earth and Planetary
  Sciences}, \textbf{53}.

\bibitem[{Tierney et~al.(2020)Tierney, Zhu, King, Malevich, Hakim,, and
  Poulsen}]{Tierney:Nature2020}
Tierney, J.~E., J.~Zhu, J.~King, S.~B. Malevich, G.~J. Hakim, and C.~J.
  Poulsen, 2020: Glacial cooling and climate sensitivity revisited.
  \textit{Nature}, \textbf{584~(7822)}, 569--573,
  \doi{10.1038/s41586-020-2617-x}.

\bibitem[{Toohey and Sigl(2017)Toohey, and Sigl}]{toohey2017volcanic}
Toohey, M., and M.~Sigl, 2017: {Volcanic stratospheric sulfur injections and
  aerosol optical depth from 500 BCE to 1900 CE}. \textit{Earth System Science
  Data}, \textbf{9~(2)}, 809--831.

\bibitem[{Tr{\'e}molet(2006)}]{tremoletAccountingImperfectModel2006}
Tr{\'e}molet, Y., 2006: Accounting for an imperfect model in 4d-var.
  \textit{Quarterly Journal of the Royal Meteorological Society},
  \textbf{132~(621)}, 2483--2504, \doi{10.1256/qj.05.224}.

\bibitem[{Valler et~al.(2022)Valler, Franke, Brugnara,, and
  Brönnimann}]{vallerUpdatedGlobalAtmospheric2022}
Valler, V., J.~Franke, Y.~Brugnara, and S.~Brönnimann, 2022: {An Updated
  Global Atmospheric Paleo‐reanalysis Covering the Last 400 Years}.
  \textit{Geoscience Data Journal}, \textbf{9~(1)}, 89--107,
  \doi{10.1002/gdj3.121}.

\bibitem[{Valler et~al.(2024)}]{valler2024mode}
Valler, V., and Coauthors, 2024: {ModE-RA: a global monthly paleo-reanalysis of
  the modern era 1421 to 2008}. \textit{Scientific data}, \textbf{11~(1)}, 36.

\bibitem[{Widmann et~al.(2010)Widmann, Goosse, van~der Schrier, Schnur,, and
  Barkmeijer}]{widmann2010using}
Widmann, M., H.~Goosse, G.~van~der Schrier, R.~Schnur, and J.~Barkmeijer, 2010:
  {Using data assimilation to study extratropical Northern Hemisphere climate
  over the last millennium}. \textit{Climate of the Past}, \textbf{6~(5)},
  627--644.

\bibitem[{Wilks(2011)}]{wilks2011statistical}
Wilks, D.~S., 2011: \textit{Statistical Methods in the Atmospheric Sciences}.
  3rd ed., Academic Press, Oxford.

\bibitem[{Wilks(2016)}]{wilks2016stippling}
Wilks, D.~S., 2016: {``The stippling shows statistically significant grid
  points'': How research results are routinely overstated and overinterpreted,
  and what to do about it}. \textit{Bulletin of the American Meteorological
  Society}, \textbf{97~(12)}, 2263--2273, \doi{10.1175/BAMS-D-15-00267.1}.

\bibitem[{Wu et~al.(2025)Wu, Gregory, Zanna,, and
  Khatiwala}]{wu2025energybudget}
Wu, Q., J.~M. Gregory, L.~Zanna, and S.~Khatiwala, 2025: Time-varying global
  energy budget since 1880 from a reconstruction of ocean warming.
  \textit{Proceedings of the National Academy of Sciences of the United States
  of America}, \textbf{122~(20)}, e2408839\,122, \doi{10.1073/pnas.2408839122}.

\bibitem[{Yukimoto et~al.(2019)}]{yukimoto2019mri}
Yukimoto, S., and Coauthors, 2019: The meteorological research institute earth
  system model version 2.0, {MRI-ESM2.0}: Description and basic evaluation of
  the physical component. \textit{Journal of the Meteorological Society of
  Japan. Ser. II}, \textbf{97~(5)}, 931--965, \doi{10.2151/jmsj.2019-051}.

\bibitem[{Zanna et~al.(2019)Zanna, Khatiwala, Gregory, Ison,, and
  Heimbach}]{zannaGlobalReconstructionHistorical2019a}
Zanna, L., S.~Khatiwala, J.~M. Gregory, J.~Ison, and P.~Heimbach, 2019: Global
  reconstruction of historical ocean heat storage and transport.
  \textit{Proceedings of the National Academy of Sciences}, \textbf{116~(4)},
  1126--1131, \doi{10.1073/pnas.1808838115}.

\bibitem[{Zhang et~al.(2025)Zhang, Li,, and Hu}]{zhang2025paleoclimate}
Zhang, H., M.~Li, and Y.~Hu, 2025: {Paleoclimate Data Assimilation: Principles
  and Prospects}. \textit{Science China Earth Sciences}, 1--18.

\bibitem[{Zhu et~al.(2019)}]{zhu2019climate}
Zhu, F., and Coauthors, 2019: Climate models can correctly simulate the
  continuum of global-average temperature variability. \textit{Proceedings of
  the National Academy of Sciences of the United States of America},
  \textbf{116~(18)}, 8728--8733, \doi{10.1073/pnas.1809959116}.

\end{thebibliography}
\end{document}